# REVIEW

# Singularity Theorems and Their Consequences

**José M. M. Senovilla**[1]

A detailed study of the singularity theorems is presented. I discuss the plausibility and reasonability of their hypotheses, the applicability and implications of the theorems, as well as the theorems themselves. The consequences usually extracted from them, some of them without the necessary rigour, are widely and carefully analysed with many clarifying examples and alternative views.

## 1. INTRODUCTION

It is now more than 30 years since the appearance of the Penrose singularity theorem [162], and more than 40 years since the fundamental Raychaudhuri equation and results were published [170]. Much work based on these important ideas was later developed and has become one of the main parts of classical general relativity: singularity theorems and related problems. It has even crossed the standard limits of general relativity and become a field of interest both for other branches of physics or mathematics and for the general public. This is due to its possible consequences concerning gravitational collapse, astrophysics and cosmology, which are very popular nowadays. The singularity theorems can be termed as one of the greatest achievements within relativity, but they contain some mathematical and physical subtleties which cause them to be, on some occasions, wrongly interpreted or not well-understood. The main conclusions of the singularity theorems are usually interpreted as providing evidence of the (classical) singular beginning of the Universe and of the singular final fate of some stars. To what extent this is truly so will be analysed here.

---

[1] Departament de Física Fonamental, Universitat de Barcelona, Diagonal 647, E-08028 Barcelona, Spain. E-mail: seno@hermes.ffn.ub.es

In fact, the singularity theorems do not exactly lead to such conclusions, but given that singularity theory in general relativity is a difficult matter, there has always been a wish to simplify the 'exact wording' of the theorems — while keeping the essentials by pinpointing the important clues — such that their assumptions and conclusions could be easily understood and applied to some simple situations. Unfortunately, all these simplifying attempts have not succeeded, even though sometimes they were devised very cleverly. I have the feeling that any other simplifying attempt yet to come is also doomed to failure, given that many explicit examples show that exact rigour — both in the assumptions and the conclusions — is absolutely necessary. Many such examples will be presented in this paper. The lesson is that one has to rely on the true actual theorems, and this is sometimes unpleasant as they need a rather good knowledge of the standard global and causality techniques in general relativity. One of the purposes of this paper is to collect all that is needed for the singularity theorems in a manner as easy and simple as possible. Everything will be treated within *classical* relativity; semi-classical theories, quantum phenomena or quantum gravity will not be taken into account, as they essentially change the physics behind the singularity theorems and thereby drastically modify their conclusions. Alternative classical gravitation theories will not be considered either, although many results are easily recovered there (see the complete review, Ref. 84).

The main questions to be treated here are: what do the singularity theorems say, and what do they not say? And, what are their actual consequences? To that end, on the one hand the paper is completely self-contained, and on the other hand it has many explicit examples illustrating the subtleties and problems behind most concepts and results. Concerning the first point, only standard basic general relativity is needed to be able to read the whole paper. The global causality theory, the concepts of maximal curves and its existence, the focusing of geodesics, and all that is basic for the singularity theorems, are fully developed step-by-step in Section 2. Of course, there are excellent references for these mathematical developments, such as [107,165]; besides, there exist very good and complete reviews, for instance [220]. Nevertheless, and even though Section 2 is essentially standard as there are no new results, I have tried to keep the level as simple as possible compatible with absolute rigour, and many of the proofs have been modified — and sometimes given in further detail — with respect to the standard references. People interested in these global techniques who nonetheless are not experts on them may find this section of some help, for most of the developments are carefully proved while keeping always the maximum simplicity. Moreover, the basic traditional structure has been partially reorganized so that every single result appears at its natural place, and some of the fundamental steps have been split into simpler smaller ones, making the way to the singularity theorems more practicable. However, readers familiar with the global techniques may find it more useful to skip Section 2 and go directly to Section 3.

The rest of the sections, with the exception of Section 5 which is devoted to the full proof of the theorems themselves, contain many original explicit and simple examples, together with a deep analysis of the concepts involved, showing how some intuitive or naive conclusions concerning singularity theorems may be false. The emphasis is placed on the alternative views with regard to what can be considered the standard opinion, and specially on the doors left open by the theorems in order to construct realistic and reasonable singularity-free spacetimes. Section 3 is devoted to the study, definition, classification and properties of singularities. Several explicit examples are shown, and the problem of choosing extensions of given incomplete spacetimes, and how to construct them, will be examined. The concept of a *singular extension* is introduced and shown to be both necessary in some situations and a non-orthodox alternative possibility in some other cases. A definition of big-bang singularity, together with the unusual properties they may have, is also provided. In Section 4 the fundamental concepts of trapped sets and closed trapped surfaces are introduced. This is mainly standard but some explicit examples are given. The consequences of their existence relevant to the singularity theorems are explicitly proved here. Again, Sections 4 and 5 may be skipped by connoisseurs.

The last two sections are completely original and may be controversial. Section 6 is devoted to a thorough analysis of the assumptions and conclusions of the theorems. A skeleton pattern-theorem for the singularity theorems is shown, and the intuitive ideas behind their proofs are explicitly described. A full detailed outspoken criticism of the singularity theorems is presented, considering each of its parts on its own, and ending with a summary of what I think their main unsolved weaknesses are. Most of this criticism is supported by many explicit and illustrative examples given in the last Section 7, with a total of nine different subsections. All the examples are simple and clear, and each of them sheds light on a particular part, or on a particular widespread — but incorrect — belief concerning the theorems; in other cases, the examples show which particular things are taken for granted, sometimes incorrectly. A tentative definition of a classical cosmological model is put forward, and many non-singular spacetimes are explicitly given. Sometimes, the existence of such examples has been considered surprising, but that is precisely the purpose: they show that many doors for constructing regular realistic spacetimes are still open. They also allow us to control the degrees of necessity of the theorem assumptions, proving manifestly that they cannot be relaxed and that they may be more demanding than is usually thought.

Before entering into the main subject, let us briefly review the notation to be used in this paper. The closure, interior and boundary of a set $\zeta$ are denoted by $\overline{\zeta}$, int $\zeta$ and $\partial \zeta$, respectively. The proper and normal inclusions are denoted by $\subset$ and $\subseteq$ as usual. Einstein's equations are written as

$$R_{\mu\upsilon} - \tfrac{1}{2} g_{\mu\upsilon} R = T_{\mu\upsilon},$$

where $T_{\mu\upsilon}$ is the energy-momentum tensor, $R_{\mu\upsilon}$ is the Ricci tensor and $R$ is the scalar curvature. The geometrized units $8\pi G = c = 1$ are used throughout the paper. An equality by definition is denoted by $\equiv$, while the end (or the absence) of a proof is signaled by ∎. Greek indices run from 0 to 3, small latin indices from 1 to 3 and capital latin indices take the values 2 and 3. The sign of the Riemann tensor $R^{\alpha}_{\beta\lambda\mu}$ is defined by eq. (26). Boldface letters and arrowed symbols indicate one-forms and vectors, respectively, and the exterior differential is denoted by $d$. Square or round brackets enclosing some indices denote the usual (anti-)symmetrization. The metric tensor field is denoted by $\mathbf{g}$, and the line-element in any local coordinate chart $\{x^\mu\}$ is the quantity

$$ds^2 = g_{\mu\upsilon} dx^\mu dx^\upsilon.$$

## 2. BASIC RESULTS IN GENERAL RELATIVITY AND CAUSALITY

The basic arena in General Relativity is the *spacetime*. A $C^k$ spacetime $(V_4, \mathbf{g})$ is a paracompact, connected, oriented, Hausdorff, 4-dimensional differentiable manifold $V_4$ endowed with a $C^k$ ($k \geq 2-$) Lorentzian metric $\mathbf{g}$ (signature $-,+,+,+$) [12,65,107,187,239]. A function $f$ is $C^{k-}$ if $f$ is $C^{k-1}$ and its $(k-1)$th derivatives are locally Lipschitz functions. The tangent space $T_p V_4$ at each point $p \in V_4$ has the typical Minkowski structure with the two-sheeted light cone, which classifies all non-zero vectors $\vec{v} \in T_p V_4$ into spacelike, timelike and null (also called lightlike) for $\mathbf{g}_p(\vec{v}, \vec{v})$ positive, negative or zero, respectively. Further, choosing an arbitrary timelike vector $\vec{t}$ as future-pointing at $p$, all timelike and null non-zero vectors in $T_p V_4$ are subdivided into future- and past-pointing according to whether $g|_p(\vec{t}, \vec{v})$ is negative or positive. If this choice can be made globally and continuously, then the spacetime is time-orientable. Not all spacetimes are necessarily so [12,41,107], but I shall assume throughout that the spacetime is time-oriented. This is equivalent to the existence of a globally defined continuous timelike vector field.

**Definition 2.1.** A hypersurface is the image of a continuous piecewise $C^3$ map $\Phi : \Sigma \longrightarrow V_4$ from an orientable 3-dimensional manifold $\Sigma$ into the spacetime. Similarly, a surface is the image of a continuous piecewise $C^3$ map $\Phi : S \longrightarrow V_4$ from an orientable 2-dimensional manifold $S$ into the spacetime. Here a continuous piecewise $C^3$ imbedding is a map $\Phi$ with inverse $\Phi^{-1}$ where both $\Phi$ and $\Phi^{-1}$ are continuous piecewise $C^3$ and such that $\Phi$ is a homeomorphism onto its image in the induced topology.

We shall identify $\Sigma$, $S$ and their images $\Phi(\Sigma)$, $\Phi(S)$ as is customary. Usually, the hypersurfaces and surfaces considered will be smooth (everywhere $C^3$), but in some cases we will need hypersurfaces with "corners" which will be explicitly stated. In any local coordinate system $\{x^\mu\}$, $(\mu, \upsilon, \ldots = 0, 1, 2, 3)$ the imbedding is simply given by the so-called parametric form, that is to say, by the equations

$$x^\alpha = \Phi^\alpha(u), \tag{1}$$

where $\{u^i\}$ (or $\{u^A\}$) are intrinsic coordinates in $\Sigma$ (or $S$), respectively $(i, j, \ldots = 1, 2, 3; A, B, \ldots = 2, 3)$, and the jacobian matrix is of rank three all over $\Sigma$ (resp. rank two over $S$). The hypersurface $\Sigma$ is also locally defined through the equation $F(x) = 0$ for some function $F$. Similarly, the surface $S$ is locally defined by two independent functions $f_1(x) = 0$ and $f_2(x) = 0$.

The intrinsic tangent vectors $\partial/\partial u^i$ in $\Sigma$ (or $\partial/\partial u^A$ in $S$) can be pushed forward to three (two) linearly independent vectors $\vec{e}_i$ ($\vec{e}_A$) on the manifold given by

$$\begin{aligned}\vec{e}_i &\equiv e_i^\mu \frac{\partial}{\partial x^\mu}\bigg|_\Sigma \equiv \frac{\partial \Phi^\mu}{\partial u^i} \frac{\partial}{\partial x^\mu}\bigg|_\Sigma, \\ \vec{e}_A &\equiv e_A^\mu \frac{\partial}{\partial x^\mu}\bigg|_S \equiv \frac{\partial \Phi^\mu}{\partial u^A} \frac{\partial}{\partial x^\mu}\bigg|_S.\end{aligned} \tag{2}$$

Note that these vectors are defined only on $\Sigma$ or $S$. We shall refer to them as the tangent vectors of $\Sigma$ or $S$. The first fundamental form of $\Sigma$ or $S$ in

$V_4$ is the pull-back of the spacetime metric $g_{\mu\nu}$ to $\Sigma$ or $S$, that is

$$\gamma_{ij} \equiv g_{\mu\nu}\bigg|_{\Sigma} \frac{\partial \Phi^{\mu}}{\partial u^i} \frac{\partial \Phi^{\nu}}{\partial u^j}, \qquad \gamma_{AB} \equiv g_{\mu\nu}\bigg|_{S} \frac{\partial \Phi^{\mu}}{\partial u^A} \frac{\partial \Phi^{\nu}}{\partial u^B}, \qquad (3)$$

which, as is obvious, gives the scalar products of the tangent vectors of $\Sigma$ or $S$ in the spacetime. Note again that the first fundamental form is defined only on $\Sigma$ or $S$. The (hyper-) surfaces are submanifolds whenever the topology induced from $V_4$ coincides with that defined by the imbedding.

Every non-zero one-form **n** defined on $\Sigma$ (or $S$) and orthogonal to all vectors tangent to $\Sigma$ (to $S$) is called a normal one-form [143]; the intrinsic characterization of any normal one-form is

$$n_\mu e_i^\mu = 0, \qquad n_\mu e_A^\mu = 0.$$

For a hypersurface the normal one-form is defined (except at corners!) up to a non-zero multiplicative factor and a sign. Locally, **n** can be written as $n_\mu \propto \eta_{\mu\nu\rho\sigma} e_1^\nu e_2^\rho e_3^\sigma$ where $\eta_{\mu\nu\rho\sigma}$ is the canonical volume 4-form in $V_4$; equivalently, **n** $\propto dF|_\Sigma$. The hypersurface is said to be timelike, spacelike or null at $p \in \Sigma$ if the normal one-form is spacelike, timelike or null at $p$. If **n** is timelike at $p$ then $\gamma_{ij}$ is positive definite at $p$. If **n** is null at $p$, then the normal vector $\vec{n}$ (defined by raising the index of **n**) is in fact tangent to $\Sigma$. In this case, the first fundamental form is degenerate at $p$ [143]. When **n** changes its character from point to point $\Sigma$ is called a general hypersurface [143]. For a surface $S$, there are two linearly independent normal one-forms defined up to non-singular linear combinations. We will assume that all surfaces are orientably embedded, in the sense that these normal one-forms can be chosen continuously over the whole surface. A surface is spacelike if there exists a normal one-form which is timelike. In this case, both independent normal one-forms can be chosen to be null everywhere on $S$. Therefore, we shall define the two null normal one-forms $\mathbf{k}^{\pm}$ as two future-pointing one-forms such that

$$k_\mu^\pm e_A^\mu = 0, \qquad k_\mu^+ k^{+\,\mu} = 0, \qquad k_\mu^- k^{-\,\mu} = 0, \qquad k_\mu^+ k^{-\,\mu} = -1, \qquad (4)$$

where the last of these relations is a condition of normalization. Of course, $\mathbf{k}^{\pm}$ are not completely determined by (4), and there still remains the freedom of changing them as follows:

$$k_\mu^+ \longrightarrow k_\mu'^+ = A^2 k_\mu^+, \qquad k_\mu^- \longrightarrow k_\mu'^- = A^{-2} k_\mu^-, \qquad (5)$$

where $A^2$ is an arbitrary non-vanishing function defined only on $S$.

For any normal one-form, the corresponding second fundamental form can be defined. This is a 2-covariant symmetric tensor $K$ defined only on $\Sigma$ (or $S$, resp.) by

$$K_{ij}(\mathbf{n}) \equiv -n_\mu e_i^\upsilon \nabla_\upsilon e_j^\mu = e_j^\mu e_i^\upsilon \nabla_\upsilon n_\mu,$$
$$K_{AB}(\mathbf{n}) \equiv -n_\mu e_A^\upsilon \nabla_\upsilon e_B^\mu = e_B^\mu e_A^\upsilon \nabla_\upsilon n_\mu, \qquad (6)$$

where the covariant derivative is that of the spacetime and the second equality is well-posed — even though the normal one-form is only defined on the submanifold — because we only need to compute the tangent derivatives.

**Definition 2.2.** A $C^k$ arc is (the image of) a $C^k$ ($k \geq 1$) map $\lambda : I \longrightarrow V_4$ from an interval $I \subseteq \mathbb{R}$ into the spacetime such that the derivative of $\lambda$ does not vanish anywhere on I. If I has compact closure the arc is called a segment. A continuous piecewise $C^k$ curve is a finite sequence of $C^k$ arcs $\{\lambda_i\}$, $i = 1, \ldots, n$, each defined on a real interval $I_i$, such that the last point of each $\lambda_i$ coincides with the first point of the next $\lambda_{i+1}$.

Segments and curves can be open, half-open or closed. However, each segment of a curve is closed except for the first and the last arcs, which may be left- and right-open, respectively. I have assumed that the derivative of each arc does not vanish anywhere, so that the tangent vector field $\vec{v}|_\lambda$ along $\lambda$ is not zero everywhere. For continuous piecewise $C^k$ curves $\gamma$, the tangent vector field is unique everywhere except at points joining differentiable arcs, called *corners*, where there may be two such tangent vectors. A curve $\gamma$ is said to be timelike (resp. null, causal, spacelike) if $\vec{v}|_\gamma$ is timelike (resp. null, non-spacelike, spacelike) everywhere. By continuity, causal curves cannot change their time-orientation in any of their arcs. Thus, a causal curve is called future- (past-) directed if the tangent vectors at all their corners are future- (past-) pointing.

In a local coordinate system $\{x^\mu\}$, any arc is expressed by the differentiable functions $x^\mu = \lambda^\mu(u)$, $u \in \mathbb{R}$. The tangent vector field is locally expressed by

$$\vec{v}|_\lambda = \left. \frac{d\lambda^\mu(u)}{du} \frac{\partial}{\partial x^\mu} \right|_\lambda.$$

**Definition 2.3.** A geodesic is a $C^2$ arc whose tangent vector field satisfies $v^\mu \nabla_\mu v^\upsilon \propto v^\upsilon$.

Equivalently, in any local chart a geodesic satisfies

$$\frac{d^2\lambda^\mu}{du^2} + \Gamma^\mu_{\upsilon\sigma}|_\lambda \frac{d\lambda^\upsilon}{du} \frac{d\lambda^\sigma}{du} = A(u) \frac{d\lambda^\mu}{du}, \qquad (7)$$

where $\Gamma^{\mu}_{\nu\sigma}$ are the connection coefficients in these coordinates and $A$ is a function. By means of a suitable change of the parameter $\tau = \tau(u)$ it is always possible to make the righthand side of (7) vanish. Such parameters are defined up to a linear transformation $\tau \rightarrow a\tau + b$ and are called affine parameters. Thus, in any affine parameter $\tau$, eq. (7) becomes (where $a \neq 0$)

$$\frac{d^2\lambda^{\mu}}{d\tau^2} + \Gamma^{\mu}_{\nu\sigma}|_{\lambda}\frac{d\lambda^{\nu}}{d\tau}\frac{d\lambda^{\sigma}}{d\tau} = 0. \tag{8}$$

It is trivial to show that the tangent vector $\vec{v}$ of an affinely parametrized geodesic has constant modulus, that is $g(\vec{v}, \vec{v})$ is constant, and hence we can speak of timelike, null and spacelike geodesics.

Looking at (8) as a system of second order linear ordinary differential equations (ODE), and given that $\Gamma^{\mu}_{\nu\sigma}$ are at least locally Lipschitz functions, from the classical theory of ODE [7] it follows that given any $p \in V_4$ and any vector $\vec{v} \in T_p V_4$ there is an interval $I \subseteq \mathbb{R}$ and a unique solution of (8) defined on I which passes through $p$ and has $\vec{v}$ as tangent vector at $p$. Collecting all these solutions they can be locally denoted by

$$x^{\mu} = G^{\mu}(\tau; p, \vec{v}), \qquad \tau \in [0, c),$$

$$G^{\mu}(0; p, \vec{v}) = x^{\mu}|_p, \qquad \frac{dG^{\mu}}{d\tau}(0; p, \vec{v}) = v^{\mu}.$$

The fundamental concept for the singularity theorems is the following

**Definition 2.4.** A geodesic from $p \in V_4$ is complete if it is defined for all values $\tau \in [0, \infty)$ of its affine parameter. A spacetime is geodesically complete at $p$ if all geodesics emanating from $p$ are complete. A spacetime is geodesically complete if it is so for all $p \in V_4$.

This notion can be refined by defining timelike, null, causal and spacelike geodesic completeness in a natural way. Most singularity theorems prove the existence of at least one incomplete causal geodesic. The relationship between this concept and that of a singularity is not clear at all, and will be studied in Section 3 in some detail.

The "exponential" map from an open neighbourhood $\mathcal{O}$ of $\vec{0} \in T_p V_4$ into a neighbourhood of $p \in V_4$ maps a given $\vec{v} \in \mathcal{O}$ into the point $G^{\mu}(1; p, \vec{v})$, provided this is defined. As the functions $G^{\mu}$ depend continuously on the initial conditions $p$ and $\vec{v}$, by choosing adequate neighbourhoods this exponential map is a homeomorphism (and a diffeomorphism if $g$ is $C^2$).

**Definition 2.5.** The functions $x^{\mu} = G^{\mu}(1; p, X^{\nu})$ define a new system of coordinates $\{X^{\nu}\}$ in a neighbourhood of $p$ which are called Riemannian normal coordinates based at $p$. Any such neighbourhood is called a normal neighbourhood of $p$. A maximal normal neighbourhood of $p$ will be denoted by $\mathcal{N}_p$ [65].

In fact, this normal neighbourhood can be chosen to be convex [243,81], and we shall do so sometimes. However, it must be realized that $\mathcal{N}_p$ is usually bigger than any convex normal neighbourhood of $p$. The above change of coordinates must be at least of class $C^1$ to keep a differentiable atlas, and consequently the functions $G^\mu$ should depend differentiably on the initial conditions. This will happen in most situations, for example when the metric is $C^2$. Nevertheless, many physical situations require the matching of two different spacetimes across a common boundary — for instance, the exterior vacuum field of a star with its interior fluid, or a wave travelling upon a background spacetime, etcetera. When this matching is performed, it is well-known that there exists a local coordinate system, called admissible in [132], in which the metric is $C^1$ piecewise $C^2$ (see e.g. Refs. 132,143). Thus, in these situations there is no guarantee that the normal coordinate neighbourhoods are well defined in the sense that the change to normal coordinates is not differentiable at the matching hypersurface. This is *crucial* for causality and, thereby, to the whole theory of singularities and singularity theorems. As far as I know, there are very few statements concerning singularities which have been proven to hold under the general case of a $C^{2-}$ metric (see, however, Refs. 37,45,46,107). In my opinion, this is a weakness of singularity theory at present and will be analysed in several places below. However, in order to proceed, the metric will be taken $C^2$ when needed, and a full list of places where this assumption is essential will be given in Section 6.

Several properties which will be needed later can be easily proved using normal coordinates $\{X^\mu\}$. First of all, we note the property [41]

$$G^\mu(m\,\tau;\, p,\, \vec{v}) = G^\mu(\tau;\, p,\, m\,\vec{v}),$$

for constant $m$. This follows because both functions satisfy the same differential equation (8) with identical initial conditions, so that they must be the same. Using this, we can write on each geodesic $x^\mu = G^\mu(\tau;\, p,\, \vec{v}) = G^\mu(1;\, p,\, \tau\vec{v})$ and then, remembering the Definition 2.5 of normal coordinates, we obtain

$$X^\mu = v^\mu\,\tau \qquad (9)$$

for the geodesics emanating from $p \in V_4$, tangent to $\vec{v}$ at $p$, and with affine parameter $\tau$. Hence, in normal coordinates the tangent vector to each geodesic is given by $v^\mu \partial/\partial X^\mu$ and then

$$g_{\mu\nu}(\vec{v}\tau)v^\mu v^\nu = g_{\mu\nu}(0)v^\mu v^\nu = \text{const.} \qquad (10)$$

from where, multiplying by $\tau^2$, it immediately follows

$$g_{\mu\nu}(X)X^\mu X^\nu = g_{\mu\nu}(0)X^\mu X^\nu, \qquad \forall X^\mu.$$

Obviously normal coordinates are not unique, but any two sets of normal coordinates based at $p$ are related by a linear homogeneous transformation of type $Y^\mu = A^\mu_\nu X^\nu$, with constant non-singular matrix $A^\mu_\nu$. By means of such a transformation the metric at $p$ can be brought into the standard Minkowskian form $g_{\mu\nu}(0) = \eta_{\mu\nu} = \text{diag}(-1,1,1,1)$. Such coordinates will be called Minkowskian normal coordinates [165]. In general normal coordinates, the combination of (8) and (9) provides $\Gamma^\sigma_{\mu\nu}(\vec{v}\tau)v^\mu v^\nu = 0$, and from this we get

$$\Gamma^\sigma_{\mu\nu}(0) = 0; \qquad \Gamma^\sigma_{\mu\nu}(X)X^\mu X^\nu = 0, \qquad \forall X^\mu.$$

**Lemma 2.1.** Within a normal neighbourhood of any $p \in V_4$, the set of all unit timelike geodesics emanating from $p$ with affine parameter $\tau$ are orthogonal to the spacelike hypersurfaces $\tau = \text{const}$. Further, in normal coordinates the Thomas formula [210,81] holds:

$$g_{\mu\nu}(X)X^\mu = g_{\mu\nu}(0)X^\mu, \qquad \forall X^\mu. \tag{11}$$

*Proof.* It has been shown before that $\Gamma^\sigma_{\mu\nu}(\vec{v}\tau)v^\mu v^\nu = 0$, which is equivalent to

$$(2\partial_\mu g_{\nu\rho} - \partial_\rho g_{\mu\nu})(\vec{v}\tau)v^\mu v^\nu = 0.$$

Using this, the derivative of (10) with respect to $v^\rho$ becomes

$$0 = \frac{\partial}{\partial v^\rho}[g_{\mu\nu}(\vec{v}\tau)v^\mu v^\nu - g_{\mu\nu}(0)v^\mu v^\nu] = 2\frac{d}{d\tau}[g_{\rho\mu}(\vec{v}\tau)v^\mu\tau - g_{\rho\mu}(0)v^\mu\tau],$$

which proves (11) as this is for arbitrary $\vec{v}$. Take now the set of all unit timelike geodesics from $p$ in normal coordinates, that is, (9) together with $g_{\mu\nu}(\vec{v}\tau)v^\mu v^\nu = g_{\mu\nu}(0)v^\mu v^\nu = -1$. This set defines a timelike congruence (see subsection 2.1 or Refs. 60,65,66,99,123,187,239) in the domain $g_{\mu\nu}(0)X^\mu X^\nu < 0$ within $\mathcal{N}_p$, whose unit tangent vector field is

$$\vec{u} = \frac{X^\mu}{\sqrt{-g_{\rho\nu}(0)X^\rho X^\nu}}\frac{\partial}{\partial X^\mu}.$$

The corresponding one-form (or covariant vector) field is then

$$\mathbf{u} = g_{\mu\nu}(X)\frac{X^\mu}{\sqrt{-g_{\rho\nu}(0)X^\rho X^\nu}}dX^\nu$$

$$= g_{\mu\nu}(0)\frac{X^\mu}{\sqrt{-g_{\rho\nu}(0)X^\rho X^\nu}}dX^\nu = -d(\sqrt{-g_{\mu\nu}(0)X^\mu X^\nu}),$$

where we have used (11). But this means that the congruence is orthogonal to the hypersurfaces $\sqrt{-g_{\mu\nu}(0)X^\nu X^\nu} = $ const., which are the hypersurfaces $\tau = $ const. ∎

One of the most important results in causality theory is that the causal properties of spacetime is *locally* equivalent to those of flat Minkowski spacetime. To be precise, let us define the future light cone $\partial C_p^+$ (resp. its interior $C_p^+$) of $p \in V_4$ as the image of the future light cone (resp. its interior) in $\mathcal{O} \subset T_p V_4$ by the exponential map (hence, it is only defined on $\mathcal{N}_p$). Then, we have the following fundamental proposition:

**Proposition 2.1.** *Any continuous piecewise $C^1$ future-directed causal curve starting at $p$ and entirely contained in the normal neighbourhood $\mathcal{N}_p$ of $p$ lies completely on the future light cone of $p$ if and only if it is a null geodesic from $p$, and is completely contained in the interior of the future light cone of $p$ after the point at which it fails to be a null geodesic.*

It must be remarked that a non-differentiable curve composed by segments of null geodesics is *not* a null geodesic. Thus, any future-directed (f-d) curve which has a segment of null geodesic from $p$ to $q$, and then changes in a corner to *another* null geodesic from $q$, lies on $\partial C_p^+$ up to $q$ and is in $C_p^+$ from $q$ on. Notice also that any f-d causal curve which is not a null geodesic at $p$ immediately enters and remains in the interior $C_p^+$. In particular, all f-d timelike curves from $p$ are completely contained in this interior. The result holds also with 'future' changed to 'past'. Our proposition is a refinement of the standard Hawking–Ellis proposition 4.5.1 (page 103 of Ref. 107) inspired by results appearing in [17], where the relationship between this causal statement and the classical Fermat principle was studied. We present here the proof of Proposition 2.1 as it is fundamental but nonetheless not easily available (see Refs. 17,107).

*Proof.* In normal coordinates, $\partial C_p$ and $C_p$ [both future (+) and past (−)] are given respectively by

$$g_{\mu\nu}(0)X^\mu X^\nu = 0, \qquad g_{\mu\nu}(0)X^\mu X^\nu < 0.$$

Let us first show that any f-d causal curve $\gamma$ which is timelike at $p$ is wholly contained in $C_p^+$. Let $X^\mu = \gamma^\mu(u)$ be the parametric form of $\gamma$ with $\gamma^\mu(0) = 0$. Its tangent vector field is

$$\vec{v}|_\gamma = \left.\frac{d\gamma^\mu}{du}\frac{\partial}{\partial X^\mu}\right|_\gamma \equiv \left.v^\mu(u)\frac{\partial}{\partial X^\mu}\right|_\gamma, \qquad (12)$$

$$g_{\mu\nu}(\gamma)v^\mu(u)v^\nu(u) \leq 0, \qquad g_{\mu\nu}(0)v^\mu(0)v^\nu(0) < 0.$$

We must show that

$$g_{\mu\nu}(0)\gamma^\mu(u)\gamma^\nu(u) < 0, \qquad \forall u > 0 \quad \text{such that } \gamma(u) \in \mathcal{N}_p. \tag{13}$$

This certainly holds for small enough values of $u > 0$, because the last relation in (12) and

$$\lim_{u \to 0} \frac{\gamma^\mu(u)}{u} = \frac{d\gamma^\mu}{du}(0) = v^\mu(0)$$

imply, by continuity, that there is a non-empty interval $(0, \bar{u}]$ in which

$$g_{\mu\nu}(0) \frac{\gamma^\mu(u)}{u} \frac{\gamma^\nu(u)}{u} < 0 \implies g_{\mu\nu}(0)\gamma^\mu(u)\gamma^\nu(u) < 0, \quad u \in (0, \bar{u}].$$

This *also* means that the future-pointing vector field $\gamma^\mu(u)\,\partial/\partial X^\mu|_\gamma$ is timelike for $u \in (0, \bar{u}]$, given that $g_{\mu\nu}(0)\gamma^\mu(u)\gamma^\nu(u) = g_{\mu\nu}(\gamma)\gamma^\mu(u)\gamma^\nu(u)$. Then (12) and the Thomas formula (11) lead to

$$\frac{d}{du}[g_{\mu\nu}(0)\gamma^\mu(u)\gamma^\nu(u)] = 2g_{\mu\nu}(0)\gamma^\mu(u)v^\nu(u)$$
$$= 2g_{\mu\nu}(\gamma)\gamma^\mu(u)v^\nu(u) < 0, \qquad u \in (0, \bar{u}],$$

where the last inequality follows because both $v^\nu(u)$ (non-spacelike) and $\gamma^\mu(u)$ (timelike) are future-pointing at the points $X^\mu = \gamma^\mu(u)$. This is also true even at corners of the curve $\gamma$ as both tangent vectors at the corner are future-pointing — there may be a jump, but *not* a change of sign, in the derivative of $g_{\mu\nu}(0)\gamma^\mu(u)\gamma^\nu(u)$. Thus, the continuous function $g_{\mu\nu}(0)\gamma^\mu(u)\gamma^\nu(u)$ is strictly decreasing in any interval $(0, u]$ in which it is negative, and (13) follows.

An immediate consequence is that any causal curve $\gamma$ that enters in $C_p^+$ remains in $C_p^+$, because given any point $q \in \gamma \cap C_p^+$ there is a timelike geodesic from $p$ to $q$. The combination of this geodesic segment together with $\gamma$ from $q$ is an f-d causal curve of the type considered before (timelike at $p$). In consequence, all points in $\gamma$ after $q$ are in $C_p^+$. Similarly, if any causal f-d curve $\gamma$ is timelike at any $q \in \gamma \cap \partial C_p^+$, then it enters and remains in $C_p^+$ from $q$ on. To prove this notice that $g_{\mu\nu}(0)\gamma^\mu(u_q)\gamma^\nu(u_q) = g_{\mu\nu}[\gamma(u_q)]\gamma^\mu(u_q)\gamma^\nu(u_q) = 0$ where $\gamma(u_q) = q$. Given that $\vec{v}|_q$ is timelike and future-pointing it follows $g_{\mu\nu}[\gamma(u_q)]\gamma^\mu(u_q)v^\nu(u_q) < 0$. Then using (11)

$$\left.\frac{d}{du}[g_{\mu\nu}(0)\gamma^\mu(u)\gamma^\nu(u)]\right|_{u_q} = 2g_{\mu\nu}(0)\gamma^\mu(u_q)v^\nu(u_q)$$
$$= 2g_{\mu\nu}[\gamma(u_q)]\gamma^\mu(u_q)v^\nu(u_q) < 0,$$

which means that $\gamma$ enters in $C_p^+$. The same reasoning serves to prove that any f-d causal curve having a corner at a point $q \in \partial C_p^+$ must enter and remain in $C_p^+$ from the corner on.

Let us then show[2] that any f-d causal curve $\gamma$ is always in the closure $\overline{C_p^+} = C_p^+ \cup \partial C_p^+$ of $C_p^+$ in $\mathcal{N}_p$, defined by the 'upper' half of $g_{\mu\nu}(0) X^\mu X^\nu \leq 0$. We keep the same assumptions and notation as above except for the last inequality in (12) which must be replaced by $g_{\mu\nu}(0) v^\mu(0) v^\nu(0) \leq 0$. Take any fixed future-pointing $C^1$ timelike vector field $\vec{w}|_\gamma$ along $\gamma$ and define the functions $\psi^\mu(u)$ as the unique $C^2$- solutions of

$$\frac{d\psi^\mu}{du} + \frac{1}{2} g^{\mu\nu}(\gamma) \partial_\rho g_{\nu\sigma}(\gamma) \psi^\rho(u) v^\sigma(u) = w^\mu(u), \qquad \psi(0) = 0. \qquad (14)$$

Let us construct the one-parameter family of curves $\{\gamma_\varepsilon\}$ starting at $p$,

$$\gamma_\varepsilon^\mu(u) \equiv \gamma^\mu(u) + \varepsilon \psi^\mu(u), \qquad \varepsilon \geq 0,$$

whose tangent vector fields $\vec{v}_\varepsilon$ are, for each $\varepsilon$, $v_\varepsilon^\mu(u) = v^\mu(u) + \varepsilon d\psi^\mu/du$. On using this and (14), an elementary calculation gives

$$\frac{d}{d\varepsilon} [g_{\mu\nu}(\gamma_\varepsilon) v_\varepsilon^\mu(u) v_\varepsilon^\nu(u)]\bigg|_{\varepsilon=0} = 2 g_{\mu\nu}(\gamma) v^\mu(u) w^\nu(u) < 0$$

and as $\gamma_0 = \gamma$ is causal this implies that there is $\bar\varepsilon > 0$ such that

$$g_{\mu\nu}(\gamma_\varepsilon) v_\varepsilon^\mu(u) v_\varepsilon^\nu(u) < 0, \qquad \varepsilon \in (0, \bar\varepsilon],$$

which means that $\{\gamma_\varepsilon\}$ are timelike, or equivalently, that the entire curves $\{\gamma_\varepsilon\}$ are in $C_p^+$ for $\varepsilon \in (0, \bar\varepsilon]$. Thus, the initial curve $\gamma$ is in $\overline{C_p^+}$ because for every point $q = \gamma(u_q)$, every neighbourhood of $q$ includes some of the points $\gamma_\varepsilon(u_q)$, which are in $C_p^+$.

It only remains to show that a causal curve lies on $\partial C_p^+$ if and only if it is a null geodesic from $p$. The direct implication holds by definition, so that we must only prove that any causal curve $\gamma$ starting at $p$ and lying on $\partial C_p^+$ must be a null geodesic. The assumptions are (same notation as above with $\gamma^\mu(0) = 0$)

$$g_{\mu\nu}(0) \gamma^\mu(u) \gamma^\nu(u) = 0, \qquad g_{\mu\nu}(\gamma) v^\mu(u) v^\nu(u) \leq 0.$$

---

[2] The proof in [107] is not completely correct as their $\beta(r, \varepsilon)$ are not timelike curves in general.

The first of these and Thomas formula (11) tell us that $g_{\mu\nu}(\gamma)\gamma^{\mu}(u)\gamma^{\nu}(u) = 0$, that is, the vector field $\gamma^{\mu}(u)\,\partial/\partial X^{\mu}|_{\gamma}$ is null. Differentiating and using again (11) we also obtain

$$g_{\mu\nu}(0)\gamma^{\mu}(u)v^{\nu}(u) = 0 \implies g_{\mu\nu}(\gamma)\gamma^{\mu}(u)v^{\nu}(u) = 0.$$

Therefore, $\vec{v}|_{\gamma}$ is null and proportional to $\gamma^{\mu}(u)\,\partial/\partial X^{\mu}|_{\gamma}$, that is $\gamma^{\mu}(u) = b(u)\,v^{\mu}(u)$ for some function $b$, with $b(0) = 0$. But this immediately implies that $\gamma$ is a null geodesic from $p$. ∎

Clearly, all definitions and properties concerning the 'future' have their corresponding counterpart for the 'past', and can be omitted in what follows.

**Definition 2.6.** A point $p \in V_4$ is a right endpoint of a curve $\gamma : I \to V_4$ if for every neighbourhood $\mathcal{U}$ of $p$ there exists $u_0 \in I$ such that $\gamma(u) \in \mathcal{U}$ for every $u \in I$ with $u \geq u_0$. Similarly for left endpoint ($u \leq u_0$). In the first case, $p$ is a future endpoint if $\gamma$ is causal and future-directed. If $\gamma$ has no future endpoint is called future endless or inextendible.

Notice that endpoints are unique for causal curves. Sometimes the single term endless is used for causal curves which are both future and past endless. The idea behind this definition is that a curve may seem to terminate at the endpoint $p$, but then there is another 'bigger' curve — an extension — containing both $p$ and the initial $\gamma$. On the other hand, if a curve is endless it runs into the 'edge' of the spacetime, be it infinity or something more awkward. These awkward places may be singularities; see Section 3.

**Definition 2.7.** Let $\{\gamma_n\}$ be an infinite sequence of curves. A point $p \in V_4$ is an accumulation point of $\{\gamma_n\}$ if every neighbourhood $\mathcal{U}$ of $p$ intersects an infinite number of the $\{\gamma_n\}$, and is a convergence point if every $\mathcal{U}$ intersects all but a finite number of the $\{\gamma_n\}$. The sequence $\{\gamma_n\}$ converges to the curve $\gamma$ if every $p \in \gamma$ is a convergence point of $\{\gamma_n\}$. Finally, $\gamma$ is a limit curve of $\{\gamma_n\}$ if there is a subsequence $\{\gamma_m\}$ converging to $\gamma$.

In general, any sequence $\{\gamma_n\}$ may have many limit curves, or no limit curve at all. Furthermore, even if all the curves in the sequence are causal, the limit curves are not necessarily so (although this is related to violations of the strong causality condition; see below). There may appear continuous curves as limits of differentiable curves, so that the more general concept of a continuous f-d causal curve is needed.

**Definition 2.8.** A continuous curve $\gamma : I \to V_4$ is said to be causal and future-directed if for every $u_0 \in \text{int } I$ there is a subinterval $J \subseteq I$ including $u_0$ and a normal neighbourhood $\mathcal{U}_{\gamma(u_0)} \supset \gamma|_J$ of $\gamma(u_0)$ such that, for every pair $u_1, u_2 \in J$ with $u_1 < u_2$, there is an f-d causal arc within $\mathcal{U}_{\gamma(u_0)}$ from $\gamma(u_1)$ to $\gamma(u_2)$.

These curves are equivalent under continuous monotonic reparametrizations. It is a trivial exercise to show that the causal f-d continuous curves of the above definition satisfy the local Lipschitz condition, so that they are differentiable almost everywhere [165].

**Proposition 2.2.** Let $p \in V_4$ be an accumulation point of a sequence of future-endless causal curves $\{\gamma_n\}$. Then there is a causal future-endless limit curve $\gamma$ of the $\{\gamma_n\}$ such that $p \in \gamma$.

*Proof.* The complete proof is rather large and technical (see Refs. 12,107), so that a sketch is given here. Let $\mathcal{N}_p$ be the normal neighbourhood of $p$ and choose Minkowskian normal coordinates $\{X^\mu\}$ in $\mathcal{N}_p$. Take the sets

$$\mathcal{B}_r \equiv \{X^\mu \mid (X^0)^2 + (X^1)^2 + (X^2)^2 + (X^3)^2 \leq r^2\} \tag{15}$$

for all rational numbers $0 \leq r \leq \tilde{r}$ and such that $\mathcal{B}_{\tilde{r}} \subset \mathcal{N}_p$. Choose any subsequence of the $\{\gamma_n\} \cap \mathcal{N}_p$ converging to $p$, which must have accumulation points in the compact set $\partial\mathcal{B}_{\tilde{r}}$. Due to the local causal structure of Proposition 2.1, any such accumulation point $p_{\tilde{r}}$ must be in $C_p \cap \partial\mathcal{B}_{\tilde{r}}$. Choose a sub-subsequence converging to $p_{\tilde{r}} \in C_p^+ \cap \partial\mathcal{B}_{\tilde{r}}$ and proceed in order in the same manner for all $r$, picking up points $p_r \in C_p^+ \cap \partial\mathcal{B}_r$ which are of accumulation for the previous subsequence. The closure of the union of all $p_r$ gives an f-d causal curve according to Definition 2.8, which has a future endpoint at $p_{\tilde{r}}$ and which is a limit curve of the sequence $\{\gamma_n\} \cap \mathcal{B}_{\tilde{r}}$. Repeating the entire process at $p_{\tilde{r}}$, and then at the next endpoints which appear in succesion, the limit curve is extended beyond any possible endpoint. ∎

### 2.1. Geodesic congruences, focal points and energy conditions

**Definition 2.9.** A congruence of curves in a domain $\mathcal{D} \subseteq V_4$ is a three-parameter family of curves such that there is one and only one curve of the family passing through each $p \in \mathcal{D}$. The vector field formed by the vectors tangent to the curves is called the tangent vector field. A congruence is timelike (resp. null) if the tangent vector field is timelike (resp. null) in $\mathcal{D}$.

The tangent vector field of a congruence is defined up to an arbitrary multiplicative non-zero function. However, for timelike congruences we can always choose the *unit* tangent vector field $\vec{u}$ satisfying

$$g(\vec{u}, \vec{u}) = -1, \tag{16}$$

and then the affine parameter $\tau$ along the curves is defined by

$$u^\mu \partial_\mu \tau = 1. \tag{17}$$

Obviously, affine parameters are defined up to the addition of an arbitrary first integral of $\vec{u}$ — this is the freedom of choosing the origin of affine parameter on each curve. Conversely, any unit timelike vector field defines locally a timelike congruence by solving the differential equations

$$\frac{dx^\mu}{d\tau} = u^\mu(x^\upsilon)$$

in any local chart. The solution depends in principle on four arbitrary constants, but one of them is spurious and can be absorbed into the affine parameter $\tau$. Thus, it is quite usual to speak of a timelike congruence by simply giving its unit tangent vector field. Traditional references on timelike congruences are [60,66,99].

Any timelike congruence provides a natural 1+3 splitting of the spacetime by means of the spatial projector associated with $\vec{u}$ defined by

$$h^\mu_\upsilon \equiv \delta^\mu_\upsilon + u^\mu u_\upsilon, \quad h^\mu_\upsilon h^\upsilon_\rho = h^\mu_\rho, \quad h^\mu_\mu = 3, \quad h^\mu_\upsilon u^\upsilon = 0, \quad h_{\mu\upsilon} = h_{\upsilon\mu}, \tag{18}$$

with $\vec{u}$ as in (16). The spatial part of any tensor with respect to the congruence is then obtained by projecting all indices with $h^\mu_\upsilon$. The line-element itself is decomposed as follows:

$$ds^2 = g_{\mu\upsilon} dx^\mu dx^\upsilon = -(u_\mu dx^\mu)^2 + h_{\mu\upsilon} dx^\mu dx^\upsilon. \tag{19}$$

The acceleration of a timelike congruence is defined by

$$a^\mu \equiv u^\rho \nabla_\rho u^\mu, \qquad a^\mu u_\mu = 0, \tag{20}$$

so that it is spatial with respect to the congruence (and spacelike). A timelike congruence is geodesic if and only if $\vec{a} = \vec{0}$, and then every curve in the congruence is a geodesic. The full covariant derivative of $\vec{u}$ can be split as follows:

$$\nabla_\upsilon u_\mu = -u_\upsilon a_\mu + h^\rho_\mu h^\sigma_\upsilon \nabla_\rho u_\sigma \equiv -u_\upsilon a_\mu + \frac{\theta}{3} h_{\mu\upsilon} + \sigma_{\mu\upsilon} + \omega_{\mu\upsilon}, \tag{21}$$

$$\begin{aligned}\sigma_{\mu\upsilon} &\equiv \frac{1}{2} h^\rho_\mu h^\sigma_\upsilon (\nabla_\sigma u_\rho + \nabla_\rho u_\sigma) - \frac{\theta}{3} h_{\mu\upsilon}, \\ \omega_{\mu\upsilon} &\equiv \frac{1}{2} h^\rho_\mu h^\sigma_\upsilon (\nabla_\sigma u_\rho - \nabla_\rho u_\sigma), \qquad \theta \equiv \nabla_\mu u^\mu,\end{aligned} \tag{22}$$

$$\sigma_{\mu\upsilon} = \sigma_{\upsilon\mu}, \qquad \omega_{\mu\upsilon} = -\omega_{\upsilon\mu}, \qquad \sigma^\mu_\mu = 0, \qquad u^\mu \sigma_{\mu\upsilon} = u^\mu \omega_{\mu\upsilon} = 0, \quad (23)$$

where $\sigma_{\mu\upsilon}$ is the *shear* tensor, $\omega_{\mu\upsilon}$ is the *vorticity* or *rotation* tensor and $\theta$ is the *expansion*, all of them relative to $\vec{u}$. The interpretation of these kinematic quantities can be found in [60,66]; they are related to the deformation and twist of a small volume element along the curves of the congruence. Let us simply remark that the necessary and sufficient condition such that **u** is integrable (proportional to a gradient) is that the vorticity tensor vanishes, that is

$$\mathbf{u} = F\, dt \iff \omega_{\mu\upsilon} = 0 \qquad (24)$$

for some functions $F, t$. In this case, the congruence is said to be *irrotational* and there exist 3-spaces orthogonal to the congruence curves defined by $t = $ const. This function $t$ is a natural time for irrotational congruences. Further, **u** is a gradient (locally an exact differential) iff the vorticity *and* the acceleration vanish

$$\mathbf{u} = -dt \iff \omega_{\mu\upsilon} = 0, \quad \text{and} \quad a^\mu = 0. \qquad (25)$$

By contracting the Ricci identities

$$(\nabla_\mu \nabla_\upsilon - \nabla_\upsilon \nabla_\mu) u^\alpha = R^\alpha_{\rho\mu\upsilon} u^\rho \qquad (26)$$

with $u^\mu$ one obtains the so-called evolution equations for the kinematic quantities. In particular, the trace produces the fundamental Raychaudhuri equation [170]

$$u^\mu \partial_\mu \theta + \frac{\theta^2}{3} - \nabla_\mu a^\mu - \omega_{\mu\upsilon} \omega^{\mu\upsilon} + \sigma_{\mu\upsilon} \sigma^{\mu\upsilon} + R_{\mu\upsilon} u^\mu u^\upsilon = 0, \qquad (27)$$

where the quantities $\omega_{\mu\upsilon}\omega^{\mu\upsilon}$ and $\sigma_{\mu\upsilon}\sigma^{\mu\upsilon}$ are non-negative, being squares of spatial tensors, and they vanish iff the corresponding full tensor $\omega_{\mu\upsilon}$ or $\sigma_{\mu\upsilon}$ vanishes, respectively.

The important concept for the singularity theory is that of focal and conjugate points along causal geodesics. To define them let us focus our attention on the irrotational and geodesic timelike congruences. The null case will be summarized later. As already seen in Lemma 2.1 the f-d timelike geodesic congruence emanating from any point $p$ is irrotational (at least in $\mathcal{N}_p$). Similarly, we have

**Lemma 2.2.** The f-d timelike geodesic congruence emanating orthogonally from any spacelike hypersurface $\Sigma$ is irrotational.

*Proof.* Let $\Sigma$ be defined locally by $\tau(x^\mu) = 0$ in a given coordinate system $\{x^\mu\}$. Take the f-d unit normal to $\Sigma$, that is $\mathbf{u} \propto d\tau|_\Sigma$ such that (16) holds at $\Sigma$. Let each point in $\Sigma$ be marked locally by three coordinates $\{y^i\}$, $(i, j, \ldots = 1, 2, 3)$, and define the unique timelike geodesic starting at each of these points with tangent vector $\vec{u}$. This provides the required timelike congruence in a neighbourhood of at least a local piece of $\Sigma$. Define now new coordinates $\{y^\mu\} = \{\tau, y^i\}$ such that on each geodesic the $y^i$ remain constant. In these coordinates it is obvious that

$$u^i = 0, \qquad u_i|_\Sigma = 0.$$

It must be shown that $\Omega_{\mu\nu} \equiv \nabla_\mu u_\nu - \nabla_\nu u_\mu = \partial_\mu u_\nu - \partial_\nu u_\mu$ vanishes. Given that $\vec{u}$ is geodesic and satisfies (16) it is immediate that

$$u^\mu \Omega_{\mu\nu} = u^\mu (\nabla_\mu u_\nu - \nabla_\nu u_\mu) = 0 \implies \Omega_{0i} = 0.$$

Using this and noting that $\partial_{[\mu} \Omega_{\nu\rho]} = 0$ there follows $\partial_0 \Omega_{ij} = 0$, and as $\Omega_{ij}|_\Sigma = (\partial_i u_j - \partial_j u_i)|_\Sigma = 0$ we get $\Omega_{ij} = 0$. Thus, $\Omega_{\mu\nu} = 0$. ∎

Incidentally, in this proof we have built what is called a Gaussian or synchronous coordinate system associated with (a piece of) $\Sigma$ [128,149]. In this system the line-element takes the simple form

$$ds^2 = -d\tau^2 + g_{ij}(y^\mu) dy^i dy^j, \tag{28}$$

as follows from (19) and (25). Here $\tau = y^0$ measures a proper time (affine parameter) along the geodesic congruence and labels the orthogonal hypersurfaces $\tau = \text{const}$. The matrix $(g_{ij})$ is positive definite for all $\tau$. For the f-d geodesic congruence emanating from a point $p \in V_4$ there is a similar system of coordinates not including $p$. Take normal coordinates $\{X^\mu\}$ in $\mathcal{N}_p$ and define new coordinates $\{y^\mu\}$ by means of

$$y^0 (= \tau) = \sqrt{-g_{\mu\nu}(0) X^\mu X^\nu}, \qquad y^i = \frac{X^i}{\sqrt{-g_{\mu\nu}(0) X^\mu X^\nu}}, \tag{29}$$

so that $y^0 = \tau$ measures proper time along the geodesics, which are given by $y^i = \text{const}$. Computing the Jacobian of the transformation (29) it is clear that the above system of coordinates is well-defined in $C_p^+$ (or $C_p^-$), where the line-element takes exactly the same form (28).

**Definition 2.10.** A point $q$ is conjugate (resp. focal) to the the point $p$ (resp. the spacelike hypersurface $\Sigma$) if the expansion of the f-d timelike geodesic congruence emanating from $p$ (resp. orthogonally from $\Sigma$) tends to $-\infty$ when approaching $q$ along a curve of the congruence.

Notice that the condition is that $\theta \to -\infty$ *along* a curve of the congruence. For example, the expansion of the timelike congruence emanating from $p$ diverges when approaching $\partial C_p^+$, as can be immediately seen from the proof of Lemma 2.1, but points in $\partial C_p^+$ are not conjugate to $p$ in general because there is no curve of the timelike congruence joining $p$ and $\partial C_p^+ - \{p\}$. Obviously, the existence of focal or conjugate points is related to the failure of the coordinate systems (28) and (29), which are not globally defined in general. The interpretation of these points can be deduced from the following reasoning. Take the canonical volume 4-form $\eta$ in $V_4$. At each hypersurface orthogonal to **u** we can define its own volume element 3-form by the usual formula $\mathbf{V} \equiv i_{\vec{u}}\eta$, where $i_{\vec{u}}$ is the interior contraction (in components, $V_{\mu\nu\rho} \equiv u^\sigma \eta_{\sigma\mu\nu\rho}$). Denoting the Lie derivative along $\vec{u}$ by $\pounds_{\vec{u}}$ it is easily seen that $\pounds_{\vec{u}}\mathbf{V} = \theta \mathbf{V}$. If $\{f^i(x)\}$ are three independent first integrals of the congruence, that is, $u^\mu \partial_\mu f^i = 0$, then $\mathbf{V} = V df^1 \wedge df^2 \wedge df^3$ and the above formula becomes simply $u^\mu \partial_\mu V = \theta V$ or equivalently

$$\theta = u^\mu \partial_\mu (\log|V|) . \tag{30}$$

In fact this formula can be more easily obtained in Gaussian coordinates (28), where the choice $f^i(x) = y^i$ is possible, and computing $\theta = \nabla_\mu u^\mu$, from where we see that $V = \sqrt{\det g_{ij}}$ with $g_{ij} = g(\partial/\partial y^i, \partial/\partial y^j)$. However, the result is general and valid beyond the domain of Gaussian coordinates. Equation (30) says that focal or conjugate points appear when $V \to 0$, which means that the volume element of the hypersurfaces orthogonal to **u** vanishes. This happens when, in the basis $\{\vec{u}, \partial/\partial f^i\}$, the three spatial vectors $\partial/\partial f^i$ are not independent or, which is the same, when there is a non-trivial linear combination of them vanishing at a point of a curve of the congruence, but not along the curve. This linear combination is called a Jacobi field along the curve [12,107], because it satisfies the Jacobi or geodesic deviation equation [128,149,239,241]. For the particular timelike geodesic irrotational congruences we are using, the vector fields $\vec{Z} \equiv A^i(f^j)\partial/\partial f^i$ satisfy both $\pounds_{\vec{u}}\vec{Z} = \vec{0}$ and $u_\mu Z^\mu = 0$, as is very easily checked, and then they coincide with the traditional Jacobi fields [12,107]. The approach used here may be preferable because there is no need to assume the continuity of the Riemann tensor, which is usually needed when the Jacobi equation is used. Note further that the particular Jacobi field $\vec{z}$ vanishing at a conjugate or focal point $q$ must have $\nabla_{\vec{u}}\vec{z}|_q \neq 0$, as otherwise $\vec{z} \equiv 0$ along the curve joining $p$ (or $\Sigma$) with $q$. Finally, let us remark that the most general $\vec{Z}$ vanishes at $p$ in the case of the geodesic congruence emanating from $p$.

From the above follows also the following result, which will be needed later:

**Lemma 2.3.** Given any f-d timelike curve $\gamma$ starting at $p$ (resp. orthogonally to $\Sigma$) there is a neighbourhood of $\gamma$ in which the Gaussian coordinate system (29) [resp. (28)] holds up to the first conjugate (resp. focal) point to $p$ (resp. to $\Sigma$).

*Proof.* Conjugate and focal points are isolated because of (30). Thus, there must be a first point conjugate to $p$ (if any) along $\gamma$, say $q$ (we concentrate on the case of $p$, the proof for $\Sigma$ is completely analogous). This means that all Jacobi fields $\vec{Z}$ orthogonal to $\gamma$ along $\gamma$ are not vanishing or, in other words, that the three vector fields $\partial/\partial f^i|_\gamma$ are independent along $\gamma$ up to $q$. These three vector fields can be trivially chosen such that they commute $[\partial/\partial f^i, \partial/\partial f^j] = \vec{0}$ and then [41] they generate locally a piece of a hypersurface orthogonal to $\gamma$. This in turn means that $\{\tau, f^i\}$ are good Gaussian coordinates in a sufficiently small neighbourhood of $\gamma$, with $\tau$ proper time along the geodesics with constant $f^i$. ∎

The question is to know when there will be conjugate or focal points in a given spacetime.

**Proposition 2.3.** If the expansion of the f-d timelike geodesic congruence emanating from $p$ (orthogonally to $\Sigma$) is negative at some point $r$ and $R_{\mu\upsilon}u^\mu u^\upsilon \geq 0$ along the timelike geodesic $\gamma$ passing through $r$, then there is a point $q$ conjugate to $p$ (focal to $\Sigma$) along $\gamma$ within a finite proper time after $r$, provided that $\gamma$ can be extended that far.

*Proof.* From Lemma 2.1 (or Lemma 2.2) the congruence emanating from $p$ (orthogonally to $\Sigma$) is irrotational and geodesic, so that Raychaudhuri's equation (27) becomes, on using (30),

$$\frac{3}{V^{1/3}} \frac{d^2 V^{1/3}}{d\tau^2} = -(\sigma_{\mu\upsilon}\sigma^{\mu\upsilon} + R_{\mu\upsilon}u^\mu u^\upsilon) \leq 0$$

which along $\gamma$ implies $V^{1/3} \leq V_r^{1/3}[1 + (\theta_r/3)(\tau - \tau_r)]$ where $V_r > 0$ and $\theta_r < 0$ are the values of $V$ and $\theta$ at $r$. Thus, $V$ vanishes before $\tau$ reaches the value $\tau_r - 3/\theta_r > \tau_r$. ∎

The condition $R_{\mu\upsilon}u^\mu u^\upsilon \geq 0$ is standard and related to the energy contents of the spacetime through Einstein's equations. However, this condition can be relaxed substantially but still maintaining the focusing effect of Proposition 2.3 (see Refs. 27,218,165). The required assumptions are usually of integral type, so that the quantity $R_{\mu\upsilon}u^\mu u^\upsilon$ is required to be non-negative in an adequate averaged sense [27,28,218]. Proposition 2.3 can be also generalized in a more relevant manner by the following standard result [12,27,72,107].

**Proposition 2.4.** If $R_{\mu\nu}u^\mu u^\nu \geq 0$ along an f-d timelike geodesic $\gamma$ and $R_{\rho\mu\sigma\nu}u^\mu u^\nu \neq 0$ at $p \in \gamma$, then $\gamma$ contains a pair of conjugate points provided that it can be extended sufficiently far.

*Proof.* The strategy of the proof (see Ref. 108) was very clearly explained in [27], and is as follows. Let $\gamma(\tau_p) = p$ and consider the f-d timelike geodesic congruences emanating from all points in $\gamma$ to the past of $p$, that is, emanating from each $\gamma(\tau)$ $\forall \tau < \tau_p$. For any of these congruences, if $\theta(\tau_p) \leq 0$, then $\theta \leq 0$ to the future of $p$ because of (27); see the reasoning in Proposition 2.3. If $\theta < 0$ for some $\tau > \tau_p$, then the result follows from the Proposition 2.3 itself. If $\theta$ vanished in $[\tau_p, \tau_1)$, then (27) and $R_{\mu\nu}u^\mu u^\nu \geq 0$ would imply $\sigma_{\mu\nu} = R_{\mu\nu}u^\mu u^\nu = 0$ for $\tau \in [\tau_p, \tau_1)$, which through (21) would provide $\nabla_\mu u_\nu|_\gamma = 0$ for $\tau \in [\tau_p, \tau_1)$. But this is impossible because the identity (26) contracted with $u^\mu$ would contradict $R_{\rho\mu\sigma\nu}u^\mu u^\nu \neq 0$ at $p$. Now, given the structure of the Raychaudhuri equation (27) it is evident that there is a finite maximum bound $\bar{\tau}$ for the values of $\tau$ at these conjugate points, because the most unfavourable case appears when $\theta(\tau_p) = 0$, which produces the conjugate point at the greater values of $\tau$, but nevertheless these are finite. Thus, take now a *past-directed* timelike geodesic congruence emanating from a point $q = \gamma(\tau_q)$ with $\tau_q > \bar{\tau}$. If there is no point past-conjugate to $q$ between $q$ and $p$, then the curves in this congruence close enough to $\gamma$ at $\tau_p$ must have $\tilde{\theta}(\tau_p) < 0$ (the tilde indicates that this expansion is 'past-directed'), as otherwise the reverse congruence would have non-positive f-d expansion at $p$ and $\tau_q$ could not be greater than $\bar{\tau}$. Then, Proposition 2.3 and $\tilde{\theta}(\tau_p) < 0$ tell us that there is a $\tilde{\tau} < \tau_p$ such that $\gamma(\tilde{\tau})$ is past-conjugate to $q$. ∎

In the above propositions, there have appeared some assumptions on components of the curvature ($R_{\mu\nu}u^\mu u^\nu$, $R_{\rho\mu\sigma\nu}u^\mu u^\nu$) which are common to *all* singularity theorems. These assumptions assure, essentially, that gravity is attractive, so that the focusing of geodesics takes place, and that there exists some matter or energy in the spacetime. They are given standard names which are collected in the following definition ($T_{\mu\nu}$ is the energy-momentum tensor).

**Definition 2.11** A spacetime satisfies:
- the timelike (resp. null) generic condition if $v_{[\alpha}R_{\rho]\mu\nu[\sigma}v_{\beta]}v^\mu v^\nu \neq 0$ at some point of each timelike (resp. null) geodesic with tangent vector $\vec{v}$.
- the null convergence condition if $R_{\mu\nu}k^\mu k^\nu \geq 0$ for all null vectors $\vec{k}$.

- the strong energy condition (SEC) if $R_{\mu\upsilon}v^{\mu}v^{\upsilon} \geq 0$ for all causal vectors $\vec{v}$.
- the weak energy condition (WEC) if $T_{\mu\upsilon}v^{\mu}v^{\upsilon} \geq 0$ for all causal vectors $\vec{v}$.
- the dominant energy condition (DEC) if WEC holds and $T_{\mu\upsilon}v^{\mu}$ is causal for all causal $\vec{v}$.

A spacetime satisfies the generic condition when both the timelike and null generic conditions hold. This condition demands that any possible causal geodesic in the spacetime must meet, sooner or later, a quantity of gravitational field (be it in the form of matter or of pure gravity) travelling in arbitrary directions. In principle, this is very reasonable for generic enough realistic spacetimes. However, some spacetimes specialized from the geometrical point of view may violate it (e.g., spherically symmetric). These will be further analysed in Section 6. The timelike generic condition is obviously equivalent to the assumption $R_{\rho\mu\sigma\upsilon}u^{\mu}u^{\upsilon} \neq 0$ in Proposition 2.4, because of the antisymmetry properties of the Riemann tensor. The quantity $Y_{\rho\sigma}(\vec{u}) \equiv R_{\rho\mu\sigma\upsilon}u^{\mu}u^{\upsilon}$ is called the first electric part of the Riemann tensor [22] with respect to the unit timelike vector $\vec{u}$, and therefore magnetic components of $R_{\rho\mu\sigma\upsilon}$ do not affect the focusing *directly*. This first electric part of the Riemann tensor is composed of the usual electric part $E_{\rho\sigma} \equiv C_{\rho\mu\sigma\upsilon}u^{\mu}u^{\upsilon}$ of the Weyl tensor $C_{\rho\mu\sigma\upsilon}$ with respect to $\vec{u}$ [22,123,167], plus terms related to the energy-momentum tensor:

$$Y_{\rho\sigma}(\vec{u}) = E_{\rho\sigma}(\vec{u}) - \frac{1}{2}h_{\rho}^{\mu}h_{\sigma}^{\upsilon}R_{\mu\upsilon} + \frac{1}{2}\left(R_{\mu\upsilon}u^{\mu}u^{\upsilon} + \frac{R}{3}\right)h_{\rho\sigma},$$

where $h_{\rho\sigma}$ is the projector (18). Note that $Y_{\mu\upsilon}$ is symmetric and spatial with respect to $\vec{u}$, that is, $Y_{\mu\upsilon}u^{\mu} = 0$. We have [12]

**Proposition 2.5.** If $Y_{\mu\upsilon}(\vec{u}) = 0$, then $R_{\mu\upsilon}u^{\mu}u^{\upsilon} = 0$. Hence, if $R_{\mu\upsilon}u^{\mu}u^{\upsilon} > 0$, then $Y_{\mu\upsilon}(\vec{u}) \neq 0$.

*Proof.* Contracting $Y_{\rho\sigma}$ with $h^{\rho\sigma}$ we readily get $R_{\mu\upsilon}u^{\mu}u^{\upsilon} = h^{\mu\upsilon}Y_{\mu\upsilon}$. Thus, if $Y_{\mu\upsilon}$ were zero, so would $R_{\mu\upsilon}u^{\mu}u^{\upsilon}$ be. ∎

This proposition means that, if the *strict* SEC holds (with the strict inequality and for non-zero vectors), then the timelike generic condition is automatically fulfilled. An analysis of how "generic" the generic condition is can be found in [13]. Concerning the null generic condition, the following formula can be easily found for any null vector $\vec{k}$

$$k_{[\alpha}R_{\rho]\mu\upsilon[\sigma}k_{\beta]}k^{\mu}k^{\upsilon} + \tfrac{1}{2}R_{\mu\upsilon}k^{\mu}k^{\upsilon}k_{[\alpha}g_{\rho][\sigma}k_{\beta]} = k_{[\alpha}C_{\rho]\mu\upsilon[\sigma}k_{\beta]}k^{\mu}k^{\upsilon}.$$

The righthand side of this relation vanishes iff $\vec{k}$ is a Debever principal null vector of the Weyl tensor (see e.g. Refs. 15,123,167). As is well known, there are at most four such vectors at each point of $V_4$, and the exact number of them is directly related to the Petrov type of the spacetime [15,123,167,168]. In consequence, the null generic condition fails to hold if any principal null direction of the Weyl tensor satisfies also $R_{\mu\nu}k^\mu k^\nu = 0$. In particular, in *vacuum* spacetimes the generic condition can only fail along the principal null geodesics of the spacetime. Further properties of the null generic condition can be found in [13]. Similarly to Proposition 2.5 we have [12]

**Proposition 2.6.** For a null vector $\vec{k}$, if $k_{[\alpha}R_{\rho]\mu\nu[\sigma}k_{\beta]}k^\mu k^\nu = 0$, then $R_{\mu\nu}k^\mu k^\nu = 0$. Thus, $R_{\mu\nu}k^\mu k^\nu > 0$ implies $k_{[\alpha}R_{\rho]\mu\nu[\sigma}k_{\beta]}k^\mu k^\nu \neq 0$. ∎

With regard to the energy conditions, the following remarks are important. First of all, SEC could have been defined equivalently for only timelike vectors $\vec{v}$, and then by continuity it follows for all causal vectors. The SEC was first introduced in [120,171], and with the same purposes as here. The WEC simply states that the energy density is non-negative as measured in any possible reference system. Similarly, the DEC states the same *plus* the causal propagation of matter and radiation as seen in any reference system. Therefore, both WEC and DEC are physically well-founded and very reasonable. Nevertheless, the focusing of neighbouring geodesics requires SEC, which is the *less* relevant and worst-founded energy condition from a physical point of view. It is certain that most known physical fields satisfy SEC, even more so in their very mild averaged version found in [27], but there is no direct simple implication concerning energy density or causal propagation as in the case of DEC. In terms of the energy-momentum tensor SEC reads $T_{\mu\nu}v^\mu v^\nu \geq v_\mu v^\mu T^\nu_\nu$, which is reasonable but not completely based on physical grounds (see Refs. 107,233). It would thus be desirable to prove the singularity theorems using DEC, or the weaker WEC, but this has not been possible so far. I will come back to this in Section 6.

All the above definitions and results can be adequately translated to the case of null congruences. Typical references on this subject are [123,150,167,180]. Let $\vec{k}$ be the tangent vector field to such a congruence. Then we have $g(\vec{k}, \vec{k}) = 0$. It is well-known that any hypersurface-orthogonal null congruence must be geodesic, because if $k_{[\mu}\nabla_\nu k_{\rho]} = 0$, on contracting with $k^\mu$ it follows that

$$k^\mu \nabla_\mu k_{[\rho}k_{\nu]} = 0 \implies k^\mu \nabla_\mu k_\rho \propto k_\rho$$

and then we can further choose an affine parameter $\tau$ such that

$$k^\mu \nabla_\mu k^\rho = 0, \qquad k^\mu \partial_\mu \tau = 1.$$

Notice also that as **k** $\propto dv$ and **k** is null, $\vec{k}$ is in fact tangent to the orthogonal hypersurfaces, because it is orthogonal to the normal one-form **k**. Nevertheless, as we saw in (4), a null vector can be truly orthogonal to a spacelike surface $S$, and this is the important concept in this case. Take then any spacelike surface $S$ and choose one of its null normal one-forms **k**. We construct all null geodesics starting at each point on $S$ with tangent vector $\vec{k}$, and this generates *not* a congruence but simply a null hypersurface orthogonal to **k**. Varying the initial surface $S$ in a direction not contained in this null hypersurface, and repeating the process for each new spacelike surface, the required null congruence is built. An appropriate basis in this case is formed by the so-called pseudo-orthonormal basis constituted by $\{\vec{k}, \vec{l}, \vec{e}_A\}$, where $k$ and $l$ are the two null normal one-forms orthogonal to the spacelike surfaces and normalized as in (4), and $\{\vec{e}_A\}$ is an orthonormal basis of the vector fields tangent to each of these surfaces. Thus, the only non-zero scalar products in this basis are

$$g(\vec{e}_2, \vec{e}_2) = g(\vec{e}_3, \vec{e}_3) = -g(\vec{k}, \vec{l}) = 1.$$

The projector orthogonal to $\vec{k}$ and $\vec{l}$ is defined by

$$N_{\mu\nu} \equiv g_{\mu\nu} + k_\mu l_\nu + k_\nu l_\mu, \qquad N^\mu_\nu N^\nu_\rho = N^\mu_\rho,$$
$$N^\mu_\mu = 2, \qquad N^\mu_\nu k^\nu = N^\mu_\nu l^\nu = 0, \qquad N_{\mu\nu} = N_{\nu\mu}.$$

By the way, it must be remarked that $N_{\mu\nu}$ at each surface $S$ is equivalent to the first fundamental form (3). Similarly, the second fundamental form (6) relative to **k** is the projected part of the covariant derivative of **k** at each $S$. In the case under consideration, this has only two independent components, which can be identified as [150,167,180]

$$\vartheta \equiv \nabla_\mu k^\mu (= N^{\mu\nu} \nabla_\mu k_\nu), \qquad (2\sigma)^2 \equiv (\nabla_\mu k_\nu + \nabla_\nu k_\mu) \nabla^\mu k^\nu - \vartheta^2, \quad (31)$$

and are called the expansion and the shear of the null congruence, respectively. There is a local coordinate system $\{\tau, v, x^2, x^3\}$ associated to these null congruences analogous to (28) which can be set up by standard methods (see e.g. Refs. 123,150,151,152,180). The line-element takes the form

$$ds^2 = -2d\tau(dv + A d\tau + B_A dx^A) + g_{AB} dx^A dx^B,$$

where $A$, $B_A$ and $g_{AB}$ depend on all four coordinates and the matrix $(g_{AB})$ is positive-definite. The surface $S$ is given by, say, $\tau = v = 0$, the null hypersurfaces $v = $ const. are orthogonal to $\mathbf{k} = -dv$, and $\tau$ measures affine parameter along the null geodesics, which are marked by constant values of $u$ and $x^A$. The limit case when $S$ is a single point $p$ is also covered by a coordinate system of this type not including $p$, by simply letting $v = 0$ denote $\partial C_p^+$.

**Definition 2.12.** A point $q$ is conjugate (resp. focal) to the the point $p$ (resp. the spacelike surface $S$) if the expansion of the f-d null geodesic family emanating from $p$ (resp. orthogonally from $S$) becomes unbounded when approaching $q$ along a curve of the family.

Instead of (30) we now have $\vartheta = k^\mu \partial_\mu (\log \mathcal{V})$, where $\mathcal{V}$ is an area element in the surfaces orthogonal to $\vec{k}$. In the above coordinates $\mathcal{V} = \sqrt{\det g_{AB}}$. Similarly, the equation replacing (27) is now [107,150,180]

$$k^\mu \partial_\mu \vartheta + \frac{\vartheta^2}{2} + 2\sigma^2 + R_{\mu\nu} k^\mu k^\nu = 0$$

which allows one to prove, similarly to Propositions 2.3 and 2.4,

**Proposition 2.7.** If the expansion of the f-d null geodesic family emanating from $p$ (orthogonally to $S$) is negative at some point $r$ and the null convergence condition holds along the null geodesic $\gamma$ passing through $r$, then there is a point $q$ conjugate to $p$ (focal to $S$) along $\gamma$ within a finite affine parameter since $r$, provided that $\gamma$ can be extended that far. ∎

**Proposition 2.8.** If the null convergence and the null generic conditions hold for an f-d null geodesic $\gamma$, then $\gamma$ contains a pair of conjugate points provided that can be extended sufficiently far. ∎

## 2.2. Maximal curves

The *length* of a piecewise differentiable curve $\gamma$ parametrized by $u$ and joining $p, q \in V_4$ is defined by

$$L(p, q; \gamma) \equiv \int_{u_p}^{u_q} \sqrt{|g(\vec{v}|_\gamma, \vec{v}|_\gamma)|} \, du \qquad (32)$$

where $\gamma(u_p) = p$, $\gamma(u_q) = q$ and the integral is taken over the differentiable segments of $\gamma$. This definition is invariant under reparametrization, and due to the comments after Definition 2.8 it can also be used for continuous causal curves. Obviously, $L$ is zero for a causal $\gamma$ iff $\gamma$ is a sequence of null segments. In general, for any pair $p, q \in V_4$ there is no lower bound on the lengths $L$ joining them through different *causal* curves, because any causal curve can be approximated by a sequence of tiny null segments in a zig-zag. However, there may be a maximum of length along some causal curve joining $p$ and $q$.

**Definition 2.13.** A causal curve from $p$ (from a hypersurface $\Sigma$) to $q$ is said to be maximal if there is no causal curve from $p$ (from some point in $\Sigma$) to $q$ with length greater than that of $\gamma$; and $\gamma$ is said to be locally maximal if there is a neighbourhood $\mathcal{U}_\gamma$ of $\gamma$ within which there is no causal curve from $p$ (from some point in $\mathcal{U}_\gamma \cap \Sigma$) to $q$ with greater length.

Evidently, if $\gamma : [a, b] \to V_4$ is locally maximal between $p = \gamma(a)$ and $q = \gamma(b)$, then it is locally maximal between any pair of its points $\gamma(u_1)$ and $\gamma(u_2)$, with $u_1, u_2 \in [a, b]$. However, a curve can be locally maximal and there may still exist another longer curve between two of its points. Similarly, maximal curves are not necessarily unique.

**Proposition 2.9.** Within $\mathcal{N}_p$, the causal geodesics from $p$ are maximal.

*Proof.* If $q \in \partial C_p^+$, from Proposition 2.1 the only causal curve from $p$ to $q$ is the null geodesic, so that the result is trivial. Take then $q \in C_p^+$. As already proved, the coordinate system (29) is valid in $C_p^+$, and the line-element takes the form (28). Any causal curve from $p$ to $q$ can be represented in parametric form by $y^0 = (\tau) = u$, $y^i = \gamma^i(u)$ so that (32) becomes

$$L(p, q; \gamma) = \int_0^{u_q} \sqrt{1 - \boldsymbol{g}_j(\gamma) \frac{d\gamma^i}{du} \frac{d\gamma^j}{du}} \, du.$$

Given that $\boldsymbol{g}_j$ is positive definite, from this relation is obvious that $L$ is a maximum only if $\gamma^i$ are constant all along. But these are the timelike geodesics from $p$. Notice further that $L(p, q; \gamma)$, when $\gamma$ is a geodesic, depends continuously on its endpoints $p$ and $q$. ∎

**Corollary 2.1.** If $\gamma$ is locally maximal, then it is a geodesic arc.

*Proof.* As $\gamma$ is locally maximal between any of its points, it must be locally (within each $\mathcal{N}_p$, for $p \in \gamma$) a geodesic. Thus, it must be a sequence of geodesic arcs. If there were a corner at some $p \in \gamma$, then there would be a geodesic from a point $s \in \gamma$ just prior to $p$ to a point $q \in \gamma$ just after $p$, and this geodesic would be longer than $\gamma$ from $s$ to $q$, which is impossible. ∎

**Corollary 2.2.** If a timelike geodesic $\gamma$ contains no pair of conjugate points between $p$ and $q$, then it is locally maximal.

*Proof.* The result is immediate from Lemma 2.3 and the proof of Proposition 2.9. ∎

The converse of this corollary also holds.

**Proposition 2.10.** If a timelike geodesic $\gamma$ from $p$ to $q_1$ is locally maximal, there is no pair of conjugate points along $\gamma$ between $p$ and $q_1$.

*Proof.* Suppose on the contrary that there is a point $r \in \gamma$, between $p$ and $q_1$, which is conjugate to $p$. By Lemma 2.3 there is a Gaussian coordinate system (28) in a neighbourhood $\mathcal{U}_\gamma$ of the portion of $\gamma$ from $p$ to $r$. In this system it is obvious that $L(p, x; \lambda) = \tau|_x$, for all $x \in \mathcal{U}$ and all geodesics $\lambda \subset \mathcal{U}$ starting from $p$. Let us denote by $\vec{u}$ the unit tangent vector field of the timelike congruence emanating from $p$. Thus, in $\mathcal{U}$ we have $\vec{u} = \partial/\partial \tau$ and $\mathbf{u} = -d\tau$. Take any point $q \in \gamma$ beyond $r$ but sufficiently close to $r$ such that $r \in \mathcal{N}_q$. Within $C_q^-$ we can set up *another* coordinate system $\{\hat{\tau}, \hat{y}^i\}$ of type (29), so that $L(x, q; \hat{\lambda}) = \hat{\tau}|_x$, for all $x \in C_q^-$ and all (past-directed) geodesics $\hat{\lambda} \subset C_q^-$ starting from $q$. Let us denote by $\hat{\vec{u}}$ the unit (past-directed) tangent vector field for the timelike congruence emanating from $q$. Hence, in $C_q^-$ we have $\hat{\vec{u}} = \partial/\partial \hat{\tau}$ and $\hat{\mathbf{u}} = -d\hat{\tau}$. Notice that both congruences $\vec{u}$ and $\hat{\vec{u}}$ contain the curve $\gamma$ within their common domain of definition $\mathcal{U} \cap C_q^-$. For every $x \in \mathcal{U} \cap C_q^-$, there are timelike geodesics $\mu$ from $p$ to $x$ and $\hat{\mu}$ from $x$ to $q$ which can be combined into a single curve $\mu\hat{\mu}$ from $p$ to $q$. From the above it follows that

$$L(p, q; \mu\hat{\mu}) = (\tau + \hat{\tau})|_x \qquad \forall x \in \mathcal{U} \cap C_q^- ;$$
$$L(p, q; \gamma) = (\tau + \hat{\tau})|_s \qquad \forall s \in \gamma \cap C_q^- \cap C_r^- .$$

We are going to show that $x$ can be chosen sufficiently close to $s$, and $s$ close enough to $r$ such that the first of the above lengths is bigger than the second.

Since $r$ is conjugate to $p$ along $\gamma$, there is a non-zero Jacobi field $\vec{z}$ along $\gamma$ such that it vanishes at $r$ but with $\nabla_{\vec{u}} \vec{z}|_r \neq 0$ ($\vec{u}$ is well-defined at $r$ being the unit tangent vector to $\gamma$). Moreover, in $\mathcal{U}$, $\pounds_{\vec{u}} \vec{z} = \vec{0}$ and $\nabla_{\vec{u}} \vec{z}$ is orthogonal to $\vec{u}$. Define $\vec{v} \equiv (\tau|_r - \tau)^{-1} \vec{z}$ within $\mathcal{U}$, so that $\vec{v}$ is orthogonal to $\vec{u}$ and $\pounds_{\vec{u}} \vec{v} = (\tau|_r - \tau)^{-1} \vec{v}$.[3] Build the unique affinely parametrized spacelike geodesic $\upsilon$ starting at $s \in \gamma \cap C_q^- \cap C_r^-$ and with initial tangent vector $\vec{v}|_s$. For points $x \in \upsilon \cap \mathcal{U}$ close enough to $s$ we have

$$(\tau + \hat{\tau})|_x = (\tau + \hat{\tau})|_s + \left.\frac{d(\tau + \hat{\tau})}{du}\right|_{u=0} u + \left.\frac{d^2(\tau + \hat{\tau})}{du^2}\right|_{u=0} \frac{u^2}{2} + O(u^3)$$

where $u$ is the affine parameter along $\upsilon$. The second term on the right-hand side vanishes, because of the initial condition $d\tau/du|_{u=0} = v^0|_s = -u_\mu v^\mu|_s = 0$, and similarly $d\hat{\tau}/du|_{u=0} = 0$. Concerning the third term, since $\upsilon$ is an affinely parametrized geodesic, the equation (8) in the Gauss-

---

[3] In fact, $\vec{v}$ can be defined along $\gamma$ beyond $\mathcal{U}$ as the unique vector field satisfying $\nabla_{\vec{u}} \vec{v} = (\tau|_r - \tau)^{-1}(\vec{v} + \nabla_{\vec{u}} \vec{z})$ with the initial condition $\vec{v}|_r = -\nabla_{\vec{u}} \vec{z}|_r$. This $\vec{v}$ is differentiable along $\gamma$ [165].

ian coordinate system implies

$$\begin{aligned}\left.\frac{d^2\tau}{du^2}\right|_{u=0} &= \left.-\Gamma^0_{ij}\right|_s \left.\frac{dy^i}{du}\frac{dy^j}{du}\right|_{u=0} = \left.-\frac{1}{2}\partial_0 g_j \, v^i v^j\right|_s \\ &= \left.-\frac{1}{2}\frac{d}{d\tau}(g_j v^i v^j)\right|_s + \left.g_j v^i \frac{dv^j}{d\tau}\right|_s \\ &= \left.-\frac{1}{2}\nabla_{\vec{u}}[g(\vec{v},\vec{v})]\right|_s + \left.g(\vec{v},\pounds_{\vec{u}}\vec{v})\right|_s.\end{aligned}$$

The equivalent calculation in the $\{\hat{\tau},\hat{y}^i\}$-system produces

$$\left.\frac{d^2\hat{\tau}}{du^2}\right|_{u=0} = \left.-\frac{1}{2}\nabla_{\hat{u}}[g(\vec{v},\vec{v})]\right|_s + \left.g(\vec{v},\pounds_{\hat{u}}\vec{v})\right|_s.$$

Hence, as $\vec{u}$ and $\hat{\vec{u}}$ are oppositely directed and using $\pounds_{\vec{u}}\vec{v} = (\tau|_r - \tau)^{-1}\vec{v}$ we arrive at

$$\left.\frac{d^2(\tau+\hat{\tau})}{du^2}\right|_{u=0} = (\tau|_r - \tau|_s)^{-1} g(\vec{v},\vec{v})|_s + g(\vec{v},\pounds_{\hat{u}}\vec{v})|_s.$$

The second term on the righthand side is well-behaved as we are within $C_q^-$. Also, $\vec{v}$ is spacelike and therefore, by choosing $s$ close enough to $r$ the lefthand side can be made positive. Fixing this $s$, there is a $\tilde{u} > 0$ depending on $s$ such that for all $u \in (0,\tilde{u})$ the corresponding points $x \in \upsilon$ have $(\tau + \hat{\tau})|_x > (\tau + \hat{\tau})|_s$ or, in other words,

$$L(p,q;\mu\hat{\mu}) > L(p,q;\gamma) \qquad \forall x \in \upsilon|_{(0,\tilde{u})}.$$

Thus, $\gamma$ cannot be locally maximal, in contradiction. ∎

It must be remarked that, even though the term $(\tau|_r - \tau|_s)^{-1} g(\vec{v},\vec{v})|_s$ seems to indicate no upper bound on the lengths beyond $r$, this is not so because, as we approach $r$, the possible values of $u$ in the geodesic $\upsilon$ *within* $\mathcal{U}$ tend to zero, and thus the total length from $p$ to $q$ remains finite. All that can be proved is what is stated in Proposition 2.10 (see Ref. 165).

Analogous proofs allow us to show the following summarized results.

**Proposition 2.11.** Within the domain covered by Gaussian coordinates (28) relative to a spacelike hypersurface $\Sigma$, the timelike geodesics orthogonal to $\Sigma$ are maximal. ∎

**Proposition 2.12.** A timelike curve $\gamma$ from $\Sigma$ to $q$ is locally maximal if and only if it is a geodesic segment orthogonal to $\Sigma$ without any point focal to $\Sigma$ between $\Sigma$ and $q$. ∎

The results concerning conjugate and focal points along null geodesics can be stated somewhat differently. To begin with, let us prove the following result.

**Proposition 2.13.** If there is an f-d causal curve $\gamma$ from $p$ to $q$ and $\gamma$ is not a null geodesic segment, then there is an f-d timelike curve from $p$ to $q$.

*Proof.* We can cover $\gamma$ by convex normal neighbourhoods centered at points of $\gamma$ and such that each of them has compact closure in the corresponding normal neighbourhood centered at the same point. If $\gamma$ passes through $q$ infinite times, we can restrict our attention to the first portion of $\gamma$ reaching $q$ which is not a null geodesic. This portion is obviously compact, so that it can be covered by a finite sub-cover $\{N_i\}_{i=1,\ldots,n}$ of the previous cover. As $\gamma$ is not a null geodesic segment then there is a point $r \in \gamma$ such that either $r$ lies in an open sub-segment which is not a null geodesic or $\gamma$ has a corner at $r$. In both cases, within a neighbourhood $N_1 \ni r$, and due to Proposition 2.1, there is an f-d timelike geodesic segment $\lambda$ joining the two endpoints $r_-$ and $r_+$ of $\gamma \cap N_1$. If the future endpoint $r_+ \neq q$, then $r_+ \in N_2$ with $N_2 \cap N_1 \neq \varnothing$. Take the curve $\lambda \cap N_2$ up to $r_+$ prolonged with the initial curve $\gamma$. Again by Proposition 2.1 and given that $\lambda$ is timelike, there is an f-d timelike geodesic segment $\mu$ joining the two endpoints $s_+$ and $s_-$ of $\lambda\gamma \cap N_2$. If $s_+ \neq q$, we proceed in the same manner until we reach $q$, and similarly to the past. This process is finite as there are only a finite number of the $N_i$. ∎

Another way of stating Proposition 2.13 is: "if there is an f-d causal curve but no f-d timelike curve from $p$ to $q$, then every f-d causal curve joining $p$ to $q$ must be a null geodesic segment." Or in yet other words, if a null curve is locally maximal (and with vanishing length!), it must be a null geodesic segment. This preliminary result can be reinforced in the same way as Corollary 2.2 and Propositions 2.10, 2.11, 2.12.

**Proposition 2.14.** Given a causal curve $\gamma$ from $p$ to $q$ (resp. from a spacelike surface $S$ to $q$), there is no neighbourhood of $\gamma$ containing a timelike curve from $p$ (resp. $S$) to $q$ if and only if $\gamma$ is a null geodesic segment from $p$ (resp. orthogonal to $S$) to $q$ without any point conjugate to $p$ (resp. focal to $S$) between $p$ (resp. $S$) and $q$. ∎

### 2.3. Causality in the large and proper achronal boundaries

In the previous subsections, the importance of Proposition 2.1 has been repeatedly manifested. Recall that Proposition 2.1 states that the *local* causal structure of the spacetime is very simple and similar to that of flat spacetime. Nonetheless, in subsection 2.2 there have appeared several new aspects due to the failure of normal coordinates globally. The exami-nation of the global causality properties is the purpose of this subsection. In the singularity theorems, there will usually be an assumption restricting the global causal properties of the spacetime. The reasonableness of these assumptions rests ultimately on the belief that any violation of causality is not physically acceptable and it would lead to difficult interpretative questions from the physical viewpoint. Standard basic references on this subject are [37,34,92,93,107,127,165].

The basic sets in causality theory are the chronological future, the causal future and the future horismos of $p \in V_4$, denoted and defined respectively by

$$I^+(p) \equiv \{x \in V_4 \mid \text{there is an f-d timelike curve from } p \text{ to } x\},$$
$$J^+(p) \equiv \{x \in V_4 \mid \text{there is an f-d causal curve from } p \text{ to } x\},$$
$$E^+(p) \equiv J^+(p) - I^+(p).$$

For any set $\zeta \subseteq V_4$ we put

$$I^+(\zeta) \equiv \bigcup_{p \in \zeta} I^+(p) \qquad J^+(\zeta) \equiv \bigcup_{p \in \zeta} J^+(p) \qquad E^+(\zeta) \equiv J^+(\zeta) - I^+(\zeta).$$

In the definition of $J^+(p)$, I have considered that the trivial geodesic (the point $p$ taken as a curve of zero extension) is a particular case of a null geodesic. Thus, $J^+(\zeta) \supseteq \zeta$. We omit dual definitions for the chronological pasts $I^-$, the causal pasts $J^-$ and the past horismos $E^-$. It is evident that if $q \in J^+(p)$, then $p \in J^-(q)$, and if $q \in I^+(p)$, then $p \in I^-(q)$. Moreover, from Proposition 2.13 it follows that: (i) if $q \in E^+(\zeta)$, then $q$ lies on an f-d null geodesic segment from $\zeta$. (ii) if $q \in J^+(p)$ and $r \in I^+(q)$ then $r \in I^+(p)$. (iii) if $q \in I^+(p)$ and $r \in J^+(q)$ then $r \in I^+(p)$. (iv) if $q \in J^+(p)$ and $r \in J^+(q)$ then $r \in J^+(p)$.

Some examples of all these sets have already appeared, namely, consider any $p \in V_4$ and its normal neighbourhood $\mathcal{N}_p$. By itself, $\mathcal{N}_p$ is a spacetime $(\mathcal{N}_p, \mathbf{g})$. In this spacetime $I^+(p) = C_p^+$, $E^+(p) = \partial C_p^+$ and $J^+(p) = C_p^+ \cup \partial C_p^+ = I^+(p) \cup E^+(p)$, as is easily seen from Proposition 2.1. Thus, $J^+(p)$ is closed as a subset of $(\mathcal{N}_p, \mathbf{g})$. This simple structure is broken in general when we leave a normal neighbourhood of $p$. Incidentally, these examples show that the sets $I^+(p)$, $J^+(p)$ are non-empty. It must be remarked that $I^+(p)$ or $J^+(p)$ relative to the spacetime $(\mathcal{N}_p, \mathbf{g})$ are *not* necessarily equal to $I^+(p) \cap \mathcal{N}_p$ and to $J^+(p) \cap \mathcal{N}_p$ in the global spacetime $V_4$, respectively; and that for a non-empty set $\zeta \subset V_4$, $E^+(\zeta)$ may certainly be empty! This curious sort of thing appears again and again in causality theory — many examples can be found in [12,37,93,107,127,165,239] — and they have led to the definition of a cascade (see Definition 2.17) of stronger and stronger restrictions on the spacetime in order to keep it "as reasonable as possible". But first of all let us see that, nevertheless, there are some basic properties kept by the global causal structure.

**Proposition 2.15.** For an arbitrary non-empty set $\zeta \subseteq V_4$ the following statements hold:
 (i) $I^+(\zeta)$ is open.
 (ii) $I^+(I^+(\zeta)) = I^+(\zeta)$; $J^+(J^+(\zeta)) = J^+(\zeta)$.
 (iii) $\underline{I^+(\bar{\zeta})} = \overline{I^+(\zeta)}$.
 (iv) $\overline{I^+(\zeta)} = \{x \in V_4 \,|\, I^+(x) \subseteq I^+(\zeta)\}$.
 (v) $\underline{J^+(\zeta)} \subseteq \overline{I^+(\zeta)}$.
 (vi) $\overline{J^+(\zeta)} = \overline{I^+(\zeta)}$; $\partial J^+(\zeta) = \partial I^+(\zeta)$; int $J^+(\zeta) = I^+(\zeta)$.

*Proof.* It is enough to prove (i) for a point $p \in V_4$, as the union of open sets is open. Take any point $q \in I^+(p)$, so that there is an f-d timelike curve $\gamma$ from $p$ to $q$. Take $\mathcal{N}_q$ and any point $r \in \gamma \cap C_q^-$. Then, by the fundamental Proposition 2.1, $C_r^+$ is an open neighbourhood of $q$ completely contained in $I^+(p)$, because we can construct an f-d timelike curve from $p$ to any point $s \in C_r^+$ by joining $\gamma$ up to $r$ with the timelike geodesic from $r$ to $s$. Statement (ii) is evident. Point (iii) follows because if there is an f-d timelike curve from $p \in \bar{\zeta}$ to $q$, and as $I^-(q)$ is open, then there are f-d timelike curves from $\zeta$ to $q$. Assertion (iv) is proved similarly in the right direction. The converse also holds because if $x$ is such that $I^+(x) \subseteq I^+(\zeta)$, then any neighbourhood of $x$, which must intersect $C_x^+ \subseteq I^+(x)$, must cut also $I^+(\zeta)$. Finally, (v) and (vi) follow immediately from this. ∎

The second property in the previous proposition is very interesting and has been used in several ways. For example

**Definition 2.14.** A set $\zeta \subseteq V_4$ is called a future set if $I^+(\zeta) \subseteq \zeta$.

From (i) and (ii) of Proposition 2.15 it is clear that $\zeta$ is an *open* future set iff $\zeta = I^+(\zeta)$; open future sets are called simply "future sets" in [94,165]. However, I use here the definition in [107], which seems standard nowadays [239].

**Definition 2.15.** A set $\zeta \subseteq V_4$ is achronal if $I^+(\zeta) \cap \zeta = \emptyset$.

In an achronal set there are no two points that can be joined by a timelike curve. Note, however, that spacelike (or null) curves, surfaces and hypersurfaces do not have to be necessarily achronal. Similarly, we can define acausal sets as those without two points related by a *non-trivial* causal curve. An important example of an achronal set is the boundary of a future set, for if $\zeta$ is a future set then from Proposition 2.15 $\emptyset = \text{int}(\zeta) \cap \partial\zeta = I^+(\zeta) \cap \partial\zeta = I^+(\bar{\zeta}) \cap \partial\zeta$ from where $I^+(\partial\zeta) \cap \partial\zeta = \emptyset$. We collect sets sharing this property in the following

**Definition 2.16.** A proper achronal boundary is the boundary of a future set.

Proper achronal boundaries are both achronal and boundaries of some set. Simple examples are the null cone of any point in flat spacetime, or $E^+(S)$ for any closed spacelike 2-sphere $S$ in flat spacetime. But there are sets which are both achronal and boundaries of a set but are *not* proper achronal boundaries. Take for instance [165] two sufficiently separated null hyperplanes in the strip $0 < x^0 < 1$ of flat spacetime.[4] Other important examples of proper achronal boundaries are the traditional 'particle' and 'event' horizons [173]. For any timelike curve $\gamma$ these can be defined, respectively, as $\partial I^+(\gamma)$ and $\partial I^-(\gamma)$ [163]. If $\gamma$ has a past (or future) endpoint $p$ ($q$), then these sets are simply $\partial I^+(p)$ ($\partial I^-(q)$). If $\gamma$ has no future endpoint, then either the event horizon is empty or it is a proper achronal boundary called the total event horizon. For example, a timelike curve with constant acceleration in flat spacetime has non-empty total event and particle horizons. Other examples are provided by de Sitter models (see Refs. 107,163).

In Definition 2.16 we could have assumed that the future set is open by simply taking its interior. Thus, every non-empty proper achronal boundary $B$ is the boundary of an open future set, say $B^+$, with $I^+(B^+) = B^+$. It is then easy to see that $B$ is also the boundary of $V_4 - B^+$, which is a past set. Let us define $B^- \equiv \text{int}(V_4 - B^+) = I^-(V_4 - B^+)$ so that $B^-$ is an open set and obviously the spacetime is the disjoint union $V_4 = B^+ \cup B^- \cup B$. This decomposition is in fact unique given that $V_4$ is connected (see Ref. 165). However, $I^+(B)$ and $I^-(B)$ do not coincide necessarily with $B^+$ and $B^-$. A very important result regarding singularity theorems is the following.

**Proposition 2.16.** Any proper achronal boundary $B$ is an imbedded three-dimensional $C^{1-}$ submanifold. In other words, $B$ is an imbedded achronal hypersurface without boundary.

*Proof.* Take any $p \in B$, its normal neighbourhood $\mathcal{N}_p$ and any point $q \in C_p^-$. Choose a small enough neighbourhood $\mathcal{U}_p$ of $p$ such that $\mathcal{U}_p \subset C_q^+ \cap \mathcal{N}_p$. As $\mathcal{U}_p$ is a proper subset of $C_q^+$, the Gaussian coordinates (29) are well-defined and the metric takes the simple form (28). Further $\mathcal{U}_p$ can be so chosen that all the f-d timelike geodesics with constant values of $y^i$ in $\mathcal{U}_p$ go from $I^-(p)$ to $I^+(p)$. As they go from $B^- \supset I^-(p)$ to $B^+ \supset I^+(p)$, all of them must cross $B$, and at a unique point, as otherwise there would be a timelike curve between two points of $B$, contradicting the achronality of $B$. Thus, there is a one-to-one map from $B \cap \mathcal{U}_p$ to $\mathbb{R}^3$, assigning to each point of $B$ (transversed by the timelike geodesic marked by $y^i = u^i$) the point in $\mathbb{R}^3$ with coordinates $\{u^i\}$. The imbedding (1) of $B$ can be now locally defined as

$$y^0 = \Phi^0(u), \qquad y^i = \Phi^i(u) = u^i$$

for some function $\Phi^0$ of the $u^i$. We must show that $\Phi^0$ is Lipschitz. To that end, choose any pair of points $r, s \in B \cap \mathcal{U}_p$ and a curve $\gamma$ in $\mathcal{U}_p$ parametrized by $t \in [0, 1]$, joining $r$ to $s$ and defined by $y^0 = \gamma^0(t)$, $y^i = \gamma^i(t) = t \Delta u^i + u^i|_r$, where $\Delta u^i \equiv u^i|_s - u^i|_r$, $\gamma^0(0) = y^0|_r$ and $\gamma^0(1) = y^0|_s$. This curve must be non-timelike in an open interval, as otherwise there would be a timelike curve joining $r, s \in B$, which is impossible. Hence, by choosing $\mathcal{U}_p$ appropriately we can assume that $\gamma$ is non-timelike, which means in our coordinates

$$\left(\frac{d\gamma^0}{dt}\right)^2 \leq g_j(\gamma) \Delta u^i \Delta u^j \leq K^2 \|\Delta u\|^2$$

for some positive constant $K$, which depends on $\mathcal{U}_p$ and $g_j$, and where $\| \|$ stands for the usual norm in $\mathbb{R}^3$. The displayed inequality implies that $|\gamma^0(t) - \gamma^0(0)| \leq K t \|\Delta u\|$ for all $t \in [0, 1]$, and therefore

$$|\gamma^0(1) - \gamma^0(0)| = \left| \Phi^0|_s - \Phi^0|_r \right| \leq K \|\Delta u\|$$

as desired. This finishes the proof. ∎

---

[4] This is why I have changed here the usual terminology: proper achronal boundaries are simply called 'achronal boundaries' in the standard literature.

Any proper achronal boundary can be divided into four disjoint subsets $B_A$, $B_N$, $B_F$, $B_P$ according to the following classification: $B_A$ is acausal; $B_N$ is the set of points through which there passes a null geodesic segment contained in $B$; $B_F$ is the set of future endpoints not in $B_N$ of null geodesic segments in $B$; and similarly $B_P$ for the past endpoints. These are fairly intuitive. The next fundamental result characterizes these subsets in a clear way [163,165]. The notation introduced above is used.

**Proposition 2.17.** Let $B$ be the boundary of the open future set $B^+$ (and of the open past set $B^-$). If there is a neighbourhood $\mathcal{U}_p$ of $p \in B$ such that $B^+ = I^+(B^+ - \mathcal{U}_p)$ then $p \in B_N \cup B_F$. And if $\mathcal{U}_p$ is such that $B^- = I^-(B^- - \mathcal{U}_p)$ then $p \in B_N \cup B_P$.

*Proof.* It is enough to prove the first part. Take any ball $\mathcal{B}_1 \subset \mathcal{U}_p \subset \mathcal{N}_p$ of the type used in the proof of Proposition 2.2. Let $\{p_n\}$ be an infinite sequence of points in $I^+(p) \cap \text{int } \mathcal{B}_1 \subset B^+$ converging to $p$. If the condition $B^+ = I^+(B^+ - \mathcal{U}_p)$ holds, then there are f-d timelike curves from $B^+ - \mathcal{U}_p \subset B^+ - \text{int } \mathcal{B}_1$ to each $p_n$. The same reasoning as in Proposition 2.2 proves that there is a causal f-d limit curve $\gamma$ joining a point in $B^+ \cap \partial \mathcal{B}_1$ to $p$. But $p$ being in $B$, $I^-(p) \subset B^-$ so that $I^-(p) \cap B^+ = \varnothing$. Then, by Proposition 2.1 $\gamma$ must be a null geodesic lying on $B$. ∎

**Corollary 2.3.** Let $B = \partial J^+(\zeta) = \partial I^+(\zeta)$ for some $\zeta \subset V_4$. Then, $B - \bar{\zeta} \subseteq B_N \cup B_F$.

*Proof.* For any $p \in B - \bar{\zeta}$ there is a neighbourhood $\mathcal{U}_p$ such that $\bar{\zeta} \cap \mathcal{U}_p = \varnothing$, and then $I^+[I^+(\zeta) - \mathcal{U}_p] = I^+(\zeta) = B^+$. Thus, the previous proposition assures that $p \in B_N \cup B_F$. ∎

In other words, $B = \partial J^+(\zeta)$ is constituted by $B_A$ and by null geodesic segments which either are past endless — so that they are not in $E^+(\zeta)$ — or have past endpoints at $\bar{\zeta}$, and dually for the past. As already explained, within the 'small' spacetimes $(\mathcal{N}_p, \mathbf{g})$ and for any compact set $\zeta \subset \mathcal{N}_p$, $\partial J^\pm(\zeta)$ are closed and equal to $E^\pm(\zeta)$, so that there are no null geodesic generators of $\partial J^+(\zeta)$ without past endpoints at $\zeta$ in $(\mathcal{N}_p, \mathbf{g})$.

This is the basic idea one wishes to keep as 'reasonable' for the causal structure of physically realistic spacetimes. Obviously, this cannot be true if there are closed or almost closed causal curves through $p \in V_4$ because these curves have the effect of making $\partial C_p^+$ to be a subset of $I^+(p)$. The hierarchy of causality conditions generally used is collected next.

**Definition 2.17.** A spacetime $(V_4, \mathbf{g})$ satisfies:
- the chronology condition at $p \in V_4$ if $p \notin I^+(p)$.
- the causality condition at $p \in V_4$ if $J^+(p) \cap J^-(p) = \{p\}$.
- the future distinguishing condition at $p \in V_4$ if $I^+(q) \neq I^+(p)$ for all $q \neq p$. Similarly for the past distinguishing condition.
- the strong causality condition at $p \in V_4$ if there are arbitrarily small neighbourhoods of $p$ which no f-d causal curve intersects in a disconnected set.
- the stable causality condition if there is a function whose gradient is timelike everywhere.
- that it is causally simple if it is distinguishing and $J^\pm(p)$ are closed for every $p \in V_4$.
- that it is globally hyperbolic if it satisfies the strong causality condition and $J^+(p) \cap J^-(q)$ is compact for all $p, q \in V_4$.

The above conditions have been given in increasing order of restriction, so that if any of them holds then all the previous hold too. A spacetime is said to satisfy the chronology condition, causality condition, etcetera, if the corresponding condition holds for every $p \in V_4$. The chronology condition means that there are no closed f-d timelike curves through $p$. Analogously, the causality condition forbids the existence of closed f-d causal curves. If the chronology condition fails at $p$, then $I^+(p) \cap I^-(p) \ni p$ and, in fact, the chronology condition fails to hold at all $q \in I^+(p) \cap I^-(p)$ as is evident. Furthermore, if $r \in I^+(p) \cap I^-(p)$ and $r \in I^+(q) \cap I^-(q)$ then $I^+(p) \cap I^-(p) = I^+(q) \cap I^-(q)$. Thus, the set of points at which the chronology condition does not hold is the disjoint union of sets of the form $I^+(p) \cap I^-(p)$ [37]. Similar results hold for the causality condition. As a curious result, the boundary of the set of points where the causality condition is violated is a proper achronal boundary [37].

**Definition 2.18.** A spacetime is called totally vicious if $I^+(p) \cap I^-(p) = V_4$ for some $p \in V_4$.

**Proposition 2.18.** If $(V_4, g)$ is totally vicious, then $I^\pm(\zeta) = J^\pm(\zeta) = V_4$ for all $\zeta \neq \varnothing$.

*Proof.* As the spacetime is totally vicious, there is a $p \in V_4$ such that $I^+(p) \cap I^-(p) = V_4$. Obviously, this condition holds for all $q \in V_4$, as $q \in I^+(p) \cap I^-(p)$, so that for all $q \in V_4$ we have $I^+(q) = J^+(q) = V_4$. ∎

There are numerous reasons for imposing the causality and chronology conditions, nonetheless there are many spacetimes which violate them. The most famous example is Gödel spacetime [95] (see also Refs. 107,179). Other examples are provided by the next proposition [10].

**Proposition 2.19.** If $V_4$ is compact, then it does not satisfy the chronology condition.

*Proof.* We can cover $V_4$ with open sets of the form $I^+(p)$, and extract a finite sub-cover $\{I^+(p_i)\}_{i=1,\ldots,n}$. So, $p_1$ must be in some of the $I^+(p_i)$, and this $p_i$ in some other, and so on. As there are only a finite number of the $p_i$, at least one of them must satisfy $p_i \in I^+(p_i)$. ∎

This poses some doubts about the interest of compact spacetimes. There are other reasons for ruling them out [107]. Another interesting result is that if the null convergence and the null generic condition hold, then the chronology and causality conditions are equivalent [107].

The problem is that not even with spacetimes satisfying the causality condition are the causal paradoxes solved. For example, one can have f-d causal curves starting at $p$ and passing arbitrarily near $p$ again, so that a particle travelling along this curve cannot affect $p$ itself, but it may influence most of the causal future of $p$ after a loop. This is the idea behind the distinguishing condition, because we have

**Proposition 2.20.** The future distinguishing condition holds at $p$ iff there are arbitrarily small neighbourhoods of $p$ which no f-d causal curve from $p$ intersects in a disconnected set.

*Proof.* First of all, let us specify what is meant by arbitrarily small neighbourhoods: essentially this means that such a neighbourhood can be found with<u>in *any*</u> neighbourhood of $p$. The proof need only treat the case with $q \in I^+ (\underline{p}), \text{as}$ otherwise there would be a neighbourhood $\mathcal{U}_q$ of $q$ not intersecting $I^+(p)$, and <u>thus,</u> $I^+(q) \cap \mathcal{U}_q \neq \emptyset$ but $I^+(q) \cap \mathcal{U}_q \cap I^+(p) = \emptyset$. In other words, if $q \notin \overline{I^+(p)}$ (implying $q \neq p$), then $I^+(q) \neq I^+(p)$. Let us assume then that $q \in I^+(p)$ with $q \neq p$ and also that there are arbitrarily small neighbourhoods of $p$ which no causal curve cuts in a disconnected a set. We must prove that $I^+(q) \neq I^+(p)$. To that end, as $V_4$ is Hausdorff there are open neighbourhoods $\mathcal{U}_q$ and $\mathcal{U}_p$ of $q$ and $p$, respectively, such that $\mathcal{U}_q \cap \mathcal{U}_p = \emptyset$. Suppose that $s \in I^+(p) \cap I^+(q) \cap \mathcal{U}_q$ (if there is no such $s$, we are finished). Then, $\underline{q \in I^-(s)}$ and there is an open neighbourhood $\mathcal{V}_q \subset \mathcal{U}_q \subset I^-(s)$. As $q \in I^+(p)$, $\mathcal{V}_q$ must cut $I^+(p)$, so that there is an $r \in \mathcal{V}_q \cap I^+(p)$. The f-d timelike curves joining $p$ and $r$ must inevitably leave $\mathcal{U}_p$, for $\mathcal{V}_q \cap \mathcal{U}_p = \emptyset$. Let $\gamma$ be one of these curves. But $r \in \mathcal{V}_q \subset I^-(s)$, so that $s \in I^+(r)$ and there are f-d timelike curves from $r$ to $s$. Let $\tilde{\gamma}$ be one of these. Joining $\gamma$ with $\tilde{\gamma}$ we get an f-d timelike curve which leaves $\mathcal{U}_p$ and then goes back to $s \in \mathcal{U}_q$. This proves that if $s \in I^+(p) \cap I^+(q) \cap \mathcal{U}_q$, there is an f-d timelike curve which leaves and re-enters $\mathcal{U}_p$. Therefore, if every $s \in I^+(q) \cap \mathcal{U}_q$ were also in $I^+(p)$, for all neighbourhoods $\mathcal{V}_p \subset \mathcal{U}_p$ there would be an f-d timelike curve intersecting $\mathcal{V}_p$ in a disconnected set, against hypothesis. Then, there must be at least an $s \in I^+(q) \cap \mathcal{U}_q$ which is not in $I^+(p)$, or equivalently, $I^+(q) \neq \underline{I^+(p)}$.

Conversely, suppose <u>that</u> for all $q \in I^+(p)$, $q \neq p$ we have $I^+(q) \neq I^+(p)$. Given that $q \in I^+(p)$, so that $I^+(q) \subseteq I^+(p)$, there must be at least a point $r_q \in I^+(p) - I^+(q)$. As $p \in I^-(r_q)$, there is a neighbourhood $\mathcal{V}_p \subset I^-(r_q)$, and this $\mathcal{V}_p$ cannot intersect $J^+(q)$, as otherwise there would be f-d causal curves from $q$ to $\mathcal{V}_p$ which combined with the f-d t<u>imelike</u> curves from $\mathcal{V}_p$ to $r_q$ would imply $r_q \in \overline{I^+(q)}$. Therefore, for all $q \in I^+(p)$, $q \neq p$, there exists a neighbourhood of $p$, $_q\mathcal{V}_p$, satisfying $_q\mathcal{V}_p \cap J^+(q) = \emptyset$. Let $\mathcal{U}_p$ be a neighbourhood of $p$ with compact <u>closure</u> within $\mathcal{N}_p$ and let $\gamma$ be an f-d causal curve <u>leaving</u> $\mathcal{U}_p$. Let $r \in \overline{\mathcal{U}_p}$ be the first point of $\gamma$ not in $\mathcal{U}_p$. Obviously, $r \in I^+(p)$, so that from the above there is an open neighbourhood $_r\mathcal{V}_p$ of $p$ with $_r\mathcal{V}_p \cap J^+(r) = \emptyset$. But $\gamma$ from $r$ to the future is, itself, in $J^+(r)$, so that once $\gamma$ has left $\mathcal{U}_p$ through $r$ it cannot come back to $_r\mathcal{V}_p$. ∎

**Corollary 2.4.** *The future (or past) distiguishing condition implies the causality condition.* ∎

Although the distinguishing condition may seem to forbid all uncomfortable direct causal violations, this is not so again. In fact, it may happen that there is no curve from $p$ which ever comes back close to $p$, but there is a curve starting *close* to $p$ and coming back close to $p$ again. The idea behind strong causality is precisely to avoid such situations. Another way of putting it is that there is a neighbourhood $\mathcal{U}_p$ such that for all $q, r \in \mathcal{U}_p$ with $r \in I^+(q)$ the sets $I^+(q) \cap I^-(r)$ are completely contained within $\mathcal{U}_p$. Evidently, strong causality implies the distinguishing condition, but it is stronger. Further, strong causality assures that any limit curve of a sequence of causal curves is causal. As a bonus, strong causality forbids another causal anomaly usually called "imprisonment" [12,37,107], and which occurs in some well-known spacetimes such as Taub–NUT or similar [107,146,179].

**Proposition 2.21.** If strong causality holds on a compact set $\mathcal{K}$, there is no future endless causal curve remaining in $\mathcal{K}$, or which enters and re-enters infinitely many times in $\mathcal{K}$.

*Proof.* Cover $\mathcal{K}$ with normal neighbourhoods of compact closure and extract a finite sub-cover $\{\mathcal{U}_i\}_{i=1,\ldots,n}$. As strong causality holds, then every f-d causal curve cannot re-enter into any of the $\mathcal{U}_i$ once it has left it. In consequence, any future endless causal curve $\gamma$ cannot remain in $\mathcal{K}$. Similarly, $\gamma$ can leave $\mathcal{K}$ and then re-enter into $\mathcal{K}$ through a $\mathcal{U}_i$ not yet crossed by $\gamma$, but this can be done only a finite number of times. ∎

Let us prove an important result to be used repeatedly in the proofs of singularity theorems and which, by itself, leads to a kind of preliminary singularity theorem as a corollary.

**Proposition 2.22.** If $(V_4, g)$ satisfies the chronology condition and every endless null geodesic contains a pair of conjugate points then strong causality holds.

*Proof.* Suppose that the strong causality condition failed at $p \in V_4$. Take the normal neighbourhood $\mathcal{N}_p$ and a sequence of nested neighbourhoods $\{\mathcal{U}_n\}$ in $\mathcal{N}_p$ converging to $p$. For each $n$, there would be an f-d causal curve $\gamma_n$ starting at $\mathcal{U}_n$, leaving $\mathcal{N}_p$ and then returning to $\mathcal{U}_n$. By Proposition 2.2 there would be an endless causal limit curve $\gamma$ passing through $p$. But this contradicts the chronology condition because, even in the case that $\gamma$ were a null geodesic, it would contain a pair of conjugate points and thus, by Proposition 2.14, there would be a timelike curve joining points of $\gamma$ and, *a posteriori*, a closed timelike curve constructed with $\gamma$ and the adequate $\gamma_n$ for large enough $n$. ∎

**Corollary 2.5.** If the null convergence, chronology, and generic conditions hold, then either the spacetime satisfies the strong causality condition or is null geodesically incomplete.

*Proof.* Immediate because from Proposition 2.8 every null geodesic either has a pair of conjugate points or is incomplete. ∎

As briefly mentioned before, most singularity theorems will prove the existence of incomplete geodesics. In this sense, the previous corollary simplifies matters considerably as one can assume the strong causality condition without restricting the problem provided that the generic, chronology, and null convergence conditions hold. It seems at first sight that strong causality is a 'strong' enough restriction on the causal violations of the spacetime. However, even though strong causality is somehow a minimal sufficient requirement, it is not completely satisfactory because we can still have f-d causal curves starting arbitrarily close to $p_1$, never passing again close to $p_1$ and going arbitrarily near $p_2$, then f-d causal curves starting arbitrarily close to $p_2$ and passing arbitrarily near $p_1$. And the same with $p_1, p_2, p_3$ and so on. In fact, as argued by Carter [37], there is an infinite number of such higher and higher conditions holding up to some degree and being violated by the next. The stable causality condition was devised by Hawking [105] to finish once and for all with the above unsatisfactory situation. A good reference on this subject is [192]. The basic idea of stable causality is that one can modify — 'open slightly' — the null cones of the spacetime at *every* point without violating the causality condition.

**Proposition 2.23.** A spacetime is causally stable if and only if there exists a continuous timelike vector field $\vec{v}$ such that the modified spacetime $(V_4, \tilde{g})$, with $\tilde{g}_{\mu\nu} = g_{\mu\nu} - v_\mu v_\nu$, satisfies the causality condition.

The modified spacetime $(V_4, \tilde{g})$ has the necessary property that every causal vector $\vec{k}$ in $(V_4, g)$ is a timelike vector in $(V_4, \tilde{g})$, because

$$\tilde{g}(\vec{k}, \vec{k}) = g(\vec{k}, \vec{k}) - [g(\vec{v}, \vec{k})]^2 < g(\vec{k}, \vec{k}) \leq 0$$

or, in other words, the null cones of $(V_4, \tilde{g})$ have been opened up.

*Proof.* Suppose that stable causality holds in $(V_4, g)$. Then, there is a differentiable function $f$ such that $-df$ is timelike everywhere (and future-pointing, without loss of generality). First of all, let us remark that $f$ is a time function, in the sense that it increases along every f-d causal curve in $(V_4, g)$. Indeed, as $-df$ is future-pointing and timelike, $k^\mu \partial_\mu f > 0$ for all future-pointing causal vectors of $(V_4, g)$. Thus, in $(V_4, g)$ there cannot be closed causal curves, and the causality condition holds. Take now the modified spacetime $(V_4, \tilde{g})$. It is easy to see that $df$ is still timelike everywhere in $(V_4, \tilde{g})$ for all choices of $\vec{v}$ such that

$$(v^\mu \partial_\mu f)^2 < -(g^{\mu\nu} \partial_\mu f \, \partial_\nu f)(1 - v^\mu v_\mu)$$

(in particular for $\mathbf{v} = df$ itself; Ref. 239). Then, all these modified spacetimes satisfy the causality condition too. For the converse, put a measure in the spacetime not related to the volume element and with total finite volume. Take the volume of the past of any point in this measure as the value of the time function at that point (see the proof of Proposition 2.25). This gives a non-continuous time function, in general, but the fact that it is also a time function in the slightly modified metrics allows its average over a range of these metrics to obtain the required continuous time function. The differentiable one can then be produced by smoothing on normal neighbourhoods. The technicalities can be looked up in the complete reference [192] (see also Refs. 105,107). ∎

**Corollary 2.6.** The stable causality condition implies the strong causality condition.

*Proof.* Take any $p \in V_4$, its normal neighbourhood $\mathcal{N}_p$ and let $f|_p$ be the value of the time function at $p$. It is very easy to see that we can choose points $r \in C_p^-$ and $q \in C_p^+$ such that the neighbourhood of $p$ defined by $C_r^+ \cap C_q^-$ has the property that all f-d causal curves enter into $C_r^+ \cap C_q^-$ at values of $f < f|_p$, and all f-d causal curves leave $C_r^+ \cap C_q^-$ at values of $f > f|_p$. As $f$ increases along every f-d causal curve, if any of these curves ever enters and leaves $C_r^+ \cap C_q^-$, then it will not enter again. ∎

In a stably causal spacetime, the hypersurfaces $f =$ const. are spacelike with future-pointing normal one-form $-df$, so that the spacetime is foliated by these hypersurfaces. In fact, as $f$ increases along all f-d directed causal curves, every causal curve can intersect each $f =$ const. at most once. In other words, the hypersurfaces $f =$ const. are acausal. The sets with these properties are called partial Cauchy hypersurfaces, see Definition 2.19. Very recent results on time functions and their porperties under cosmological conditions can be found in [5]. In particular, the relationship between the so-called regular cosmological time functions and global hyperbolicity is studied.

Stable causality is recognized as the proper condition avoiding any possible causal violation or paradox. Still, it does not recover the simple causal structure of normal neighbourhoods shown before, i.e., the closedness of $J^\pm(p)$. For example, Minkowski spacetime with point $(1, 1, 0, 0)$ deleted is stably causal, but $J^+(0, 0, 0, 0)$ is not closed. This problem is not related to causal violation but rather to the global causal structure of the spacetime. Whether or not such 'awkward' behaviours should be forbidden on physical grounds is not clear to me. In any case, the conditions eliminating them are causal continuity and causal simplicity. Causal continuity is an intermediate condition between stable causality and causal simplic-ity [109], but it will not be considered here. Causally simple spacetimes were defined above as those distinguishing spacetimes such that $J^{\pm}(p)$ are closed for every $p \in V_4$. It must be noticed that the requirement of the distinguishing condition is necessary (compare Ref. 107), as otherwise totally vicious spacetimes are causally simple because of Proposition 2.18 (see Ref. 215).

**Proposition 2.24.** A distinguishing spacetime is causally simple if and only if $\partial J^+(p) = E^+(p)$ and $\partial J^-(p) = E^-(p)$ for every $p \in V_4$.

*Proof.* If $J^+(p)$ is closed then $\overline{J^+(p)} = J^+(p)$, so that $\partial J^+(p) \equiv \overline{J^+(p)} -$ int $J^+(p) = J^+(p) - I^+(p) = E^+(p)$. The converse is similar. ∎

The strength of causal simplicity is defined by the next result, where the necessity of assuming the distinguishing condition appears again.

**Proposition 2.25.** Any causally simple spacetime satisfies the stable causality condition.

*Proof.* There are several ways to prove this. One which will be helpful later is as follows. First of all, let us remark that in any causally simple $(V_4, g)$ we have $q \in J^+(p) = \overline{J^+(p)}$ iff $p \in J^-(q) = \overline{J^-(q)}$. This implies that the spacetime is reflecting [109], i.e. $I^+(q) \subset I^+(p)$ iff $I^-(p) \subset I^-(q)$, because of statement (iv) in Proposition 2.15. In fact, $V_4$ is strictly reflecting, that is, $I^+(q) \subsetneq I^+(p)$ iff $I^-(p) \subsetneq I^-(q)$ because it is distinguishing ($I^{\pm}(p) = I^{\pm}(q)$ iff $p = q$). For the same reasons, $J^+(q) \subset J^+(p)$ iff $J^-(p) \subset J^-(q)$ for $p \neq q$. These properties are more than needed here, but are interesting for the global hyperbolicity of next subsection. Put an additive measure $\mathcal{M}$ in $V_4$ such that the volume of each open set is positive and the total volume of $V_4$ is finite in $\mathcal{M}$ (see Ref. 91). For each $p \in V_4$, define $t^-(p)$ as the volume of $J^-(p)$ in the measure $\mathcal{M}$. Evidently, $t^-$ is bounded and, from the above properties, strictly increasing along every f-d causal curve $\gamma$ and continuous along $\gamma$. To see this last part, take any infinite sequence of points $\{p_n\}$ converging to $p \in \gamma$ and such that $p_n \in J^-(p)$ for all $n$. Every $q \in J^-(p_n)$ for all $n$ must be in $J^-(p)$, because all $p_n \in J^+(q)$ which implies by causal simplicity $p \in J^+(q) = \overline{J^+(q)}$. Thus $t^-(p_n) \to t^-(p)$ and the same happens for any infinite sequence $\{p_n\}$ converging to $p \in \gamma$ and such that $p_n \in J^+(p)$ for all $n$. In fact, $t^-$ can be smoothed to a differentiable time function using any standard procedure (see Ref. 192). ∎

Finally, the last given condition, global hyperbolicity, includes all others.

**Proposition 2.26.** Any globally hyperbolic spacetime is causally simple.

*Proof.* If there were a point $q \in \overline{J^+(p)} - J^+(p)$ for some $p \in V_4$, it would follow that $I^+(q) \subseteq I^+(p)$. For any $s \in I^+(q) \subseteq I^+(p)$ define $J^+(p) \cap J^-(s)$. Then $q$ would be in $\overline{J^+(p) \cap J^-(s)}$ but not in $J^+(p) \cap J^-(s)$. But this is impossible as $J^+(p) \cap J^-(s)$ is compact. ∎

However, global hyperbolicity is much more than a causality condition as is intimately related to the existence of maximal curves in the spacetime, and to the existence of acausal 3-dimensional hypersurfaces which *every* causal curve meets. The next subsection is devoted to a summary of all these fundamental properties. The basic references are [91,191].

### 2.4. Global hyperbolicity and maximal curves

First of all, let us define the past Cauchy development or past domain of dependence of a set $\zeta$ and its past Cauchy horizon, respectively, as follows

$$D^-(\zeta) \equiv \{x \in V_4 \,|\, \text{every f-d endless causal curve from } x \text{ meets } \zeta\},$$
$$H^-(\zeta) \equiv \overline{D^-(\zeta)} - I^+[D^-(\zeta)]$$

and dually for the future domain of dependence $D^+(\zeta)$ and future Cauchy horizon $H^+(\zeta)$. Naturally, $D^-(\zeta) \supset \zeta$. We also denote $D(\zeta) \equiv D^+(\zeta) \cup D^-(\zeta)$ and $H(\zeta) \equiv H^+(\zeta) \cup H^-(\zeta)$, called the total domain of dependence and total Cauchy horizon, respectively. The Cauchy horizons are the causal boundaries of the domains of dependence and, by their definition, are always closed. As an example, in a normal neighbourhood $\mathcal{N}_p$ of $p$ we can take any spacelike hypersurface $\tau = $ const. as defined in Lemma 2.1. Then, $H^-(\tau = \text{const.}) = \partial C_p^+$ within $(\mathcal{N}_p, g)$. In general, if $\zeta$ is achronal, then $D^-(\zeta) \cap I^+(\zeta) = \emptyset$. The set $D^+(\zeta)$ is important on physical grounds because if energy and matter propagate causally, then all possible physical events in $D^+(\zeta)$ are influenced *exclusively* by the matter and energy on $\zeta$. Thus, $H^+(\zeta)$ is the future boundary where this happens. In fact, it can be proved that if DEC holds on $D(\zeta)$ and the energy-momentum tensor vanishes at $\zeta$ then it vanishes on the whole of $D(\zeta)$ [107]. Recently, the same result has been found for the curvature tensor in vacuum [23].

**Lemma 2.4.** For any closed achronal set $\zeta$, $H^+(\zeta)$ is achronal and int $D^+(\zeta) = I^+(\zeta) \cap I^-[D^+(\zeta)]$.

*Proof.* From its definition, $I^-[H^+(\zeta)] \subset I^-[D^+(\zeta)] \subset V_4 - H^+(\zeta)$, so $H^+(\zeta) \cap I^-[H^+(\zeta)] = \emptyset$. For the second part, if $p \in $ int $D^+(\zeta)$ then obviously $p \in I^+(\zeta)$ and also there is a neighbourhood $\mathcal{U}_p$ of $p$ within int $D^+(\zeta)$, so that there is a $q \in C_p^+ \cap \mathcal{U}_p \subset $ int $D^+(\zeta)$ which implies that $p \in I^-(q) \subset I^-[D^+(\zeta)]$. Conversely, if $p \in I^+(\zeta) \cap I^-[D^+(\zeta)]$, then there is a neighbourhood $\mathcal{U}_p$ such that $\mathcal{U}_p \subset I^-(q) \cap I^+(r)$ for $q \in D^+(\zeta)$ and $r \in \zeta$. For all $s \in \mathcal{U}_p$, $r \in I^-(s) \subset I^-(q)$ so that all endless past-directed causal curves from $s$ must meet $\zeta$. ∎

By similar methods one can prove [107,165]

**Lemma 2.5.** For any closed achronal set $\zeta$, $\overline{D^+(\zeta)}$ is the set of $p \in V_4$ such that every past-directed endless timelike curve from $p$ meets $\zeta$. Then, $I^+[H^+(\zeta)] = I^+(\zeta) - D^+(\zeta)$ and $\partial D^+(\zeta) = H^+(\zeta) \cup \zeta$. ∎

The Cauchy horizons will have properties very similar to those of proper achronal boundaries, except for the possibility of having an 'edge'. The precise definition is

$$\text{edge}(\zeta) \equiv \{x \in \bar{\zeta} \mid \text{every neighbourhood } \mathcal{U}_x \text{ of } x \text{ contains}$$
$$\text{points } p \in C_x^-\text{ and } q \in C_x^+ \text{ and}$$
$$\text{an f-d timelike curve from } p \text{ to } q \text{ not meeting } \zeta\}.$$

Notice that curves or surfaces have an edge equal to themselves. Points in the edge $(\zeta)$ are points in $\bar{\zeta} - \zeta$ together with those in which $\zeta$ is not a continuous 3-dimensional manifold. Evidently, proper achronal boundaries have no edge. If any closed achronal set has no edge (called edgeless), then an argument identical to that in Proposition 2.16 proves that it is an imbedded 3-dimensional $C^{1-}$ submanifold. In particular, $H^+(\zeta)$ will satisfy this whenever edge$[H^+(\zeta)] = \emptyset$. But this happens if and only if edge$(\zeta) = \emptyset$ because of the general result:

**Lemma 2.6.** For a closed achronal set $\zeta$, $I^+[\text{edge}(\zeta)] \cap \overline{D^+(\zeta)} = \emptyset$ and edge$[H^+(\zeta)] = $ edge$(\zeta)$.

*Proof.* For the first part it is not necessary to assume that $\zeta$ is closed. If $s \in I^+(p)$ for some $p \in $ edge$(\zeta)$, there is a neighbourhood $\mathcal{U}_s$ of $s$ in $I^+(p)$. Given that for every neighbourhod of $p$ there are points $q \in C_p^-$, $r \in C_p^+$ and an f-d timelike curve $\gamma$ from $q$ to $r$ not meeting $\zeta$, then we can join up a past-directed timelike curve from any point in $\mathcal{U}_s$ to $r$ — not meeting $\zeta$ because $\zeta$ is achronal — with $\gamma$ and then extend it to the past of $q$ indefinitely. This curve does not intersect $\zeta$, so that $\mathcal{U}_s \cap D^+(\zeta) = \emptyset$, hence $s \notin \overline{D^+(\zeta)}$. For the second part take any $p \in $ edge$(\zeta)$ so that $p \in D^+(\zeta)$. The first part of the lemma implies $I^+(p) \cap \overline{D^+(\zeta)} = \emptyset$ so that $p \notin I^-[D^+(\zeta)]$, and thus $p \in \overline{D^+(\zeta)} - I^-[D^+(\zeta)] = H^+(\zeta)$. To see that in fact $p \in $ edge$[H^+(\zeta)]$, remember that every neighbourhood $\mathcal{U}_p$ of $p$ contains points $q \in C_p^-$, $r \in C_p^+$ and an f-d timelike curve $\gamma$ from $q$ to $r$ not meeting $\zeta$. This very $\gamma$ cannot meet $H^+(\zeta)$, because every past-directed inextendible timelike curve from $H^+(\zeta)$ intersects $\zeta$ (Lemma 2.5 *when $\zeta$ is closed*), $\zeta$ is achronal and $q \in I^-(\zeta)$. The converse is similar. ∎

**Proposition 2.27.** If $\zeta$ is a closed achronal set, $H^+(\zeta)$ is generated by null geodesics segments which are either past endless or have a past endpoint at edge $(\zeta)$.

*Proof.* The idea is the same as in Proposition 2.17 and its Corollary 2.3. Take the set $B^+ \equiv I^+(\zeta) - D^+(\zeta)$, which by Lemma 2.5 is an open future set because $B^+ = I^+[H^+(\zeta)]$. Take its boundary $B \equiv \partial B^+$, which is a proper achronal boundary by Definition 2.16. As $H^+(\zeta)$ is achronal and $B^+ = I^+[H^+(\zeta)]$, we have $H^+(\zeta) \subseteq \overline{B^+} - B^+ = B$, so that $H^+(\zeta)$ is a closed subset of $B$. If $p \in H^+(\zeta) - \zeta$, then there is an f-d timelike curve $\gamma$ from $\zeta$ to $p$. Take any point $q \in \gamma \cap C_p^-$ and a neighbourhood $\mathcal{U}_p$ of $p$ such that $\mathcal{U}_p \subset C_q^+$. From every $r \in \mathcal{U}_p \cap B^+$, and given that $B^+ = I^+(\zeta) - D^+(\zeta)$, there exists a past-directed endless timelike curve not meeting $\zeta$, and therefore not meeting $H^+(\zeta)$ either. This curve must intersect $\mathcal{U}_p$ at $s$, say. There is also an f-d timelike curve from $q$ to $s$ not passing through $\mathcal{U}_p$ but necessarily crossing $H^+(\zeta)$, as $p \notin \text{edge}[H^+(\zeta)]$. Joining these two curves, an f-d timelike curve from $H^+(\zeta) - \mathcal{U}_p$ to any $r \in \mathcal{U}_p \cap B^+$ is constructed and thus $\mathcal{U}_p \cap B^+ \subset I^+[H^+(\zeta) - \mathcal{U}_p]$. This means the condition of Proposition 2.17 holds, and hence $p \in B_N \cup B_F$. Analogously, if $p \in (H^+(\zeta) \cap \zeta) - \text{edge}(\zeta)$, choose the neighbourhood $\mathcal{U}_p$ within $C_q^+ \cap C_r^-$ for some points $q \in I^-(\zeta)$ and $r \in I^+(\zeta)$, and such that every timelike curve in $C_q^+ \cap C_r^-$ meets $H^+(\zeta)$ and $\zeta$. Then, the same reasoning as before proves that $p \in B_N \cup B_F$. ∎

**Definition 2.19.** Every edgeless closed acausal set $\Sigma$ is called a partial Cauchy hypersurface. If also $D(\Sigma) = V_4$, then $\Sigma$ is a global Cauchy hypersurface.

Global Cauchy hypersurfaces are referred to simply as Cauchy hypersurfaces. Evidently, $\Sigma$ is a Cauchy hypersurface iff $H(\Sigma) = \varnothing$. These definitions can be weakened by relaxing the acausality to mere achronality. A very simple way to know whether any given $\Sigma$ is a Cauchy hypersurface is by using the next result [91,165].

**Proposition 2.28.** A closed acausal set $\Sigma$ is a Cauchy hypersurface if and only if every endless null geodesic meets $\Sigma$.

*Proof.* If $\Sigma$ is a Cauchy hypersurface then every f-d endless causal curve meets $\Sigma$, in particular every endless null geodesic. Conversely, let us prove first that $\Sigma$ is edgeless. If every f-d endless null geodesic $\gamma$ from $p \notin \Sigma$ meets $\Sigma$, then we can take $q \in \gamma$ before $\gamma$ meets $\Sigma$ and another f-d endless null geodesic from $q$ which by hypothesis must also meet $\Sigma$. By Proposition 2.13 there is an f-d timelike curve from $p$ to $\Sigma$, so that $p \in I^-(\Sigma)$. Similarly, if every endless past-directed null geodesic from $p \notin \Sigma$ meets $\Sigma$, then $p \in I^+(\Sigma)$. As $\Sigma$ is acausal, the only remaining possibility is $p \in \Sigma$. Thus, $\partial I^+(\Sigma) = \partial I^-(\Sigma) = \Sigma$, so that $\Sigma$ is a proper *acausal* boundary, whence edgeless. Now, as $\Sigma$ is acausal and edgeless, by Proposition 2.27 all hypothetical null generators of the would-be $H(\zeta)$ cannot meet $\Sigma$, against hypothesis. Hence, $H(\zeta) = \emptyset$.                                                                        ∎

In fact, as is obvious, every null geodesic not merely intersects $\Sigma$ but crosses it and enters into $I^+(\Sigma)$ or $I^-(\Sigma)$. This result can be established in full generality.

**Lemma 2.7.** Let $\zeta$ be any closed achronal set. Then
  (i) if $p \in D^-(\zeta) - H^-(\zeta)$, every f-d endless curve from $p$ meets $\zeta - H^-(\zeta)$ and $I^+(\zeta)$.
 (ii) if $p \in \text{int } D(\zeta)$, then every past and future endless curve through $p$ intersects both $I^\pm(\zeta)$.
(iii) strong causality holds on $\text{int } \overline{D(\zeta).}$
(iv) if strong causality holds on $J^+(\zeta)$, then $H^+[\overline{E^+(\zeta)}]$ is non-compact or empty.

*Proof.* For the first part, let $\gamma$ be any f-d endless causal curve from $p$. Take any point $q \in D^-(\zeta) \cap C_p^-$, which always exists as $p \notin H^-(\zeta)$. Construct an f-d endless causal curve $\lambda$ from $q$ such that for each $r \in \lambda$ there is an $r_\lambda \in \gamma \cap I^+(r)$. This can be easily done by covering $\gamma$ with a locally finite system of normal neighbourhoods and building $\lambda$ step by step (see Ref. 165; alternatively, see Refs. 91,239). Obviously, $\lambda$ must meet $\zeta$ at some point $s$, and $s$ cannot be in $H^-(\zeta)$ because $I^-[H^-(\zeta)] \cap D^-(\zeta) = \emptyset$ by Lemma 2.5. Further, there is a point $s_\lambda \in \gamma \cap I^+(s)$ by construction, so that $\gamma$ enters into $I^+(\zeta)$. Statement (ii) is immediate from the first because $\text{int } D(\zeta) = D(\zeta) - H(\zeta)$ and $\zeta \subseteq D^+(\zeta) \cap D^-(\zeta)$. To prove (iii), notice that the chronology condition holds in $\text{int } D(\zeta)$, as otherwise there would be a closed timelike curve crossing $\zeta$ inevitably and thus violating its achronality. Now suppose that strong causality condition failed at $p \in \text{int } D(\zeta)$ and proceed exactly as in Proposition 2.22 to build an endless causal limit curve $\gamma$ through $p$ of a sequence of f-d causal curves $\{\gamma_n\}$ starting and coming back closer and closer to $p$. According to the second part, there would be points of $\gamma$ in both $I^+(\zeta)$ and $I^-(\zeta)$. So, if $p \in J^+(\zeta)$ there would also be some f-d curves $\gamma_n$, for $n$ big enough, starting at $I^+(\zeta)$ and then entering into $I^-(\zeta)$, violating again the achronality of $\zeta$. If $p \in J^-(\zeta)$ the reasoning is identical by following the $\gamma_n$ in the past direction. Finally, point (iv) is an important result found in [108]. To prove it recall that, by Corollary 2.3, through any point $p \in \partial J^+(\zeta) - \zeta$ there passes a past-directed null segment lying in $\partial J^+(\zeta)$ which is either past endless (if $p \in \partial J^+(\zeta) - E^+(\zeta)$) or has a past-endpoint at $\zeta$ (if $p \in E^+(\zeta)$). In the first case, from statement (i) it follows that $p \notin D^+[\partial J^+(\zeta)] - H^+[\partial J^+(\zeta)]$, because the past-endless null segment remains in $\partial J^+(\zeta)$ and thus it cannot enter into $I^-[\partial J^+(\zeta)]$. In other words,

$$\partial J^+(\zeta) - \overline{E^+(\zeta)} \subseteq H^+[\partial J^+(\zeta)]$$

from which it follows that

$$\overline{D^+[\partial J^+(\zeta)]} - \overline{D^+[\overline{E^+(\zeta)}]} = \partial J^+(\zeta) - \overline{E^+(\zeta)} \subseteq H^+[\partial J^+(\zeta)],$$

because this is the set of points from which there is a past-directed endless timelike curve $\gamma$ intersecting $\partial J^+(\zeta) - \overline{E^+(\zeta)} \subseteq H^+[\partial J^+(\zeta)]$, and this $\gamma$ must in fact start at $\partial J^+(\zeta) - \overline{E^+(\zeta)} \subseteq H^+[\overline{\partial J^+(\zeta)}]$ itself due to the achronality of $H^+[\partial J^+(\zeta)]$ and Lemma 2.5. Immediately we also deduce

$$\operatorname{int} D^+[\overline{E^+(\zeta)}] = \operatorname{int} D^+[\partial J^+(\zeta)],$$
$$H^+[\partial J^+(\zeta)] - H^+[\overline{E^+(\zeta)}] = \partial J^+(\zeta) - \overline{E^+(\zeta)}.$$

Now, suppose that $H^+[\overline{E^+(\zeta)}]$ were compact. If edge$[\overline{E^+(\zeta)}] = \emptyset$ this would already be impossible due to Propositions 2.21 and 2.27. In general, however, there may be a non-empty edge (there is an example in Ref. 108). In any case, we could cover $H^+[E^+(\zeta)]$ with a finite number of normal neighbourhoods $\{\mathcal{U}\}_{i=1,\ldots,n}$ with compact closure, so that the closure $\mathcal{K}$ of their union would still be compact and $\mathcal{K} \supset H^+[E^+(\zeta)]$. Note that every past-directed curve from any $q \in J^+(\zeta)$ to $\zeta$ must cross $D^+[E^+(\zeta)]$, because of the above properties and the fact that $I^-[\partial J^+(\zeta)] \cap \zeta = \emptyset$. If $H^+[E^+(\zeta)]$ were also non-empty, then $J^+(\zeta) - D^+[\partial J^+(\zeta)] \cap \mathcal{K} \neq \emptyset$ and there would be a past-endless causal curve $\gamma_1$ from this set not crossing $\partial J^+(\zeta)$, and thus remaining in $J^+(\zeta)$, nor crossing $D^+[E^+(\zeta)]$. If this curve remained in $\mathcal{K}$, it would contradict strong causality by Proposition 2.21. Suppose $\gamma_1$ left $\mathcal{K}$ and take any $q_1 \in \gamma_1 - \mathcal{K}$. As $q_1 \in J^+(\zeta)$, there would be a past-directed causal curve $\tilde{\gamma}_1$ from $q_1$ to $\zeta$, and thus $\tilde{\gamma}_1$ would have to cross $D^+[E^+(\zeta)]$ which in turns means that $\tilde{\gamma}_1$ would have previously entered into $J^+(\zeta) - D^+[\partial J^+(\zeta)] \cap \mathcal{K}$ again. Choose then another curve $\gamma_2$ having the same properties as $\gamma_1$ and repeat the procedure — if $\gamma_2$ ever leaves $\mathcal{K}$ — constructing the corresponding $\tilde{\gamma}_2$, $\gamma_3$, etcetera. The combination of all these curves would produce a past-endless causal curve entering and re-entering, or remaining, into $\mathcal{K}$, in contradiction with Proposition 2.21 again. ∎

**Proposition 2.29.** For any closed achronal set $\zeta$, $(\operatorname{int} D(\zeta), \mathbf{g})$ is globally hyperbolic.

*Proof.* Let $p, q \in \operatorname{int} D(\zeta)$ and take any infinite sequence of points $\{p_n\}$ in $J^+(p) \cap J^-(q)$. We must find an accumulation point of the $\{p_n\}$ in $J^+(p) \cap J^-(q)$. To that end, choose any sequence of f-d causal curves $\{\gamma_n\}$ from $p$ to $q$ such that $p_n \in \gamma_n$ for each $n$. First, let us prove that there is an f-d causal limit curve of the sequence $\{\gamma_n\}$ which goes from $p$ to $q$. The $\{\gamma_n\}$ are future endless in the spacetime $(V_4 - \{q\}, \mathbf{g})$, and as $p$ is an accumulation point of the sequence, by Proposition 2.2 there is an f-d causal limit curve $\gamma$ passing through $p$ and future-endless *in* the excised spacetime $(V_4 - \{q\}, \mathbf{g})$. Assume $p \in D^-(\zeta)$ (if $p \in D^+(\zeta)$ the argument is dual) and consider the two possibilities: (i) $q \in D^-(\zeta)$ or (ii) $q \in I^+(\zeta) \cap \text{int } D(\zeta)$. In case (i), $\gamma$ cannot enter into $I^+(\zeta)$ because $q \notin I^+(\zeta)$ and $C_q^-$ intersects all the $\gamma_n$. As a consequence of Lemma 2.7 the limit curve $\gamma$ is not future endless. As $\gamma$ is future endless in $V_4 - \{q\}$, $\gamma$ necessarily has a future endpoint at $q$. In case (ii), we can consider $p \in I^-(\zeta)$, as otherwise $p \in \zeta \subset D^+(\zeta)$ and the dual reasoning of case (i) applies, interchanging $q$ and $p$. Now, $\gamma$ does enter into $I^+(\zeta)$, because $q \in I^+(\zeta)$. Take a point $r \in \gamma \cap I^+(\zeta)$ and choose a subsequence $\{\gamma_m\}$ of $\{\gamma_n\}$ converging to the piece of $\gamma$ which goes from $p$ to $r$. Reversing the argument, the subsequence $\{\gamma_m\}$ has $q$ as accumulation point so that there is a past-directed causal limit curve $\tilde{\gamma}$ passing through $q$, which is past endless *in* $(V_4 - \{p\}, \mathbf{g})$ and enters into $I^-(\zeta)$. Obviously then, $\tilde{\gamma}$ cannot remain to the future of $r$ [which is in $I^+(\zeta)$] so that $\tilde{\gamma}$ must pass through $r$, because $r$ is a convergence point of the subsequence $\{\gamma_m\}$. Then, the combination of $\gamma$ from $p$ to $r$ together with $\tilde{\gamma}$ from $r$ to $q$ provides the required limit curve. Summarizing, in all cases there is an f-d causal curve $\lambda$ (say) from $p$ to $q$ which is a limit curve of the $\{\gamma_n\}$. Choose an appropriate subsequence $\{\hat{\gamma}_k\}$ converging to $\lambda$ and the corresponding subsequence $\{\hat{p}_k\}$ of the initial sequence of points. Take any neighbourhood $\mathcal{U}$ of $\lambda$ with compact closure. $\mathcal{U}$ contains all the $\hat{\gamma}_k$, whence all the $\hat{p}_k$, but a finite number, so that there is an accumulation point $\hat{p} \in \overline{\mathcal{U}}$ of the subsequence $\{\hat{p}_k\}$. Evidently, $\hat{p}$ must lie on $\lambda$ because every neighbourhood of $\hat{p}$ contains an infinite number of the $\hat{p}_k$, and then of the $\hat{\gamma}_k$, and $\{\hat{\gamma}_k\}$ converges to $\lambda$. Thus, the arbitrary initial sequence of points $\{p_n\}$ accumulates at $\hat{p} \in \lambda \subset J^+(p) \cap J^-(q)$, so that $J^+(p) \cap J^-(q)$ is compact. ∎

By similar or different [165] methods, one can prove:

**Proposition 2.30.** For a closed achronal set $\zeta$, if $p \in \text{int } D(\zeta)$ then $J^+(\zeta) \cup J^-(p)$ is compact. ∎

A corollary of Proposition 2.29 is that the existence of a Cauchy hypersurface for the spacetime implies global hyperbolicity. As proved by Geroch [91], the converse also holds.

**Proposition 2.31.** A spacetime is globally hyperbolic if and only if it contains a Cauchy hypersurface.

*Proof.* One implication is Proposition 2.29 itself. For the converse, assume the spacetime is globally hyperbolic, and hence $(V_4, \mathbf{g})$ is causally simple by Proposition 2.26 so that the results at the beginning of the proof of Proposition 2.25 and the construction of the time function $t^-$ hold. Analogously, we can construct the dual time function $t^+$, which decreases along every f-d causal curve. Both $t^\pm$ are positive by construction. By Proposition 2.21 every future endless causal curve cannot remain (or indefinitely re-enter) within any compact set, in particular it must leave any set of the form $J^+(p) \cap J^-(q)$, which is compact by hypothesis. This means that $t^+$ must tend to zero along every such $\gamma$, and similarly $t^-$ must tend to zero along any past endless causal curve. Consider the differentiable function $t \equiv t^-/t^+$. The sets $\Sigma_t : \{t = \text{const.}\}$ are acausal, because $t$ strictly increases along every f-d causal curve. In fact, $t$ takes all possible values in $(0, \infty)$ along every causal curve, for $t^-$ goes to zero to the past and $t^+$ tends to zero to the future. Thus, for an arbitrary fixed value $t_0$, the acausal set $\Sigma_{t_0}$ is crossed by all endless causal curves so that $\Sigma_{t_0}$ is a Cauchy hypersurface. ∎

**Corollary 2.7.** If the spacetime contains a Cauchy hypersurface $\Sigma$, then $V_4$ is homeomorphic to $\mathbb{R} \times \Sigma$ by a map such that for all $c \in \mathbb{R}$, the image of $\{c\} \times \Sigma$ in $V_4$ is a Cauchy hypersurface.

*Proof.* Simply take a global timelike vector field $\vec{u}$ and its timelike congruence. Define the projection $\Pi : V_4 \to \Sigma$ by taking each point $x \in V_4$ along the curve of the congruence $\vec{u}$ passing through $x$ until it reaches $\Sigma$ in a unique point $\Pi(x) \in \Sigma$. The required homeomorphism can be simply defined by combining $\Pi$ with the real-valued function $\log t : V_4 \to \mathbb{R}$. Sometimes, this homeomorphism may be improved to a diffeomorphism [192]. ∎

Global hyperbolicity was introduced somewhat differently by Leray [131] (cited, for instance, in Refs. 41,81,91,107,165), and its importance depends on the fact that it allows for the existence and uniqueness of solutions of wave-type equations [81] and leads to the existence of maximal geodesics. This second part is of tremendous importance in singularity theorems. Define the Lorentzian distance $d_{p,q}$ as the least upper bound of $L(p, q; \gamma)$ for all f-d causal $\gamma$ when $q \in J^+(p)$ and zero otherwise. In principle, $d_{p,q}$ may be $\infty$ and is obviously non-symmetric in $p$ and $q$. As we already know, $d_{p,q}$ is continuous when $q \in \mathcal{N}_p$ (see the proof of Proposition 2.9). This is not true in general, but easily provable results are

**Lemma 2.8.** Let $(V_4, g)$ be strongly causal.
(i) Let $\gamma$ be any f-d causal curve from $p$ to $q \in J^+(p)$. Given any $\varepsilon > 0$, there is a neighbourhood $\mathcal{U}$ of $\gamma$ such that no other curve $\lambda \subset \mathcal{U}$ has length bigger than $L(p, q; \gamma) + \varepsilon$.
(ii) If the spacetime is globally hyperbolic, then $d_{p,q}$ is continuous in $p$ and $q$.

[see, e.g., Refs. 107,165,239 for point (i), and Refs. 12,107 for (ii)]. ∎

**Proposition 2.32.** If the spacetime is globally hyperbolic then for all $p, q \in V_4$ with $q \in J^+(p)$ there exists a maximal geodesic arc joining $p$ to $q$.

*Proof.* If $q \in E^+(p)$, then the result is already known (Proposition 2.13). Take then $q \in I^+(p) (= J^+(p) - E^+(p))$ (Propositions 2.26 and 2.24). We first show that $d_{p,q}$ is finite in globally hyperbolic spacetimes. As $J^+(p) \cap J^-(q)$ is compact, we can cover it with a finite number of convex normal neighbourhoods $\{\mathcal{U}\}_{i=1,\ldots,n}$ with compact closure such that none of them contains any f-d causal curve of length greater than any fixed bound $k$. Strong causality holds, so that all f-d causal curves can cross each $\mathcal{U}$ at most once. Thus, $d_{p,q} \leq nk$. It remains to see that there is a causal curve $\gamma$ such that $L(p, q; \gamma) = d_{p,q}$. From Lemma 2.8, $d_{p,q}$ is continuous as a function of $q$ or $p$. Then, by classical results in analysis, $d_{p,q}$ attains its maximum when varying over a compact set. Of course, this does not prove the existence of $\gamma$ yet. However, a maximal $\gamma$ can be easily built as follows. By its definition, $d_{p,q} \geq d_{p,x} + d_{x,q}$ for all $x \in J^+(p) \cap J^-(q)$. Let $p \in \mathcal{U}$ (say with $q \notin \mathcal{U}$), and define the function $f(x) \equiv d_{p,x} + d_{x,q}$. As $f(x)$ is finite and continuous, it attains its maximum value, which is $d_{p,q}$, at some point $s$ in the compact set $\partial \mathcal{U} \cap J^+(p) \cap J^-(q)$, so that $d_{p,s} + d_{s,q} = d_{p,q}$. Let us take the timelike geodesic $\lambda$ starting at $p$ and passing through $s$, which is maximal at all its points within $\mathcal{N}_p$ by Proposition 2.9, so that $d_{p,x} = L(p, x; \lambda)$ for all $x \in \mathcal{N}_p$, which in turn means that the relation $d_{p,x} + d_{x,q} = d_{p,q}$ holds for all $x \in \lambda$ previous to $s$, because $d_{p,q} = d_{p,s} + d_{s,q} = d_{p,x} + d_{x,s} + d_{s,q}$. If there were a last point $r \in \lambda \cap J^-(q) - \{q\}$ such that $\lambda$ is maximal from $p$ to $r$ ($d_{p,r} = L(p, r; \lambda)$ by continuity), then we would have $d_{p,y} > d_{p,r} + d_{r,y}$ for every $y \in \overline{C_r^+} \cap \overline{\mathcal{U}}$ (say), because $d_{r,y} = L(r, y; \lambda)$, $\lambda$ being the unique maximal geodesic from $r$ to $y$ in $C_r^+$. But then

$$d_{p,y} + d_{y,q} > d_{p,r} + d_{r,y} + d_{y,q}, \qquad \forall y \in \overline{C_r^+} \cap \overline{\mathcal{U}},$$

which is impossible because the function $d_{r,y} + d_{y,q}$ must also attain its maximum value $d_{r,q} = d_{r,z} + d_{z,q}$ at some point $z$ in the compact set $\partial U_2 \cap J^+(r) \cap J^-(q)$, and the previous displayed inequality would then imply at $z$ the absurd result $d_{p,z} + d_{z,q} > d_{p,r} + d_{r,q} = d_{p,q}$. Therefore, $\lambda$ is maximal within $J^+(p) \cap J^-(q)$ and the relation $d_{p,x} + d_{x,q} = d_{p,q}$ holds for all $x \in \lambda \cap J^-(q)$. To see finally that $\lambda$ reaches $q$, note that $\lambda$ must leave the compact set $J^+(p) \cap J^-(q)$ (by Proposition 2.21) through some point $\tilde{q} \in \partial J^-(q)$, which actually it has to be $q$ itself as otherwise $\tilde{q} \in E^-(q) - \{q\}$ (by causal simplicity) so that $d_{\tilde{q},q} = 0$ implying $d_{p,\tilde{q}} = d_{p,q}$ (as $\tilde{q}$ lies on $\lambda$), which is absurd because the combination of the maximal timelike geodesic $\lambda$ with the f-d null geodesic from $\tilde{q}$ to $q$ cannot be maximal (Corollary 2.1). ∎

It must be stressed that the converse is not true in general: there are non globally hyperbolic spacetimes such that there is a maximal curve between any pair of causally related points. Further, the maximal geodesic curve in this proposition is not necessarily unique. Similarly, we also have

**Proposition 2.33.** Let $\Sigma$ be a partial Cauchy hypersurface. Then, for all $q \in D^+(\Sigma)$ there is an f-d maximal geodesic from $\Sigma$ to $q$. ∎

Once the existence of maximal curves has been established in globally hyperbolic spacetimes, many of the key results for singularity theorems can be readily deduced. For instance, the following property, closely related to the so-called causal disconnection [12], and leading to a corollary which is, by itself, the fundamental and main argument used in the proof of the celebrated and very powerful Hawking–Penrose singularity theorem [108] (Lemma 5.1).

**Proposition 2.34.** Let $(V_4, g)$ be globally hyperbolic and take two diverging sequences of points $\{p_n\}$ and $\{q_n\}$ such that, for each $n$, $q_n \in J^+(p_n) - \{p_n\}$. Let $\gamma_n$ be a maximal geodesic from $p_n$ to $q_n$. If all such $\{\gamma_n\}$ intersect a compact set $\mathcal{K}$, then there is an endless f-d maximal causal curve in the spacetime which also meets $\mathcal{K}$.

By diverging sequence $\{p_n\}$ is meant any infinite sequence such that every compact set contains only a finite number of the $p_n$. Of course, this does not mean that the points $p_n$ approach infinity for large $n$, because they can also 'diverge' towards some other edge of the spacetime, for example, a singularity. The relation $q_n \in J^+(p_n)$ must hold only for each $n$, and $q_n$ or $p_n$ do not have to be necessarily in $J^+(p_k)$ for $k \neq n$. The existence of each $\gamma_n$ is assured by Proposition 2.32. Moreover, $\mathcal{K}$ is *any* compact set and needs not be achronal or spacelike.

*Proof.* Extend the geodesics $\gamma_n$ indefinitely to the past and future and still call them $\gamma_n$. Of course, these endless $\gamma_n$ may fail to be maximal for points on $\gamma_n$ not between $p_n$ and $q_n$. By assumption, all these geodesics meet $\mathcal{K}$. Choose points $r_n \in \gamma_n \cap \mathcal{K}$ for each $n$. The sequence $\{r_n\}$ has an accumulation point $r \in \mathcal{K}$ and, by Proposition 2.2, the sequence $\{\gamma_n\}$ has an f-d endless causal limit curve $\gamma$ passing through $r$. This is the desired curve and intuitively $\gamma$ is clearly maximal, whence geodesic by Corollary 2.1. To prove it rigorously, let $\{\gamma_m\}$ be a subsequence of $\{\gamma_n\}$ converging to $\gamma$ and choose any pair of points $x, y \in \gamma$, with $y \in J^+(x)$. Take sequences $\{x_m\}, \{y_m\}$ converging to $x$ and $y$, respectively, such that $x_m, y_m \in \gamma_m$ for each $m$. Given any $\varepsilon > 0$, let $\mathcal{U}$ be a neighbourhood with compact closure of the portion of $\gamma$ from $x$ to $y$ such that no f-d causal curve in $\mathcal{U}$ has length greater than $L(x, y; \gamma) + \varepsilon$ (point (i) in Lemma 2.8). There is an $m_1$ such that the portion of $\gamma_m$ from $x_m$ to $y_m$ lies entirely in $\mathcal{U}$ but $p_m, q_m \notin \mathcal{U}$ for all $m > m_1$. This means that $x_m \in J^+(p_m)$ and $y_m \in J^-(q_m)$, so that $\gamma_m$ are maximal between $x_m$ and $y_m$ for all $m > m_1$. But then, from the continuity of $d_{x,y}$ (point (ii) in Lemma 2.8) we have the following chain of inequalities for all $\varepsilon > 0$:

$$\forall m > m_1, \qquad d_{x,y} \leq d_{x_m, y_m} + \varepsilon = L(x_m, y_m; \gamma_m) + \varepsilon \leq L(x, y; \gamma) + 2\varepsilon.$$

This immediately implies $L(x, y; \gamma) = d_{x,y}$, and hence $\gamma$ is maximal. ∎

**Corollary 2.8.** *Let $(V_4, g)$ be globally hyperbolic. If there exist a future-endless causal curve $\gamma$ and a past-endless causal curve $\lambda$, and all f-d causal curves from $\lambda$ to $\gamma$ intersect a compact set $\mathcal{K}$, then there is an endless f-d maximal causal curve.*

*Proof.* Simply choose two diverging sequences $\{q_n\}$ and $\{p_n\}$ on the future- and past-endless curves, respectively. ∎

As a trivial example of the above, take Minkowski spacetime and two diverging sequences $\{q_n\}$ and $\{p_n\}$ lying on any endless f-d null geodesic $\gamma$ (here the compact set is any $p \in \gamma$). More interesting is the case of a globally hyperbolic spacetime with a compact Cauchy hypersurface. However, the true importance of the above result is that it can be applied to any region of the form int $D(\zeta) \subset V_4$ due to Proposition 2.29, which will lead to the stronger singularity theorems of wider application (see Section 5).

## 3. DEFINITION, TYPES AND EXAMPLES OF SINGULARITIES

There is no widely accepted definition of singularity in General Relativity. Even though there is some consensus on considering causal geodesic incompleteness as a signal of the existence of a 'singularity', the related question of the extendibility and which type of extendibility of the spacetime allows one to maintain reasonable doubts on how robust and well-posed such a definition is. Further, there are examples of compact spacetimes containing incomplete causal geodesics, and therefore singularities by the usual definition (see Example 3.4 and Refs. 146,179). It seems that the only unanimously accepted idea regarding singularities is that the Friedman–Lemaître–Robertson–Walker (FLRW) cosmological models [82,83,129,174,240] contain a universal singularity in the finite past as long as they are expanding and do not violate SEC [66,112,124,128, 149, 179,241]. Not even something as simple as the singularity appearing in the maximal analytical *extension* of Schwarzschild vacuum solution [188] has a common accepted view (see Example 3.3, Refs. 1,16,24,25,114 and references therein). The fact that this singularity is 'spacelike' and in the future of every possible particle hitting it is, to say the least, something strange and hitherto unexplained. There are many other controversial examples, and I shall try to give some account of them in what follows.

The difficulty in recognizing what a singularity is can be appreciated by looking at any history of this concept in General Relativity, e.g. the review [220]. Singularities, horizons, extensions and infinities in the curvature have been mixed, confused and inadequately combined many times. Fortunately, today there is some agreement on what a horizon and an extension are; also on what some diverging curvatures can mean. Nevertheless, there is no complete consensus on *when* an extension must be performed, and in that case *which* extension as they are highly non-unique [46,190]. Besides, properties of singularities are not clear, and this leads to several unrelated classifications which will be described presently.

Key references on the definition of singularity are [90,179], where the main difficulties arising from most simple approaches are clearly explained. Many of the ideas presented here were inspired by the paper [190], in which a more up-to-date view and an excellent introduction can be found. The intuitive idea one has of a singularity is some 'place' where something goes wrong. However, making this precise encounters impressive logical problems. Let us consider some simple illustrative examples.

**Example 3.1 (The FLRW models).** These are spacetimes where the manifold is $V_4 = I \times \Sigma$ with $I \subseteq \mathbb{R}$ an open interval and $\Sigma$ is either $\mathbb{R}^3$ or $S^3$. The metric is locally characterized by the existence of a group of isometries of six parameters acting transitively on three-dimensional spacelike hypersurfaces. Therefore, they are *the* spatially homogeneous and isotropic models. The line-element is given by [112,124,128,149,159,241]

$$ds^2 = -dt^2 + a^2(t)[d\chi^2 + \Sigma^2(\chi, k)(d\vartheta^2 + \sin^2\vartheta d\varphi^2)]$$

where $a(t)$ is called the scale factor and $\Sigma(\chi, k)$ is defined by

$$\Sigma(\chi, k) \equiv \begin{cases} \sin\chi & \text{if } k = 1, \\ \chi & \text{if } k = 0, \\ \sinh\chi & \text{if } k = -1. \end{cases} \quad (33)$$

Here, $k$ is called the *curvature index* and $k = 1, 0, -1$ for the so-called closed, flat or open models, respectively. The range of $\vartheta$ and $\varphi$ is the standard one on the 2-sphere and the range of $\chi$ is $0 < \chi < \infty$ for cases $k = 0, -1$ and $0 < \chi < \pi$ for the closed case $k = 1$. For later use, we write the *Friedman equations* for the above metric, which give the energy density $\varrho(t)$ and isotropic pressure $p(t)$ in the preferred comoving observer with velocity vector $\vec{u} = \partial/\partial t$,

$$\varrho = 3\frac{a_{,t}^2 + k}{a^2}, \qquad \varrho + 3p = -6\frac{a_{,tt}}{a}, \qquad \varrho_{,t} + 3(\varrho + p)\frac{a_{,t}}{a} = 0, \qquad (34)$$

where commas indicate partial derivative. Only two of these three equations are independent and they are fully equivalent to the Einstein equations. In fact, the second is Raychaudhuri's equation (27) for these models. The energy-momentum tensor takes the form of a perfect fluid, the Weyl tensor is zero (conformally flat) and the shear, vorticity (22) and acceleration (20) of the fluid velocity vector $\vec{u}$ vanish ($\sigma_{\mu\nu} = \omega_{\mu\nu} = 0$, $a_\mu = 0$). The expansion takes the simple form $\theta = 3a_{,t}/a$. The possibility of geodesically complete FLRW spacetimes has been analysed in [186] and in wider generality in [176]. Very recently, a full detailed analysis of geodesics in FLRW spacetimes has been carried out in [182]. Nevertheless, application of the general result shown in Proposition 2.3 proves that if $\theta(t_0) > 0$ for some $t_0 > 0$ and $\varrho + 3p \,(= R_{\mu\nu}u^\mu u^\nu) \geq 0$, then $a$ was zero in the finite past (say at $t = 0$) and $\theta$ and $\varrho$ are unbounded there. This is a physical singularity, as the matter quantities blow up due to (34). Furthermore, this singularity is *universal*, in the sense that it appears at a finite length along any endless past-directed causal curve. This is why the names 'initial' or 'big-bang' singularity are used. When talking about 'cosmological singularities', most people refer to such a type of singularity, even though there is no justification for such an unfortunate name, as we will see. Of course, the singularity is not part of the spacetime (the interval I does not include the value $t = 0$), but anyway one loosely says that there is a singularity *at* $t = 0$. The characteristic feature of the missing $t = 0$ region is that all possible past-endless causal curves reach it with finite length, so that the spacetime is not geodesically complete according to Definition 2.4. In physical terms, all possible particles and photons suddenly appear at the singularity, out of 'nowhere'. This simple example shows the idea of what one wishes to consider a singularity: the curvature and physical quantities diverge and the physical curves are incomplete. Unfortunately, most cases are not so simple.

**Example 3.2 (Flat spacetimes).** The traditional Minkowski spacetime is the manifold $V_4 = \mathbb{R}^4$ with line-element

$$ds^2 = -dt^2 + dx^2 + dy^2 + dz^2$$

in Cartesian cordinates $\{t, x, y, z\}$. This spacetime is flat, in the sense that the Riemann tensor vanishes $R^\rho_{\sigma\mu\nu} = 0$, and is geodesically complete. In fact, the normal neighbourhood $\mathcal{N}_p$ of any point $p$ is the whole spacetime. This is the paradigmatic example of what one wishes to be a singularity-free spacetime: no diverging curvatures and no incomplete curves. It is well-known that any spacetime with vanishing Riemann tensor has *locally* the metric above [65,187], but the global properties can certainly change providing relevant new examples. For example, consider the manifold $V_4 = \mathbb{R}^4 - \{O\}$, where $O$ stands for the origin in Cartesian coordinates, with the above metric. The curvature is still trivial, but now many geodesics and other curves are incomplete. It may seem obvious that this is rather an avoidable problem, as the spacetime can be *extended* to a larger one including the cut out point and making it complete. Nevertheless, this is not clear in general for a) the extension may not be so obvious, is not unique and in some cases may lead to other singularities, b) sometimes there are incomplete curves and no regular extension, and c) physical arguments may require the avoidance of new unphysical regions which appear in the possible extensions.

As an easy example of a), take the manifold $V_4 = \mathbb{R}^4 - \mathcal{B}_1$, where $\mathcal{B}_1$ is the solid unit 3-sphere centered at the origin given by (15). Again the new spacetime is geodesically incomplete, and there is a trivial extension which makes it complete. But there are other 'unfortunate' extensions, such as the one defined by imbedding $\mathbb{R}^4 - \mathcal{B}_1$ into $\mathbb{R}^4$ by means of $(t', x', y', z') = (\mathcal{R} - 1)(t, x, y, z)$ with $\mathcal{R}^2 \equiv t^2 + x^2 + y^2 + z^2$ (see Ref. 190). The whole 'missing' 3-sphere has been mapped to a single missing point, and the metric can be seen to be flat everywhere except at the origin in which it has a singularity. One may be tempted to discard this second extension because of the singularity in the metric components, but this is not a good reason in general, as in most physical cases the extensions lead to the appearance of singularities. More interesting cases with these problems are treated in Example 3.3. For the time being let us remark that these unfortunate extensions, which do not add any boundary regular point, are examples of what we will call singular extensions. On the other hand, as an example of b) take Minkowski spacetime and identify the region $y - ax = 0$, $x > 0$ with the symmetric region $y + ax = 0$, $x > 0$ in the natural way (cutting out the part between these two identified hyperplanes) for some positive constant $a$. At each 2-plane with $t, z$ constants, this produces a cone with vertex at $x = y = 0$ so that the new spacetime is geodesically incomplete but with vanishing Riemann tensor. However, there are no possible *regular* extensions now (see the very interesting discussions in Refs. 71,179). In other words, *all* possible extensions are singular. The metric in cylindrical coordinates $\{t, \rho, \varphi, z\}$ reads

$$ds^2 = -dt^2 + d\rho^2 + \rho^2 d\varphi^2 + dz^2 \tag{35}$$

with $\rho > 0$ and $(\varphi + \arctan a)$ identified with $(\varphi - \arctan a)$. Singularities of this type are called conical. Their main problem is that the elementary flatness condition does not hold at the would-be axis (see e.g. Ref. 142) and they are supposed to describe some physical situations of interest. For instance, the above example is usually considered as the spacetime of a cosmic string (see, e.g. Refs. 230–232, and for a more mathematical treatment Ref. 49). All in all, finite curvature can certainly happen in some good candidates for singularities. A much more problematic geodesically incomplete spacetime with regular curvature is given by Misner–Taub–NUT-like metrics; see Example 3.4. Finally, for c) see the next example.

**Example 3.3 (Schwarzschild).** Schwarzschild spacetime [188] is locally the unique spherically symmetric vacuum solution of Einstein's equations. The manifold is taken usually as that part of $\mathbb{R}^4$ with $x^2 + y^2 + z^2 > 2M$. In typical spherical coordinates $\{t, r, \vartheta, \varphi\}$ the line-element reads

$$ds^2 = -\left(1 - \frac{2M}{r}\right)dt^2 + \left(1 - \frac{2M}{r}\right)^{-1} dr^2 + r^2(d\vartheta^2 + \sin^2 \vartheta d\varphi^2)$$

and $M$ can be interpreted as the total mass-energy of the spacetime [123,128,149,241], which is supposed to describe the gravitational field outside a finite spherically symmetric body. It is a simple exercise to see that there are incomplete geodesics approaching the region $r = 2M$. For example, the f-d null geodesics $\gamma$ with affine parameter $\tau$ given by constants $\vartheta$ and $\varphi$ with

$$t = \gamma^0(\tau) = t_0 + \tau - 2M \log\left(1 - \frac{\tau}{r_0 - 2M}\right), \qquad r = \gamma^1(\tau) = r_0 - \tau,$$

start at any point $(t_0, r_0, \vartheta_0, \varphi_0)$ for $\tau = 0$ and have affine parameters constrained to values $\tau < r_0 - 2M$, even though they are endless. All other geodesics approaching $r = 2M$ are incomplete too. The curvature tensor is well-behaved at values of $r$ close to $2M$. Actually, the curvature invariant $R_{\mu\nu\rho\sigma}R^{\mu\nu\rho\sigma}$ is proportional to $M^2/r^6$. Thus, the situation is as before — no curvature problem but incomplete geodesics — and one can try to perform extensions. The question is that there is a variety of such extensions at hand, and one should decide the matter contents of the extended part and the new manifold. Some commonly accepted vacuum extensions are those found by Eddington–Finkelstein and Kruskal (see e.g. Refs. 107,149). In the first case the new manifold may be taken as $\mathbb{R} \times \mathbb{R}^3 - \{O\}$ with primed spherical coordinates and the imbedding of

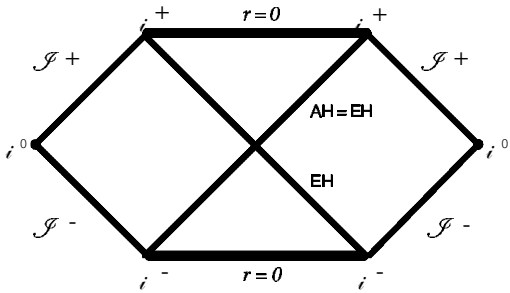

Figure 1. This is the Penrose conformal diagram of the Kruskal extension of Schwarzschild spacetime [163]. The conventions are standard (see Ref. 107): each point in the diagram represents a 2-sphere, the future direction is upwards, null lines are at 45°; conformal infinity is represented by $i^{\pm}$ (future and past timelike), $i^0$ (spacelike) and $\mathscr{I}^{\pm}$ (future and past null). Sometimes the event horizon is denoted by EH. In this case it corresponds to $r = 2M$, which are null hypersurfaces. Singularities are denoted by either thick solid lines or zig-zagging lines. In this case, there appear two singularities ($r = 0$), one in the future and one in the past. Half of this picture (say to the right of each of the diagonals) represents the Penrose diagram of the Eddington–Finkelstein extensions (advanced and retarded). The right 'square' with vertices at the centre, $i^+$, $i^-$ and $i^0$ is the diagram of the original Schawarzschild spacetime, so that we can see that the null geodesics reach $r = 2M$ in finite affine parameter. This square is asymptotically flat and no particle living here for ever can receive any information from the blach hole region. There appears *another* symmetric asymptotically flat region to the left. If the matter creating the Schwarzschild field is taken into account this second asymptotically flat region and the singularity in the past disappear (see Fig. 2).

the original Schwarzschild spacetime is given by $r' = r$, $\vartheta' = \vartheta$, $\varphi' = \varphi$ and
$$t' = t \pm r \pm 2M \log\left(\frac{r}{2M} - 1\right)$$
(each sign gives an extension, called advanced and retarded, respectively). However, in the new region $r' < 2M$ the coordinate $r'$ is timelike, the coordinate $t'$ is spacelike, and the border $r' = 2M$ is a null hypersurface called the horizon. Furthermore, the problems partially remain, as there are still incomplete geodesics, in this case of two types: those approaching values of $r' \to 0$ and those near values of $t' \to \pm\infty$. In the first case the curvature becomes unbounded when $r' \to 0$ so that this seems a singularity with no possible extension. In the second case the Riemann tensor is again regular when $t' \to \pm\infty$ and thus one can try to find a bigger extension. The Kruskal extension achieves this and maintains the previous extension as a subextension by combining two Eddington–Finkelstein spacetimes in an appropriate way. The price is that the manifold acquires strange topological properties [149] and there appear a new asymptotically flat region not causally related with the initial one, a new singularity and no possible further extension. This is better understood through the Penrose conformal diagram [163] of the Kruskal extension (Figure 1). Of course, all this can be avoided by matching Schwarzschild with an appropriate interior solution (see Figure 2), but even in this case it seems that the appearance of part of the Eddington–Finkelstein extension can be realistic in collapsing stars or black holes. Also, the singularity 'at' $r' = 0$ in both extensions is not what one expects naively. It has nothing to do with 'a point at the centre', but rather is in the future of observers entering into the $r' < 2M$ region. Consequently, a possible traveller could enter into the region, never stop seeing the star which creates the gravitational field and nonetheless suddenly disappear in the future singularity. This rather strange possibility is accepted today as the prediction of classical relativity for compact collapsed objects.

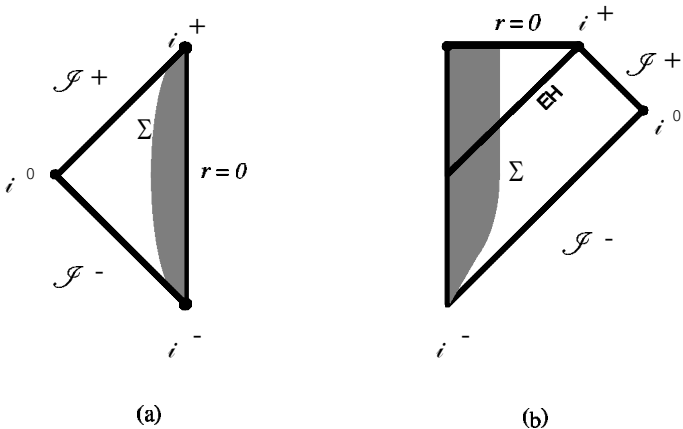

Figure 2. If a part of the vacuum Schwarzschild spacetime is matched with an appropriate interior metric which is regular at the centre $r = 0$ and such that the matching hypersurface $\Sigma$ is timelike going from $i^-$ to $i^+$ [case (a)], then there are no incomplete curves and no singularities. The shadowed zone represents the interior object. However, if the matching hypersurface reaches the value $r = 2M$ then it can cross this horizon and create a black hole region with a future singularity [case (b)]. Therefore, this part of the Eddington–Finkelstein extension may be realistic. Notice that an external observer living in the asymptotically flat region will receive information *only* from the object before it collapses to the black hole. However, if any particle enters into this region, then it sees the star all along until suddenly it hits the future singularity and disappears.

Other similar extensions, non-analytical and with matter, are also possible (see for instance Example 4.3 and subsection 7.9). A completely different one worth mentioning is analogous to that given in Example 3.2 when the solid 3-sphere was cut out. Take Schwarzschild spacetime and imbed it into $\mathbb{R} \times \mathbb{R}^3 - \{O\}$ by means of $\tilde{t} = t$, $\tilde{r} = r - 2M$, $\tilde{\vartheta} = \vartheta$ and $\tilde{\varphi} = \varphi$. Now one can look at $\tilde{r} = 0$ as a point at each instant of time. The line-element becomes (see also Ref. 63)

$$ds^2 = -\frac{\tilde{r}}{\tilde{r} + 2M} d\tilde{t}^2 + \frac{\tilde{r} + 2M}{\tilde{r}} d\tilde{r}^2 + (\tilde{r} + 2M)^2 (d\tilde{\vartheta}^2 + \sin^2 \tilde{\vartheta} d\tilde{\varphi}^2)$$

and has a singularity at the line $\tilde{r} = 0$ — this is a singular extension — similar to the conical one of the string in Example 3.2, and which is not noticeable by inspection of the Riemann tensor. Of course, this resembles what one expects to be the singularity of a point-like particle, although it may lead to problems when considering the collapse of some stars. Whether or not values of $r < 2M$ can be reached by physical objects is debatable (see Section 7, Ref. 141, and references therein). Of course, the above construction is not accepted by the majority of the relativistic community nowadays. Nevertheless, it has been cleverly argued by several authors that something of this type may be the correct interpretation (in fact, this is Schwarzschild's original interpretation). I refer the reader to [1–3,160], and to the papers [16,63,114] where the point-like structure of the $r = 2M$ region in the Schwarzschild spacetime has been claimed from quite different points of view.

In summary, one faces the following problem: in the vacuum spherically symmetric spacetime, either the singularity is spacelike, noticeable by the Riemann tensor and in the future of the observers and of the star itself, or it is a conical-type singularity not allowing for spheres with area less than $2M$, hence not allowing for collapses which seem reasonable. Neither of the two possibilities is satisfactory and, in my opinion, the problem is related to the necessity of giving a consistent theory of a singular axis of symmetry [144]. Thus, not even in the simple case of the Schwarzschild spacetime are the definition and properties of the singularity clear.

**Example 3.4 (Misner).** Here we briefly consider a simple example due to Misner [146] which mimics most unusual properties of the Taub–NUT vacuum spacetime (see, e.g. Ref. 107). The manifold is $V_4 = \mathbb{R}^3_+ \times S^1$, where $\mathbb{R}^3_+$ is the upper half of $\mathbb{R}^3$, with coordinates $\{t, x, y, \phi\}$ and the line-element is

$$ds^2 = -t^{-1} dt^2 + dx^2 + dy^2 + t d\phi^2.$$

Take the null geodesics parametrized by affine $\tau$ with $x = x_0$, $y = y_0$, $t = t_0 + \tau$ and $\phi = \phi_0 \pm \log(1 + \tau/t_0)$. Obviously, all these geodesics are incomplete and they spiral around into the past without ever reaching $\tau =$ $-t_0$, corresponding to $t \to 0$. The Riemann tensor vanishes everywhere again. Take then an extension to the whole of $\mathbb{R}^3 \times S^1$ with coordinates $\{t', x', y', \phi'\}$ so that the imbedding of the original $V_4$ is $t' = t$, $x' = x$, $y' = y$ and $\phi' = \phi \pm \log t$ — actually, two different extensions, one for each sign. The line-element of the extensions is

$$ds^2 = \mp 2 dt\, d\phi' + dx^2 + dy^2 + t\, d\phi'^2$$

and half of the previous geodesics (those with the sign corresponding to that of the extension) are now complete, but the other half are not, as they spiral around the regular points $t' = 0$ without ever reaching them but with total finite affine parameter. Somehow, it seems that a particle travelling on this geodesic would tend to *many* different points at some finite value of its affine parameter. Furthermore, the extended metrics are completely inextendible: there are no possible extensions, neither regular *nor singular*. The teaching is therefore twofold: firstly, there are two inequivalent *analytical* extensions for the original spacetime; and secondly, there can be incomplete causal geodesics in regular completely *inextendible* spacetimes. Incidentally, let us remark that the extended spacetimes violate the chronology condition so that there appear imprisoned geodesics in the sense of Proposition 2.21. I shall consider any of the two extended spacetimes as non-singular, in agreement with [190], even though there are incomplete geodesics. Compare with [107], where a complete study of this behaviour is performed.

Hitherto, the cases with regular curvature but incomplete geodesics have been considered. But there is also the opposite possibility, that is, non-regular curvature components but complete geodesics. Of course, this may happen when the problem lies in a bad choice of basis for computing the components. Therefore, the logical thing is to consider curvature invariants such as $R$, $R_{\mu\nu}R^{\mu\nu}$, $R_{\mu\nu\rho\sigma}R^{\mu\nu\rho\sigma}$, etcetera. Unfortunately, this does not characterize all possible cases that may be considered as singularities, as shown by plane waves [8,21,62,100,123,161] which have all curvature invariants vanishing but non-zero Riemann tensor. This happens for Petrov type N and III [15,123,168] vacuum solutions. In these cases there may be problems with the curvature which must be detected in an appropriate basis. The correct thing to do is to use orthonormal bases parallelly propagated along a curve, because the curvature components in these bases cannot be badly behaved as long as the curve has endpoints. With all these examples and ideas at hand, we can try to give a sensible definition of singular spacetime and singularity. Two essential ingredients are needed, extensions and incomplete curves. Let us start with the first [190] ($\Phi^*$ denotes the pull-back of a map $\Phi$ of manifolds; Refs. 41,107).

**Definition 3.1.** An envelopment of the spacetime $(V_4, g)$ is an imbedding $\Phi$ of $V_4$ into another connected manifold $\hat{V}_4$ with $\Phi(V_4) \subset \hat{V}_4$. Let $\hat{\partial} V_4$ be the boundary of $\Phi(V_4)$ in any given envelopment $\hat{V}_4$. An extension of $(V_4, g)$ is any spacetime $(\hat{V}_4, \hat{g})$ with $\hat{V}_4$ as base manifold and such that $(\Phi^{-1})^* g = \hat{g}|_{\Phi(V_4)}$. It is called a $C^k$ regular extension if the metric $\hat{g}$ is $C^k$ at $\hat{\partial} V_4$. Otherwise, if the metric $\hat{g}$ is not $C^1$ at $\hat{\partial} V_4$, or if the envelopment has no $\hat{g}$ well-defined at $\hat{\partial} V_4$, the extension is called singular.

This definition may seem more complicated than necessary. However, it is needed if all cases appearing in the above examples are to be covered. Thus, the extension of flat spacetime without a point to the whole Minkowski spacetime is $C^\infty$ regular, but only $\hat{\partial} V_4$ is added (a point). The singular extensions are included because they may be needed on physical grounds. An example was given by the non-standard extension of Schwarzschild of Example 3.3. And there are many other cases: for instance, any axially symmetric metric with a singular axis needs the concept of singular extension. It might be argued that the existence of incomplete curves will do the job just the same in these cases, but the exceptional Example 3.4 proves that this is not so. In this case, the incomplete geodesics do not signal any 'missing points' as there cannot be any. Somehow, the manifold by itself — without the metric — is complete (it cannot be properly imbedded as an open set into any other spacetime), and thus the notion of singularity has no meaning [190]. Finally, let us remark that the concept of singular extension allows one to add the 'singular points' to the envelopment $\hat{V}_4$, as for instance the big-bang singularity of Example 3.1, and also the points 'at infinity', even though some different interpretations may arise for the added points depending on the extension. In fact, most Penrose conformal diagrams can be thought of as singular extensions.

Of course, extensions are not unique. Not even analytical extensions are unique as already seen in Example 3.4. In fact, there are usually infinitely many inequivalent extensions of a given extendible spacetime. Some of them may be singular, some other may be $C^k$ regular for some $k$ but leading to other extendible spacetimes with either only singular extensions or regular ones, and so on. This is a fundamental problem from the physical point of view, because if there is an extendible spacetime such that one suspects that the extension may have physical relevance, this extension has to be *invented*, and the matter content, the symmetry group, the Petrov type, and all other physical properties may be given *ad hoc*. Even if maintaining the same type of energy-momentum tensor, symmetry group, etcetera, the extended spacetimes cannot be guessed. The next is an illustrative example of physical relevance [75].

**Example 3.5 (Vaidya).** The Vaidya spacetime [225,226] is the unique spherically symmetric solution to Einstein's equations for a pure radiation energy-momentum tensor. The manifold is $\mathbb{R}^4 - \{O\}$ and its line-element is given by

$$ds^2 = -\left(1 - \frac{2M(t)}{r}\right)dt^2 + 2\varepsilon dt\, dr + r^2(d\vartheta^2 + \sin^2\theta d\varphi^2),$$

where the mass function $M(t)$ is assumed to be non-negative. If $M(t)$ is a constant, we obtain the retarded ($\varepsilon = -1$) or advanced ($\varepsilon = 1$) Eddington–Finkelstein extensions seen in Example 3.3. Penrose's conformal diagram for $\varepsilon = -1$ is shown in Figure 3. As can be seen, radial ingoing null geodesics are incomplete, as they reach the future event horizon $r = 2M(t \to \infty)$ for a finite value of their affine parameter. There is no problem with the curvature unless at $r = 0$, so extensions may be sought.

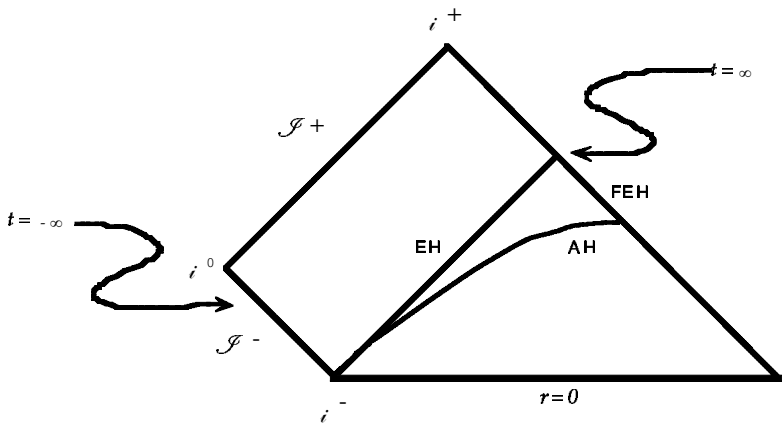

Figure 3. This is the conformal diagram of Vaidya's spacetime in the case that $M(t)$ is a decreasing function of $t$. There appears an apparent horizon AH (see Example 4.2) which is spacelike and deviates from the EH. Null geodesics are incomplete to the future as they approach the future event horizon FEH with finite affine parameter. This metric is therefore extendible through FEH. There is a curvature singularity at $r = 0$.

The way to extend this metric can be seen in [75] and references therein. The important points are: first of all, the *type* of energy-momentum tensor beyond the horizon must be specified. A possible natural choice is to keep the same form of the unextended spacetime, including vacuum as a particular case. Second, one can also demand the fulfilment of WEC or DEC. With these assumptions, the remaining task is finding a mass function which extends $M(t)$ beyond the horizon. Unfortunately,[5] even with all these restrictions there are infinite possible prolongations for the mass function. The analytical continuation must be given up for this case, as argued in [75], where the following example can be found. Imagine that $M(t)$ arrives at the horizon with all derivatives vanishing. Then, two quite different continuations would be, for instance, the one with $M$ constant in all the extended part, and the one which is "symmetric" with respect to the horizon. Both extensions satisfy DEC and give $C^\infty$ mass functions, but they provide completely different physical spacetimes. The Penrose conformal diagrams for both inequivalent cases are presented in Figure 4. In the first case, the metric becomes the Eddington–Finkelstein extension of Schwarzschild beyond the horizon, while in the second case the radiation of null particles goes on after crossing it. The question of which is the 'correct' extension has no meaning whatsoever, unless further information is at hand.

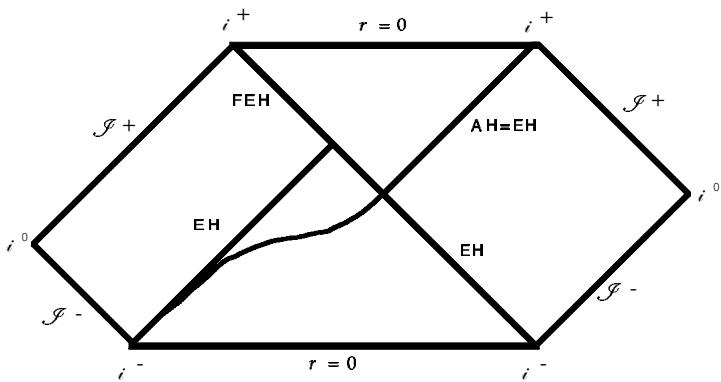

Figure 4(a).

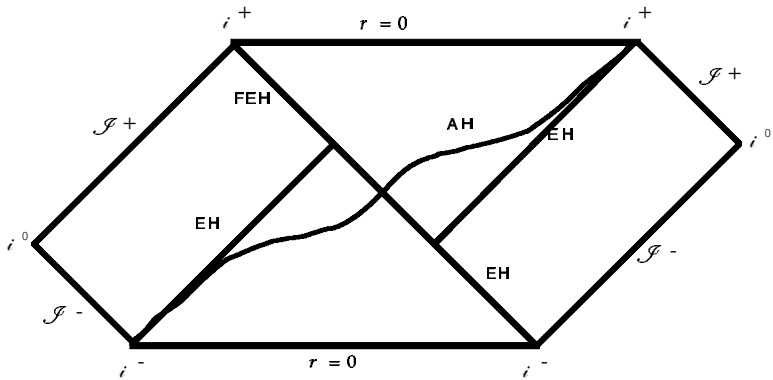

Figure 4(b).

Figure 4. Consider the case that $M(t)$ arrives at FEH with all derivatives vanishing. Then, the AH becomes null there. Two completely different but quite logical extensions can then be performed. In the first one (a), the extended part is simply the Eddington–Finkelstein extension of Schwarzschild spacetime, so that $M$ = const. throughout the extended region and the AH becomes null and coincides with EH there. In the second case, the mass $M(t)$ behaves in a time-symmetrical way with respect to the horizon, so that the extended region is a time-reversal copy of the original spacetime.

---

[5] This is unfortunate because the extension is not defined, but it is certainly fortunate from the physical point of view, for it allows to describe the many feasible different behaviours of $M$. In case the extension were unique, the physics would be very dull.

Sometimes, the different concept of *local extension* has been also used [12,42,71,107]. However, I shall not consider it here because even Minkowski spacetime is locally extendible [12]. Local extendibility may be of interest in understanding conical singularities [42,71].

Let us pass to the analysis of incomplete curves. The existence of incomplete curves may indicate either the existence of a singularity or the possibility of an extension (or none; see Example 3.4). Intuition leads one to think that if a spacetime is timelike geodesically incomplete (say), then it will also be null and spacelike geodesically incomplete. Unfortunately, this is not the case as shown by the maximal extension of the Reissner–Nordström solution [107]; other examples can be seen in [11,88]. Actually, all three types of geodesic completeness are independent and inequivalent. However, one may try to think the other way round, and intuition again says that if the spacetime is geodesically complete in all three senses, then there cannot be timelike curves (say) which are incomplete. Well, wrong again. There is a celebrated example by Geroch [90] of a geodesically complete spacetime containing an endless timelike curve of bounded acceleration and finite total proper time (17). No doubt, a particle travelling along this path has the same right to say it runs into a singularity as freely falling particles have. Even more, Beem [11] has constructed an example of a geodesically complete globally hyperbolic spacetime containing such incomplete curves with bounded acceleration. Thus, the sensible thing to do is to consider incompleteness of every possible endless curve as indication of a singularity. Nevertheless, in general there is no concept of proper time or affine parameter, and hence the definition of incomplete general curves is not clear yet. As an example, consider the curve $\gamma$ defined by $t = \gamma^0(u) = u$, $x = \gamma^1(u) = \sin u$, $\gamma^2 = \gamma^3 = 0$ in Minkowski spacetime, whose tangent vector is $\vec{u} = \partial_t|_\gamma + \cos u\, \partial_x|_\gamma$. This curve is obviously timelike everywhere except at the points with $u = k\pi$ for any integer $k$, so that the concept of affine parameter (17) is not well-defined. This is solved by taking a completely general affine parameter as follows (see e.g. Refs. 46,184). Let $\gamma$ be any $C^1$ curve parametrized by $u$ and let $p \in \gamma$. Choose any orthonormal basis $\{\vec{e}_\mu\}$ parallelly propagated along $\gamma$. Let $v^\mu(u)$ be the components of the tangent vector $\vec{v}|_\gamma$ in this basis. The *generalized affine parameter* $\tau$ is defined by

$$\tau \equiv \int_{u_p}^{u} \sqrt{\delta_{\mu\nu} v^\mu v^\nu}\, du \tag{36}$$

where $\delta \equiv \mathrm{diag}(1,1,1,1)$. Evidently, $\tau$ depends on the initial conditions for the basis (chosen at $p$, say). But the important property is that $\tau$ relative to a basis $\{\vec{e}_\mu\}$ is finite at any point in $\gamma$ if and only if any other generalized affine parameter relative to another basis is also finite [107,184]. Analougously to Definition 2.4 we have

**Definition 3.2.** A $C^1$ endless curve from $p \in V_4$ is complete if the generalized affine parameters are defined for all $\tau \in [0,\infty)$. A spacetime is b-complete at $p$ if all $C^1$ curves emanating from $p$ are complete. A spacetime is b-complete if it is so for all $p \in V_4$.

Naturally, b-completeness implies that all timelike curves with bounded acceleration are complete, which in turn implies timelike geodesic completeness. The relation between incompleteness and extensions is partially given by the next result.

**Proposition 3.1.** If the spacetime is timelike, or null, or spacelike geodesically complete then it has no possible regular extension.

*Proof.* Suppose there were a $C^1$ extension. Then there would be a nonempty boundary $\hat{\partial} V_4$ at which the extended metric $\hat{g}$ is $C^1$. The geodesics starting at some point $p \in \Phi(V_4)$ and approaching $\hat{\partial} V_4$ would have bounded affine parameters. But this is impossible as $\Phi$ is an isometry between $V_4$ and $\Phi(V_4)$ and thus leaves affinely parametrized geodesics invariant.  ∎

**Corollary 3.1.** If the spacetime is b-complete, then it has no regular extension.  ∎

Hence, if a spacetime is b-complete, the only possible extensions are singular, and they *only* add points at infinity. These points, unreachable by any geodesic, are not to be considered singularities. The converse of Proposition 3.1 is not true, as already explained. Now a definition of singularity can be given.

**Definition 3.3.** A singularity of $(V_4, g)$ relative to a singular extension is the endpoint in $\hat{V}_4$ of a curve incomplete within $(V_4, g)$. A spacetime is singularity-free if it has no singularities.

The definition of singularity depends on the extension. There is no other possibility, because singular points are not part of the spacetime. Thus, a candidate for singularity may be a singularity relative to one extension, and a regular point of an alternative regular extension. For example, the horizon $r = 2M$ is regular in the Kruskal extension, and a singularity in the singular extension of Example 3.3. This kind of singularity has been called *removable* in [190], as one can always choose the regular extension. Nonetheless, a removable singularity does not have to be necessarily removed on physical grounds, as argued before with the singular axes and the Example 3.3. Other possible examples of physical interest are those spacetimes with Cauchy horizons, such as Reissner–Nordström or similar (see e.g. Refs. 29,107,140 and references therein). For, if one looks at the Cauchy development int $D(\Sigma)$ of an asymptotically flat hypersurface $\Sigma$ as the spacetime, then there are incomplete timelike and null geodesics ending at the Cauchy horizon $H(\Sigma)$. Of course, one can make the typical regular extension beyond $H(\Sigma)$, but it has been claimed repeatedly that $H(\Sigma)$ is unstable against all possible realistic perturbations (see, e.g., Refs. 38,102,103,217, 245). Further, the extension usually needs a change in the topology of the spacetime — this is in fact necessary for completely regular extensions [29]. Thus, perhaps what should be done is to perform simply a singular extension making the whole of $H(\Sigma)$ a null singularity. Actually, it has been recently claimed that this may be the generic picture for realistic spacetimes [156].

A non-removable singularity is called an essential singularity. On the other hand, notice that if the spacetime is b-complete, then it has no singularity. Similarly, if the spacetime has no singular extension, or if it does but only adding points at infinity, then there are no singularities either. Consequently, Example 3.4 is geodesically incomplete but without singularities. As a matter of fact

**Proposition 3.2.** Any compact spacetime is singularity-free.

*Proof.* A compact spacetime has no envelopment, because $\Phi(V_4)$ would be a compact *open* proper subset of $\hat{V}_4$, which is impossible for connected $\hat{V}_4$. ∎

Of course, we already know that compact spacetimes have little physical interest due to Proposition 2.19. Nevertheless, they illustrate the fact that there can be b-incomplete singularity-free spacetimes, as in the case of Example 3.4. In fact, this example can be modified to construct a truly compact spacetime with incomplete geodesics (see Refs. 146,179).

In any case, the existence of an incomplete endless curve is necessary for the existence of singularities. Thus, assume that there is an incomplete curve $\gamma$ and choose any of their generalized affine parameters $\tau$. Obviously, there is a minimum value $\hat{\tau}$ of $\tau$ such that $\tau \in [0, \hat{\tau})$. The question of the behaviour of the curvature approaching the singularity along $\gamma$ acquires a precise meaning: take the pertinent function $F$ and simply compute the limit of $F|_\gamma$ when $\tau \to \hat{\tau}$. This has led to the standard classification of singularities described in [71].[6]

**Definition 3.4.** The essential singularities of a spacetime can be classified as follows:
 (i) $C^k$ quasi-regular singularities if all the components of the $k$-th covariant derivative of the Riemann tensor computed with respect to a parallelly propagated orthonormal basis are locally bounded when approaching the singularity along any incomplete curve.
 (ii) $C^k$ non-scalar curvature singularities if they are not $C^k$ quasi-regular but all the scalar curvature invariants remain well-behaved when approaching the singularity.
 (iii) $C^k$ scalar curvature singularities otherwise.

The conical singularity of the cosmic string in Example 3.2 is quasi-regular, while the singularity in the FLRW models of Example 3.1 is scalar. The quasi-regular singularities can be further subclassified into specialized, primeval and holes [122,220], and the question of their stability has been addressed several times [71,121,122]. Further, it has been shown that there is always a local extension making them locally removable [42]. The non-quasi-regular singularities are called *matter* singularities if the problem arises with some component of the Ricci tensor. If the problem appears for some component of the Weyl tensor but not for the Ricci tensor they are called *Weyl* or *pure gravitational* singularities. Examples of non-scalar singularities are explicitly presented in [50,68,71,110,119,199]. Their instability against generic matter perturbations has been claimed in [110,119]. In the scalar singularities, there is a curvature scalar — polynomial scalar constructed with the Riemann tensor, the volume 4-form, the metric and the covariant derivatives — which behaves badly when approaching the singularity. It has been shown that, in fact, a singularity is of scalar type iff there are badly-behaved curvature components in every orthonormal tetrad along the incomplete curve [201]. However, the scalar singularities can be *directional* singularities, in the sense that the diverging curvature scalar may remain locally bounded for some directions of approach. The paradigmatic example of a directional scalar singularity is that of the Curzon metric (see e.g. Refs. 123,189), as was first noticed in [87]. It was soon claimed [203] that this type of behaviour is due to the non-pointlike nature of the singularity, and that the singularity appeared not along some 'directions' of approach but rather along some trajectories, that is, along some incomplete curve [52]. Other singularities of this type were considered in [111]. A thorough analysis of the different trajectories and the possible extensions of Curzon's spacetime was given in [189], and the directional singularity was finally interpreted as a ring.

---

[6] Here, I shall not enter into the important subject of *naked singularities* and the related problem of cosmic censorship [164,166] (see, e.g., Refs. 47,116,220 and references therein).

This leads to the fundamental question: is there any way to define local properties — shape, character, strength, etcetera — of singularities? There have been several constructions trying to solve this question, all of them attaching a boundary to the spacetime. We summarize them briefly now (see Ref. 179). The causal boundary (c-boundary) was put forward in [94] by using some open future sets (Definition 2.14). The idea is that, for any point $p \in V_4$, $I^+(p)$ is a future set not decomposable into proper open future subsets. However, there are other future sets with these properties which are not the chronological future of any point. It can be proven that such sets are the chronological future of a past-endless timelike curve. Thus, they somehow signal singularities or points at infinity. The set of all indecomposable past and future sets can be thought of as containing all points in $V_4$ plus the c-boundary (see also Ref. 107). Somewhat differently, Geroch introduced the geodesic boundary (g-boundary) by constructing equivalence classes of endless incomplete geodesics and a notion of proximity between them [89]. Unfortunately, the g-boundary is not determined in general and does not consider non-geodesic curves so that it cannot contain all singularities. Sometimes, the g-boundary can be given a metric structure and thus some local properties of certain singularities are defined. The most complete structures are the bundle-boundary (b-boundary) due to Schmidt [184,185], and the abstract boundary (a-boundary) recently put forward in [190]. The b-boundary uses the frame bundle of $V_4$ with a Riemannian metric and provides a correspondence between incomplete curves in $V_4$ and incomplete curves in the metric sense in the frame bundle. The difficulties for applying this construction are tremendous and the b-boundary has been explicitly calculated only in very simple idealized cases [33,115,220], not always with the expected result. An important attempt to overcome all these difficulties is the a-boundary, which in fact is defined for any manifold independently of being a spacetime or even having a connection. Essentially, the a-boundary collects all possible boundary points arising in all the envelopments of a given manifold. In the case of spacetimes, the boundary points can be classified by using appropriate families of curves with definite properties (such as geodesics or others), leading to (possibly directional) singularities, points at infinity, and some other cases [190].

Regarding the 'strength' of a singularity, obviously this can only be meaningful for non-quasi-regular singularities. The singularity theorems of Section 5 predict the existence of incomplete causal geodesics, and sometimes they say something about their location and strength (see Section 5). Finally, with respect to the character of the singularity, most simple known cases are termed 'spacelike', 'null', or 'timelike' relying only on intuition. No-one doubts that the singularity in the Kruskal extension (Fig. 1) is spacelike, but making this precise in general is rather difficult. For the purposes of this work, the following tentative definition is to be used. This may not cover all cases and is admittedly vague, but it will be enough for us.

**Definition 3.5.** A set $S \subseteq \hat{\partial} V_4$ relative to a singular extension is said to be spacelike (resp. null, timelike, general, achronal, acausal, k-dimensional, etcetera) if there is a metric $\tilde{g}$ preserving the relevant properties of $g$ such that $S$ is spacelike (resp. null, ...) in $(\hat{V}_4, \tilde{g})$.

This definition tries to incorporate the intuitive ideas behind the character of singularities and is reminiscent of the Penrose conformal diagrams [163]. By 'relevant properties of $g$' is meant those properties which must be kept concerning the characteristic one whishes to assign to the singularity (usually this will include causal properties). The most common and best-founded situation is when $\tilde{g}$ is conformally related to $g$, because the causal structure is invariant by conformal transformations of the metric. Furthermore, the definition can be applied to points at infinity appearing in $\hat{\partial} V_4$. In this way, the traditional timelike $i^{\pm}$, spacelike $i^0$, or null $\mathscr{I}^{\pm}$ infinities [107,149,163] are recovered with their corresponding properties. In general, the incomplete geodesics predicted by the singularity theorems have no definite properties, so that one does not know whether the singularity is in the past or the future, or if it is spacelike in the cosmological models, or whether is a matter singularity or not, scalar or not, etcetera. This is important regarding the possible cosmological singularity, which is usually expected to be universal, spacelike and at least in the past. For example, if the singularity is timelike, then it can be avoided and perhaps not seen by most observers, so that it would be a very mild breakdown. In fact, many known examples do not satisfy the desired properties and explicit spacetimes with all possibilities are explicitly known [50,68,124,178,198,199,204]. Regarding the types of black-hole singularities, no definite expectation exists yet. Thus, the question of which type of singularity (if any) is generic in cosmological models, including both spatially homogeneous and inhomogeneous, and in collapsed black holes has no answer yet [50,156,194]. I will come back to this subject in Sections 6 and 7.

I have used the name 'universal' singularity in Example 3.1 and the previous paragraph. The idea is to capture the intuitive notion of *big-bang* singularity, at which everything has its origin. The paradigmatic example is that of standard FLRW models and some homogeneous models [50,68,147,148]. There have been several studies on this subject from different angles, ranging from the classical papers [18,19] (see also Ref. 9), to the Penrose conjecture concerning the relationship between a hypothetic entropy of the Universe and the initial singularity [166,236]. The *Weyl tensor hypothesis* assures that the appropriate thermodynamic boundary condition at cosmological singularities is the vanishing of the Weyl tensor. A milder version only requires that at any cosmological singularity $S$

$$\lim_{x \to S} \frac{C_{\mu\nu\rho\sigma}C^{\mu\nu\rho\sigma}}{R_{\mu\nu}R^{\mu\nu}} = 0$$

so that matter dominates at the singularity. Obviously, this is satisfied by the FLRW models, in which $C_{\mu\nu\rho\sigma} = 0$ throughout the spacetime. However, it has been partially proved [221,222,154] that for realistic perfect fluids with an *isotropic singularity* the Weyl tensor hypothesis only allows for FLRW models. The concept of isotropic singularity was precisely defined in [98,237], and essentially requires that the singularity be spacelike in a singular extension with an appropriate conformally related metric $\tilde{g}$ of Definition 3.5 (see also Refs. 59,96). This is also related to the following definition, which was put forward elsewhere [195] (see also the very recent related work, Ref. 5).

**Definition 3.6.** A singularity set $S \subseteq \hat{\partial} V_4$ relative to a singular extension is called a big-bang (or initial) singularity if every past-endless causal curve approaches $S$ at a finite generalized affine parameter.

Big-crunch singularities can be defined analogously. Here, there is no assumption about the behaviour of the Weyl tensor, but somehow we demand that the big-bang singularity is a kind of 'singular Cauchy hypersurface' for the spacetime. Using Definition 3.6, most known big-bang singularities are spacelike, but this is not a general property and they can change character sometimes — they do not even have to be achronal; see Example 3.7 — as explicit examples using the Lemaître–Tolman [20,130,223] models show [124]. They have been further classified according to the behaviour of the shear eigendirections for perfect or similar fluids into 'point-like', 'cigars' and 'pancakes' [211], and also the term velocity-dominated is used for irrotational models (see Ref. 234), if the three-spaces orthogonal to the velocity vector of the fluid have curvature divergences less strong than that of their second fundamental forms [59]. A good account of the possible big-bang singularities and their properties in spherically symmetric and Szekeres models [209,206,207] can be found in [97,124,204] and references therein. To illustrate these points and Definitions 3.5 and 3.6, let us consider two final examples.

**Example 3.6 (Null big bangs in FLRW models).** Let us consider the flat ($k = 0$) FLRW models with a barotropic equation of state $p = \gamma \varrho$ for $-1 < \gamma \leq 1$. The solution of (34) is

$$a(t) = Ct^{2/(3(1+\gamma))}, \qquad C = \text{const}. \tag{37}$$

As always, all past-endless causal curves approach $t = 0$ with finite values of their generalized affine parameter. However, some of them do so with $\chi \to \infty$. Thus, this will not be a big-bang singularity for the singular extensions which do not cover $\chi = \infty$. One might think naively that the singularity is *always* spacelike, though. This is not so for some important cases. To see all this, define the traditional parametric time [128] $\eta$ by means of $d\eta = dt/a(t)$ so that the line-element takes the explicitly conformally flat form

$$ds^2 = a^2(\eta)[-d\eta^2 + d\chi^2 + \chi^2(d\vartheta^2 + \sin^2\vartheta d\varphi^2)].$$

Obviously, the range of $\eta$ is as follows: $0 < \eta < \infty$ for $\gamma > -\frac{1}{3}$; $-\infty < \eta < \infty$ for $\gamma = -\frac{1}{3}$; and $-\infty < \eta < 0$ for $\gamma < -\frac{1}{3}$. Therefore, by using now the traditional Penrose conformal diagram of Minkowski spacetime, the following singular extension is obtained

$$2\eta = \tan\left(\frac{\hat{t}+\hat{r}}{2}\right) + \tan\left(\frac{\hat{t}-\hat{r}}{2}\right), \qquad 2\chi = \tan\left(\frac{\hat{t}+\hat{r}}{2}\right) - \tan\left(\frac{\hat{t}-\hat{r}}{2}\right),$$

where the ranges of $\hat{t}$ and $\hat{r}$ (the coordinates in $\hat{V}_4$) are to be chosen adequately for each case. The three different cases have been drawn in Figure 5 (see Ref. 76), and in all three cases $t = 0$ is a big-bang singularity according to Definition 3.6. As we can see, for $\gamma \leq -\frac{1}{3}$ this big-bang singularity is null. This is related to the fact that there is no particle horizon in these FLRW, which in turn might seem related to their inflationary character for $\gamma < -\frac{1}{3}$; nonetheless, the extreme case $\gamma = -\frac{1}{3}$ satisfies SEC. Note also that future null infinity may also appear as spacelike when $\gamma < -\frac{1}{3}$. In summary, the big-bang singularities may be null, *even for* FLRW *models. Let us remark that DEC holds for all these models.*

Of course, we could have tried the naive singular extension by simply adding $t = 0$ to the original range of coordinates. But then $t = 0$ would be approached only by timelike geodesics so that past-endless null geodesics, for instance, will not approach $t = 0$ because they will do so only for $\chi = \infty$. For this singular extension $t = 0$ is not a big-bang singularity and, furthermore, it still contains incomplete curves: null geodesics and timelike curves. The previous singular extension, which is the standard Penrose diagram, is obviously preferable.

**Example 3.7 (Unusual big bangs).** The character and structure of big-bang singularities can be somewhat complicated, as the following simple examples show. Let us use some subclass of a new family of spatially inhomogeneous algebraically general dust spacetimes recently presented in [198]. In the first case, the manifold $V_4$ is any connected part of $\mathbb{R}^4$ with $\sin[a(t-x)] + e^{-a(t+x)} < 0$ in Cartesian coordinates $\{t, x, y, z\}$ and for any constant $a \neq 0$. The line-element reads

$$ds^2 = -dt^2 + dx^2 + \{\sin[a(t-x)] + e^{-a(t+x)}\}^2 dy^2 + e^{2a(t-x)} dz^2.$$

This is a solution of Einstein's equations for an energy-momentum of dust, that is, $T_{\mu\upsilon} = \varrho u_\mu u_\upsilon$ with unit $\mathbf{u} = -dt$ and its energy density is

$$\varrho = \frac{-4a^2 \, e^{-a(t+x)}}{\sin[a(t-x)] + e^{-a(t+x)}} > 0.$$

The expansion of **u** is

$$\theta = a \, \frac{\sin[a(t-x)] + \cos[a(t-x)]}{\sin[a(t-x)] + e^{-a(t+x)}}$$

so that all the dust geodesic congruence is initially expanding and then recontracts (Figure 6). Making the natural singular extension to $\mathbb{R}^4$, the whole boundary $\hat{\partial} V_4$ is $\sin[a(t-x)] + e^{-a(t+x)} = 0$ and is obviously a matter singularity. This is a big-bang singularity for $a > 0$ according to Definition 3.6, but with three important features: firstly, it has a general character (that is, it is partly spacelike, partly timelike and partly null). Secondly, the part of the singularity which is *to the future* of the dust is in fact a necessary part of the big-bang singularity, as otherwise there would be past-endless causal curves not approaching the singularity. And thirdly, the singularity is not of big-crunch type, for even though all the timelike curves of the dust congruence approach the singularity, many f-d causal curves can avoid it completely. In the case $a < 0$, the singularity is of big-crunch type, and even though all the dust originates at the singularity, there are many past-endless causal curves which do not intersect the singularity. All these properties are clearly seen in the diagram of Fig. 6.

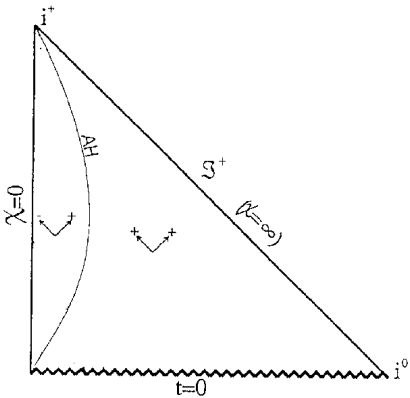

**Figure 5(a)**

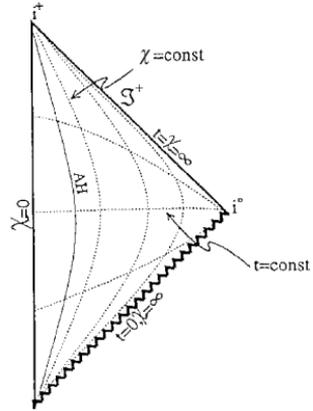

**Figure 5(b)**

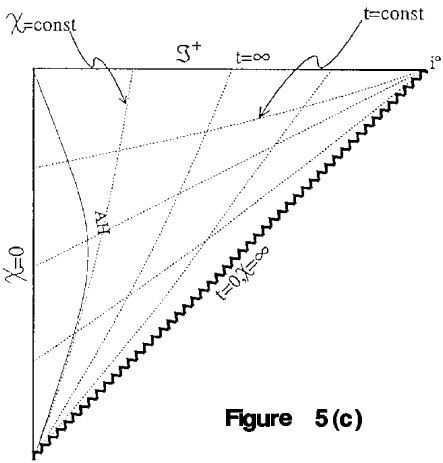

**Figure 5(c)**

Figure 5. These are the Penrose diagrams for the flat $p = \gamma\varrho$ FLRW spacetimes, with $-1 < \gamma < 1$. For $\gamma > -1/3$ [case (a)], $t = 0$ is a big-bang spacelike singularity. The apparent horizon — see Example 4.2 — has been shown here for $-1/3 < \gamma < 1/3$, in which case it is a timelike hypersurfce. For $\gamma = 1/3$ the AH is a null hypersurface which can be draw going from the left lower corner to $\mathscr{I}^+$ diagonally. If $\gamma > 1/3$ the AH is a spacelike hypersurface and can be shown as a line going from the lower left corner to $i^0$. On the other hand, for $\gamma = -1/3$ [case (b)] and $\gamma < -1/3$ [case (c)], the $t = 0$-singularities are of big-bang type but obviously *null*. In case (c), $\mathscr{I}^+$ is spacelike. In both cases (b) and (c), all the lines of the perfect-fluid congruence ($\chi$ = const.) start from the bottom point, but null geodesics and other causal curves can emerge from the rest of the null singularity.

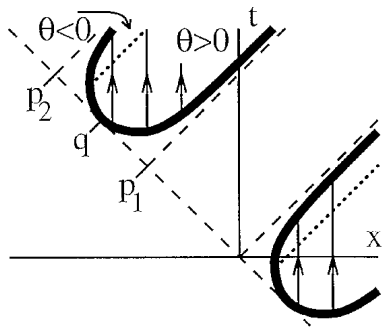

Figure 6. This is a possible diagram for the first spacetime of Example 3.7 with $a > 0$. Here and in Figs. 7, 8, and 9, only the $\{t, x\}$-plane is shown, because the whole spacetime is the product of these with the group orbits. Thus, each point may be thought to represent a 2-dimensional orbit. The singularities (thick lines) split the whole plane into infinite disconnected regions with positive energy density $\varrho$. Only one of those can be considered as the spacetime. The arrowed lines indicate the dust flow in these regions. The dotted line represents the hypersruface where the expansion of the dust congruence vanishes, and the regions with positive and negative expansion are also indicated. The 'points' $p_1$, $q$ and $p_2$ are $t - x = \pi/a$, $3\pi/2a$ and $2\pi/a$, respectively, and the singularity is of big-bang type but with a general character, being spacelike between $p_1$ and $q$, null at $q$ and timelike in the rest. The part of the singularity to the future of the dotted line cannot be avoided by the dust particles, but it can certainly be avoided by photons and other causal curves. More interestingly, this future part is an *essential* part of the big-bang singularity, as otherwise there would be particles travelling indefinitely to the past without reaching the singularity. The case $a < 0$ is obtained by inverting the arrowed lines and the sign of $\theta$. Thus, for $a < 0$ the singularity is not a big bang but it is a big crunch.

Consider also the second case in [198]. The line-element is given by

$$ds^2 = -dt^2 + dx^2 + \{F_b(t - x) + [a(t + x)]^b\}^2 dy^2 + [a(t - x)]^{2(1-b)} dz^2,$$

where $a > 0$ and $b$ are constants, and $F_b(t - x)$ is a function whose explicit form depends on the value of $b$, giving the following three different

possibilities:

$$F_b(t-x) = \begin{cases} \sqrt{a(t-x)}\{c_1[a(t-x)]^Q + c_2[a(t-x)]^{-Q}\} \\ \qquad\qquad\qquad\qquad\qquad\text{(a) for } b \in (b_-, b_+), \\ \sqrt{a(t-x)}\{c_1 - c_2 \log[a(t-x)]\} \\ \qquad\qquad\qquad\qquad\qquad\text{(b) for } b = b_\pm, \\ \sqrt{a(t-x)}\, c_1 \cos\{Q \log[a(t-x)] + c_2\} \\ \qquad\qquad\qquad\qquad\qquad\text{(c) for } b \notin [b_-, b_+], \end{cases}$$

where $2Q = \sqrt{|(2b+1)^2 - 8b^2|}$, $2b_\pm = (1 \pm \sqrt{2})$, and $c_1, c_2$ are arbitrary constants. The dust velocity vector and corresponding energy density are

$$\mathbf{u} = \frac{1}{\sqrt{t^2 - x^2}}(-t\,dt + x\,dx), \qquad \varrho = -\frac{4b(b-1)\,[a(t+x)]^b}{(t^2-x^2)\{F_b + [a(t+x)]^b\}},$$

while the expansion of the dust geodesic congruence is

$$\theta = \frac{(2-b)F_b + 2[a(t+x)]^b + (t-x)\dot{F}_b}{(t^2-x^2)^{1/2}\{F_b + [a(t+x)]^b\}},$$

so that the dust congruence expands initially and then, either remains expanding forever or recollapses to a future singularity. Choosing always the natural singular extension and noting that the possible singularities may appear at $t - x = 0$, or at $t + x = 0$, or at $F_b(t-x) + [a(t+x)]^b = 0$, a very rich variety of possibilities arises [198]. Among them, some illustrative examples are presented here in Figures 7, 8 and 9. In Fig. 7 the big-bang singularity is null but all the dust congruence only approaches a 2-plane of the whole null singularity. Anyway, the rest of the singularity is essential for it to be a big bang. In Fig. 8, the whole singularity is, so to speak, 'closed' and both of big-bang and big-crunch type. The singularity is general, with spacelike, timelike and null subsets. All this is curious enough, but also the null part is a pure Weyl singularity, the matter quantities being regular there. Finally, in Fig. 9 the spacetime has a big-crunch and no big-bang, even though all the curves in the dust congruence approach the singularity to the past. But now, the null part $t + x = 0$ is completely regular, so that the spacetime is further extendible to the past. Here, there even appears the possibility of extending the various disconnected spacetimes shown in the figure and making them part of a single spacetime. This example and that of Figure 6 with $a < 0$ further show that the singularity predicted by the Raychaudhuri singularity theorem (see Section 5, Theorem 5.1) does not necessarily have to be of the big-bang type. All in all, once again the intuitive ideas, in this case about the big bang, are shaken by some explicit and reasonable examples.

774                                                                                                    Senovilla

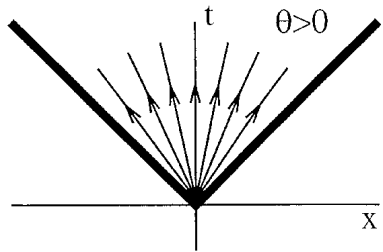

Figure 7. Here the notation is as in Fig. 6. The singularity is null everywhere with a corner. All the dust congruence starts from the corner, but other past-endless causal curves can hit the rest of the singularity.

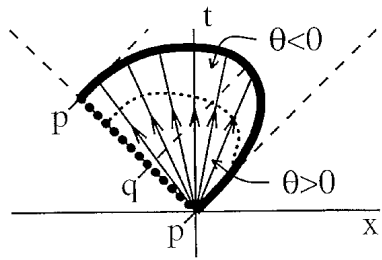

Figure 8. Notation again as in Fig. 6, and lines marked by $p$ and $q$ stand for the zeros of $F_b$ and its derivative, respectively. In this case the singularity 'closes over itself' and is of both big-bang and big-crunch type. There is a null part of the singularity which is avoided by the dust, but from where other particles emerge. This part is denoted by a dotted line because it is a pure Weyl singularity (the matter quantities are regular there).

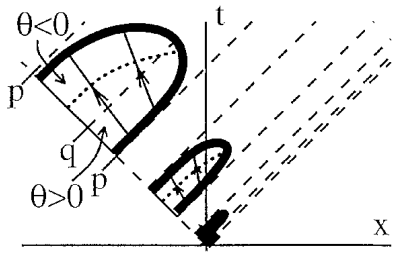

Figure 9. The notation is as in Figs. 6 and 8. Only one of the disconnected regions is the spacetime. Now, the singularity has a general character, but the continuous null line to the left is not an essential singularity and the spacetime can be extended across it to the past. In fact, an extension may be sought which combines the different disconnected sets to form part of a single spacetime. It seems that a logical past-extension might be a vacuum plane wave [198]!

## 4. TRAPPED SETS, TRAPPED SURFACES AND RELATED CONCEPTS

Two of the three basic ingredients for singularity theorems have been already studied: causality and energy conditions. In this Section the third fundamental ingredient, which usually consists in an appropriate boundary or initial condition, is analysed. In general this condition tries to express the fact that some *finite region* of the space (not spacetime) is *trapped* within itself and nothing can escape from it, at least initially. To start with, let us present the concept of a trapped surface.

**Definition 4.1.** A trapped surface is a spacelike surface $S$ in which the traces of the two null f-d second fundamental forms have the same sign. When both traces are negative the surface is future trapped, while if they are positive it is past trapped.

The two null f-d second fundamental forms are the second fundamental forms defined in (6) corresponding to the two null f-d normals (4), that is

$$K_{AB}^{\pm} \equiv K_{AB}(\mathbf{k}^{\pm}) \equiv -k_{\mu}^{\pm} e_A^{\upsilon} \nabla_{\upsilon} e_B^{\mu} = e_B^{\mu} e_A^{\upsilon} \nabla_{\upsilon} k_{\mu}^{\pm} .$$

Obviously, under a change of type (5) the two null second fundamental forms transform as

$$K_{AB}^{+} \longrightarrow K_{AB}^{'+} = A^2 K_{AB}^{+} , \qquad K_{AB}^{-} \longrightarrow K_{AB}^{'-} = A^{-2} K_{AB}^{-} .$$

The traces of the two null second fundamental forms are given by

$$K^{\pm} \equiv \gamma^{AB} K_{AB}^{\pm} , \qquad (38)$$

where $\gamma^{AB}$ is the contravariant metric on $S$, that is to say, the inverse of the first fundamental form: $\gamma^{AC} \gamma_{CB} = \delta_B^A$. Thus, a spacelike surface is trapped if and only if the scalar

$$\kappa \equiv K^{+} K^{-} \qquad (39)$$

is positive everywhere on $S$. From the above it is evident that the scalar $\kappa$ is invariant under transformations (5), so that there is no need to care about the normalization factor of the null normals. The physical interpretation of the concept of trapping is simple. As is obvious from (31) and the remarks preceding it, the scalars (38) are equivalent to the expansions $\vartheta^{\pm}$ of the two families of null geodesics which are tangent to $\vec{k}^{\pm}$ at $S$ [107]. Hence, a 2-surface $S$ is trapped if, given the two families of f-d null geodesics orthogonal to $S$, both of them are converging or diverging all over $S$.

The definition of trapped surfaces can be slightly improved by distinguishing between the different cases that may appear. In fact, sometimes only the sign of one of the traces is important (see Ref. 107), because there may be physical reasons to expect this particular trace having a definite sign if the 'trapping' does not take place. In these cases, identifying the trace $K^+$ (resp. $K^-$) with the one hoped to be positive (resp. negative) let us define a future (+)-trapped surface as a spacelike surface in which $K^+$ is non-positive, and a future (−)-trapped surface when $K^-$ is non-negative. When there is an intrinsic definition of outgoing (+) and ingoing (−) null normal forms of $S$, then the future (+)-trapped surfaces are called outer trapped, and the future (−)-trapped surfaces are called inner trapped. Another important possibility is the case of almost trapped surfaces:

**Definition 4.2.** A marginally (+)-trapped surface is a spacelike surface $S$ with the trace $K^+$ vanishing. Similarly for marginally (−)-trapped surface. $S$ is absolutely non-trapped if the scalar $\kappa$ of (39) is negative.

Notice that a marginally (±)-trapped surface is an extreme case of a (±)-trapped surface. For a marginally trapped surface the scalar $\kappa$ vanishes, while for an absolutely non-trapped surface, one of traces $K^\pm$ is strictly positive and the other is strictly negative, so that none of them can vanish anywhere on $S$. Before proceeding any further, let us see some simple examples.

**Example 4.1 (Trapped surfaces in Minkowski spacetime).** Let us consider Minkowski spacetime (Example 3.2) and the following spacelike surfaces. Firstly, take the surface $S$ with imbedding (1)

$$t = \Phi^0(u) = \log(\cosh(u^3)), \qquad x = \Phi^1(u) = 0,$$
$$y = \Phi^2(u) = u^2, \qquad z = \Phi^3(u) = u^3,$$

so that the first fundamental form (3) is $\gamma_{22} = 1$, $\gamma_{23} = 0$ and $\gamma_{33} = \cosh^{-2}(u^3)$. A straightforward calculation leads to the two null f-d normal one-forms

$$\sqrt{2}\,\mathbf{k}^\pm = (-\cosh(u^3)dt \pm dx + \sinh(u^3)dz)|_S,$$

the corresponding null second fundamental forms

$$K^\pm_{22} = K^\pm_{23} = 0, \qquad \sqrt{2}\,K^\pm_{33} = \cosh^{-1}(u^3),$$

and the traces

$$\sqrt{2}\,K^\pm = \cosh(u^3) \implies 2\kappa = \cosh^2(u^3).$$

Thus, this simple surface is past trapped. This particular surface is non-compact and extends to infinity. A trapped surface compactified in one direction can be built in Minkowski as follows. Choose cylindrical coordinates so that the line-element takes the form (35) but with the usual $2\pi$-periodicity for $\varphi$. Define the imbedding of a surface $S$ as

$$t = \Phi^0(u) = \log(\cosh(u^3)), \qquad \rho = \Phi^1(u) = \rho_0,$$
$$\varphi = \Phi^2(u) = u^2, \qquad z = \Phi^3(u) = u^3,$$

where $\rho_0$ is constant. A computation analogous to the previous one leads to

$$\sqrt{2}K^\pm = \cosh(u^3) \pm \frac{1}{\rho_0} \implies 2\kappa = \cosh^2(u^3) - \frac{1}{\rho_0^2}.$$

Therefore, all these surfaces are past trapped if $\rho_0 > 1$. Again, these surfaces are non-compact and extend to infinity. Looking for a compact trapped surface in Minkowski spacetime one readily realizes that there is going to be trouble. Take, for instance, the simplest case of a 2-sphere, which can be imbedded as follows:

$$t = \Phi^0(u) = T, \qquad x = \Phi^1(u) = R\sin(u^2)\cos(u^3),$$
$$y = \Phi^2(u) = R\sin(u^2)\sin(u^3), \qquad z = \Phi^3(u) = R\cos(u^3)$$

for some constants $T, R$. A straightforward calculation (easier in spherical coordinates) provides

$$\sqrt{2}K^\pm = \pm\frac{1}{R} \implies 2\kappa = -\frac{1}{R^2}$$

so that the 2-spheres are absolutely non-trapped in Minkowski. The idea here is very simple. If the surface is compact, it seems reasonable that the *outgoing* null geodesics expand while the *ingoing* ones must contract. In fact, using this result it can be proved that there is no compact trapped surface without boundary $S$ in Minkowski. To see it, the homogeneity of spacetime can be used, and the fact that $S$ is compact implies that it must be osculant (bi-tangent) to some 2-sphere somewhere. At this place, the surface is non-trapped as can be explicitly shown [196]. Finally, trivial examples of marginally trapped surfaces are the 2-planes $t = T$, $x = X$ for constant $T, X$. Here, both traces $K^\pm$ vanish.

Thus, in Minkowski spacetime one cannot find a finite region in space such that its future is 'contained within a finite region', so to speak. But this is the important concept for the development of singularities in a spacetime, which was introduced by Penrose [162] in order to show one of the first modern singularity theorems (see also Ref. 163). The precise definition is:

**Definition 4.3.** A closed trapped surface is a compact without boundary trapped surface.

Of course, analogously we can also define closed ($\pm$)-trapped surfaces closed marginally trapped surfaces, and closed absolutely non-trapped surfaces. So far, there has been no need to assume that the surfaces are connected. In fact, this assumption is superfluous for our purposes, because all the definitions of this Section, as well as all the theorems we shall prove in the next one, apply to each *connected component* of any surface. Nevertheless, in order to fix ideas and to avoid unnecessary complications in the proofs of the theorems, we shall assume from now on that all surfaces are connected.

**Example 4.2 (Closed trapped surfaces).** Here we present two typical examples of closed trapped surfaces. First, let us take the Vaidya metric (Example 3.5) which, in particular, includes the Eddington–Finkelstein extension of Schwarzschild spacetime (Example 3.3) for constant $M$. Take any 2-sphere imbedded in the spacetime by means of

$$t = \Phi^0(u) = T, \quad r = \Phi^1(u) = R, \quad \vartheta = \Phi^2(u) = u^2, \quad \varphi = \Phi^3(u) = u^3,$$

for constant $T, R$, so that the first fundamental form is $\gamma_{22} = R^2$, $\gamma_{23} = 0$ and $\gamma_{33} = R^2 \sin^2(u^2)$. The two f-d null normal one-forms are obviously

$$\mathbf{k}^+ = -dt|_S, \qquad \mathbf{k}^- = \left(\varepsilon dr - \frac{1}{2}\left(1 - \frac{2M(T)}{R}\right)dt\right)\bigg|_S.$$

A simple calculation gives then

$$K^+ = \frac{2\varepsilon}{R}, \quad K^- = -\frac{\varepsilon}{R}\left(1 - \frac{2M(T)}{R}\right) \implies \kappa = -\frac{2}{R^2}\left(1 - \frac{2M(T)}{R}\right).$$

Therefore, these 2-spheres are trapped iff $R < 2M(T)$, and are absolutely non-trapped iff $R > 2M(T)$. Notice that, in the case of Eddington–Finkelstein metric ($M = $ const.), the region with closed trapped surfaces is precisely the region added to Schwarzschild with the extension. Further, the 2-spheres with $R = 2M(T)$ are marginally trapped, and have $K^- = 0$. The set of all such marginally trapped 2-spheres constitute a hypersurface defined by $r - 2M(t) = 0$. This hypersurface is called the apparent horizon (AH), and separates the zones with and without closed trapped 2-spheres [126]. The normal one-form to the AH is $\mathbf{n} = (dr - 2M_{,t}dt)|_{\text{AH}}$, and its modulus is $g(\vec{n},\vec{n}) = 4\varepsilon M_{,t}|_{\text{AH}}$, so that the AH is a null hypersurface at a point $p \in \text{AH}$ iff $M_{,t}|_p = 0$. In particular, in the Eddington–Finkelstein case the AH is null and, in fact, it is called the event horizon EH (see Ref. 107). In the general case, the AH is a non-timelike hypersurface iff $\varepsilon M_{,t} \leq 0$, which is precisely the condition such that WEC and DEC are satisfied. Hence, for physically realistic Vaidya metrics the AH is spacelike or null.

The second example is the FLRW spacetime (Example 3.1). Take the 2-spheres

$$t = \Phi^0(u) = T, \quad \chi = \Phi^1(u) = R, \quad \vartheta = \Phi^2(u) = u^2, \quad \varphi = \Phi^3(u) = u^3,$$

for constants $T$, $R$, so that the first fundamental form is $\gamma_{22} = R^2$, $\gamma_{23} = 0$ and $\gamma_{33} = R^2 \sin^2(u^2)$. The two f-d null normal one-forms are $\sqrt{2}\mathbf{k}^\pm = (-dt \pm a(T)d\chi)|_S$ and their respectives traces read

$$\sqrt{2}K^\pm = \frac{1}{a}\left(a_{,t} \pm \frac{\Sigma_{,\chi}}{\Sigma}\right)\bigg|_S.$$

Therefore, these 2-spheres will be trapped if and only if

$$\kappa = K^+ K^- > 0 \iff a_{,t}^2 + k - \frac{1}{\Sigma^2} > 0 \iff \frac{\varrho}{3}a^2 - \frac{1}{\Sigma^2} > 0,$$

where in the last equivalence we have used Friedman's equations (34). Thus, $S$ is past trapped for all values of $R$ such that

$$\Sigma(R,k) > \sqrt{\frac{3}{\varrho(T)}} \frac{1}{a(T)},$$

which is always possible if $\varrho > 0$ *provided* that $a_{,t} \neq 0$ and that the FLRW extends that far. The interpretation of this result is simple. Take any of these trapped 2-spheres. The out- and in-going null congruences orthogonal to $S$ form new 2-spheres a little time after they leave $S$. Of course, the outgoing ones form a 2-sphere with a bigger value of $\chi$, and thus its area is bigger than that of the initial one if the universe expands ($a_{,t} > 0$). The ingoing ones form a new 2-sphere with a smaller value of $\chi$, but as the universe expands its area can still be bigger than that of the initial one *at the initial time*. Thus, both new 2-spheres have increased their area. Again the AH can be defined as the hypersurface separating the trapped from the non-trapped 2-spheres. This AH is given now by $\varrho\Sigma^2 a^2 = 3$, and therefore its normal one-form is

$$\mathbf{n} = [(\varrho a^2)_{,t}\Sigma^2 dt + 2\varrho a^2 \Sigma\Sigma_{,\chi}d\chi]|_{\text{AH}}$$

whose modulus is [on using (34)] $3a^2\Sigma^2\Sigma^2_{,\chi}(\varrho + p)(\varrho - 3p)|_{AH}$, so that the AH in physical FLRW models ($\varrho + p > 0$) is a timelike, null or spacelike hypersurface if $3p$ is less than, equal to or greater than $\varrho$, respectively.

The importance of the concept of closed trapped surface rests on the following result [12].

**Proposition 4.1.** If the null convergence condition holds and there exists a closed future (resp. past) trapped surface $S$, then either $E^+(S)$ [resp. $E^-(S)$] is compact or the spacetime is null geodesically incomplete to the future (resp. past), or both.

*Proof.* Assume that $(V_4, g)$ is null geodesically complete and that $S$ is future trapped (the past case is identical). Then both traces $K^{\pm}$ are negative, which in turn means that the expansions $\vartheta^{\pm}$ of the null geodesics emanating orthogonally from $S$ are initially negative all over $S$. Let $\vartheta_M$ be the maximum value of $\vartheta^{\pm}$ at the compact $S$. By Propositions 2.7 and 2.14 it follows that all the null geodesics emanating orthogonally from $S$ enter into $I^+(S)$ at or before a finite affine parameter $\leq -2/\vartheta_M$. Define $\mathcal{K}$ as the set of points reached by all these null geodesics from $S$ up to the affine parameter $-2/\vartheta_M$ inclusive, so that $\mathcal{K}$ is compact. Evidently $J^+(S) - I^+(S) = E^+(S) \subseteq \mathcal{K}$, so that it is enough to show that $E^+(S)$ is closed. Let $\{p_n\}$ be an infinite sequence of points $p_n \in E^+(S)$ and let $p$ be their accumulation point in the compact $\mathcal{K}$. By construction $\mathcal{K} \subset J^+(S)$ so that $p \in J^+(S)$. If $p$ were in $I^+(S)$, then there would be a neighbourhood $\mathcal{U}_p$ of $p$ in $I^+(S)$, which is open [Proposition 2.15, point (i)], and thus there would be some $p_n \in \mathcal{U}_p \subset I^+(S)$, which is impossible as $p_n \in E^+(S)$. In consequence, $p \in J^+(S) - I^+(S) = E^+(S)$, proving that $E^+(S) \subseteq \mathcal{K}$ is closed and thus compact. ∎

In a completely analogous manner, by using Propositions 2.7 and 2.14 it is easy to prove

**Proposition 4.2.** If the null convergence condition holds and there exists a point $p \in V_4$ such that the expansion of the f-d null geodesic family emanating from $p$ becomes negative along every curve of the family then either $E^+(p)$ is compact or the spacetime is null geodesically incomplete to the future, or both. ∎

Let us remark that $E^+(S)$ or $E^+(p)$ in both previous results may still be empty, because we have not assumed any causality condition. In any case, this is the idea that one wishes to keep for the general singularity theorems, that is, sets $\zeta$ such that $E^+(\zeta)$ or $E^-(\zeta)$ are compact. The precise definition is [108]:

**Definition 4.4.** A non-empty achronal set $\zeta$ is called future (resp. past) trapped if $E^+(\zeta)$ [resp. $E^-(\zeta)$] is compact.

Notice that a closed trapped surface is not necessarily a trapped set, because a trapped surface need not be *achronal*. But even if it is achronal, $E^+(S)$ may still be non-compact if the spacetime is null geodesically incomplete, due to Proposition 4.1. Concerning points such as those of Proposition 4.2, they are trapped sets provided that the spacetime is null geodesically complete. Actually, no example of trapped set has yet been presented explicitly here. The next example shows some.

**Example 4.3 (Trapped sets).** Consider the closed ($k = 1$) FLRW models of Example 3.1. Choose any hypersurface $\Sigma_T \equiv \{t = T\}$ = const., which is obviously compact (it is $S^3$). The computation of $E^+(\Sigma_T)$ is simple. If $\gamma$ is any null geodesic starting at $\Sigma_T$, $\gamma$ reaches points to the causal future of $\Sigma_T$, say a point $(t, \chi_1, \vartheta_1, \varphi_1)$ with $t > T$, which can *always* be also reached by the timelike geodesic defined by $t = \tau$, $\chi = \chi_1$, $\vartheta = \vartheta_1$ and $\varphi = \varphi_1$, where $\tau$ is proper time. Thus $E^+(\Sigma_T) = J^+(\Sigma_T) - I^+(\Sigma_T) = \Sigma_T$, so that $E^+(\Sigma_T)$ is compact. Similarly, $E^-(\Sigma_T) = \Sigma_T$. Hence, each $\Sigma_T$ is a future and past trapped set. The idea is better understood by looking at the Penrose diagram of these spacetimes. In Figure 10 we have drawn the case for dust (pressure $p = 0$). The trapping of the set in this case occurs simply because the FLRW model is closed, in the sense that each spacelike hypersurface $\Sigma_T$ is compact without boundary. This is a general property, and *any* compact achronal set $\Sigma$ without edge in any spacetime is both a future and past trapped set. This follows because, $\Sigma$ having no edge, there are no possible generator segments of $E^\pm(\Sigma)$, and thus $E^+(\Sigma) = E^-(\Sigma) = \Sigma$, which is compact by assumption.

To look for a non-trivial trapped set the non-singular black holes may be used [28,29,140]. For example, take the simple non-singular black hole satisfying WEC presented in [140]. Incidentally, these spacetimes are alternative extensions of the original Schwarzschild spacetime allowing for non-vacuum interior regions. Thus, this is another example of the non-uniqueness of the extensions of a given spacetime. This extension keeps the whole exterior original Schwarzschild region but it leads to no singularities in the added regions with $r < 2M$. In Eddington–Finkelstein-like coordinates, the line element is given by

$$ds^2 = -e^{4\beta}\left(1 - \frac{2m}{r}\right)dt^2 + 2e^{2\beta}dt\,dr + r^2(d\vartheta^2 + \sin^2\vartheta\,d\varphi^2),$$

where $m(r)$ and $\beta(r)$ take the following explicit form:

$$m(r) = \begin{cases} (r^3/16M^4)(20M^2 - 15rM + 3r^2) & \text{for } 0 < r \leq 2M, \\ M & \text{for } r \geq 2M, \end{cases}$$

$$\beta(r) = \begin{cases} (5r^2/192M^4)(24M^2 - 16rM + 3r^2) - 5/12 & \text{for } 0 < r \leq 2M, \\ 0 & \text{for } r \geq 2M, \end{cases}$$

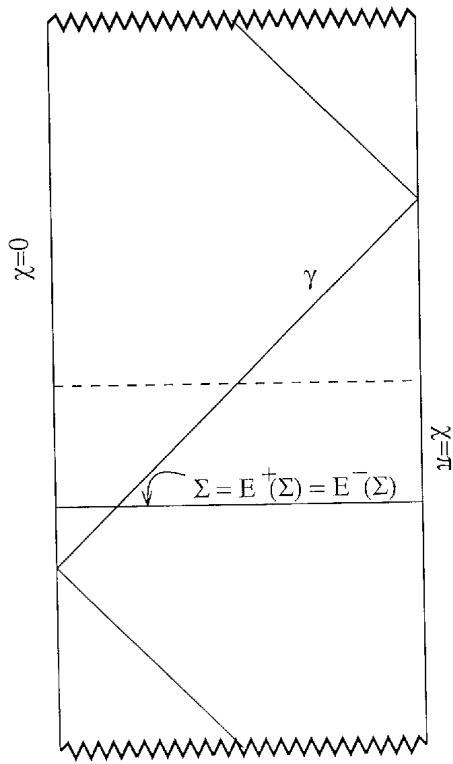

Figure 10. This is the Penrose conformal diagram of the closed ($k = 1$) dust FLRW model (compare with Figure 21(ii) in Ref. 107). I am grateful to Raül Vera for helping me with the drawing of this figure. The zig-zagging lines are the big-bang and big-crunch singularities. The two vertical lines represent $\chi = 0$ and its antipodes $\chi = \pi$. The dust expands from the big bang up to the middle dashed line, which represents the recollapsing time, and then contracts up to the big crunch. A compact edgeless acausal hypersurface is represented by $\Sigma$, which has the $S^3$ topology. This particular $\Sigma$ has timelike normals diverging everywhere. When a null line reaches one of the origins $\chi = 0, \pi$ then it 'rebounds' and follows its journey as indicated with the null geodesic $\gamma$. As is obvious, $I^+(\Sigma)$ and $I^-(\Sigma)$ are the parts over and below $\Sigma$ in the diagram, respectively, and therefore $J^+(\Sigma) - I^+(\Sigma) = E^+(\Sigma) = \Sigma$ and $J^-(\Sigma) - I^-(\Sigma) = E^-(\Sigma) = \Sigma$. Thus, $\Sigma$ is future and past trapped, because $E^+(\Sigma) = E^-(\Sigma) = \Sigma$ are compact. Incidentally, from the diagram follows that a photon emanating from the big bang can go round the whole universe and come back to the same position in space *just* when arriving at the big crunch. As an exercise, the reader may try to draw the AH in this spacetime.

and where $M$ is the mass in the Schwarzschild spacetime. Thus, for $r \geq 2M$ this is just Schwarzschild spacetime in these coordinates. There is a unique value $r_1 < 2M$ of $r$ such that $r_1 = 2m(r_1)$. A simple calculation analogous to that in Example 4.2 shows that the 2-spheres with constant

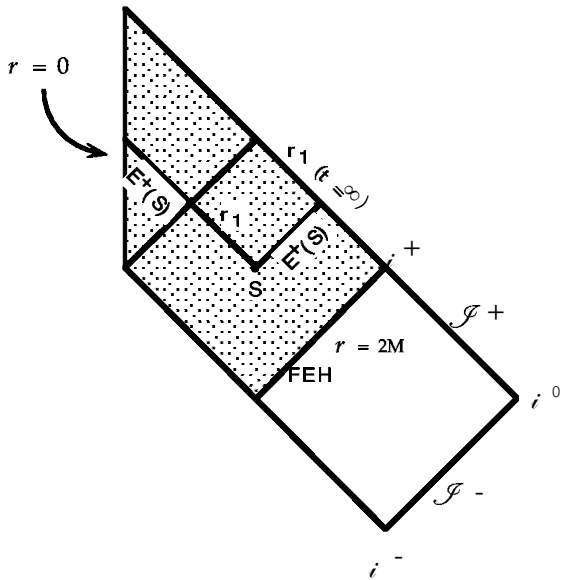

Figure 11. Penrose conformal diagram of the spherically symmetric spacetime of Example 4.3. There is an asymptotically flat Schwarzschild region with event horizon at $r = 2M$, and then a matter-filled interior zone with another null hypersurface $r = r_1$ such that $2m(r_1) = r_1$. Finally, $r = 0$ is the origin of coordinates and there is no singularity there (thus, the $r = 0$ points are real points and not 2-spheres). Every point of the 'square' between $r_1 < r < 2M$ represents a closed trapped 2-sphere, such as the $S$ shown. However, half of the f-d null geodesics emanating from $S$ reach $r = r_1$ with $t \to \infty$ but finite affine parameter, and thus they are incomplete. It follows that $E^+(S)$ is not compact whence $S$ is not a trapped set. This is due to the g-incompleteness of the spacetime. This spacetime is extendible across $r = r_1$, and a possible maximal extension is given in the next figure.

$t$ and $r$ are trapped if and only if $r_1 < r < 2M$. The curious property of this spacetime is that it is regular at $r = 0$ [140]. The Penrose conformal diagram of this spacetime is shown in Figure 11, where the incompleteness of the f-d null geodesics is also manifested, as they reach $t \to \infty$ with finite proper time and $r \to r_1$. Then, the mentioned closed trapped 2-spheres are *not* trapped sets in the sense of Definition 4.4. Nevertheless, a regular extension of this spacetime can be performed, and the new spacetime thus obtained has no incomplete curves [140]. The global structure of the new spacetime is similar to that of the maximally extended Reissner–Nordström metric [107,140], but with the $r = 0$–regions completely regular. This is shown in Figure 12. Consequently, the topology of the spacetime changes in the intermediate regions and the spacelike hypersurfaces are $S^3$ there (see Refs. 28,29). Further, the above cited trapped 2-spheres $S$ are true trapped sets, as their $E^+(S)$ is now compact, reaching $r = 0$ in all possible directions and then entering into $I^+(S)$. This is shown in Fig. 12. This possibility arises due to the change of topology, so that the null generating segments of $E^+(S)$ can meet after travelling round the closed $S^3$.

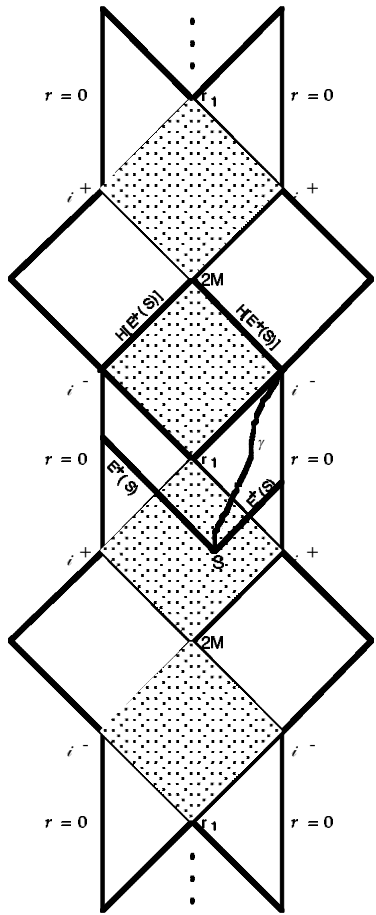

Figure 12. This is the Penrose diagram of a maximal extension of the spacetime of Fig. 11. The resulting whole spacetime extends indefinitely upwards and downwards, and is inextendible and singularity-free. The global structure resembles that of Reissner–Nordström [107], but the $r = 0$ regions are completely regular. Therefore, there is a change in the topology of the slices, and a 'horizontal' slice going from $r = 0$ to the other $r = 0$ has the $S^3$ topology, as in the case of Fig. 10. All points in the shadowed squares represent closed trapped 2-spheres. Choosing a particular one, $S$, their future horismos have been drawn, showing that they are compact and thus $S$ is a trapped set in this case. The future Cauchy horizon $H^+[E^+(S)]$ is also represented. It can be checked that $H^+[E^+(S)]$ is non-compact (due to the fundamental Lemma 2.7). Then, there are many future-endless timelike curves completely contained in $\mathrm{int}D^+[E^+(S)]$, as Lemma 4.1 proves. One possibility is indicated by the curve $\gamma$. Another one is given by the $r = 0$ timelike geodesic itself.

Thus, the closed trapped surfaces can fail to be trapped sets because either the spacetime is not null geodesically complete or they are not achronal. In fact, this second possibility is of little relevance due to the following result [12].

**Proposition 4.3.** If the null convergence and the strong causality conditions hold and there exists a closed future-trapped surface $S$, then either $E^+(S) \cap S$ is a trapped set or the spacetime is null geodesically incomplete, or both.

*Proof.* First, the achronality of $E^+(S) \cap S$ follows from that of $E^+(S)$. Assume that $(V_4, g)$ is null geodesically complete. Then, from Proposition 4.1 it follows that $E^+(S)$ is compact, so that $E^+(S) \cap S$ is also compact. Further, $E^+(S) \cap S$ is non-empty, because if it were, then it would be $S \subset I^+(S)$, in evident contradiction with Proposition 2.21 as $S$ is compact. Let us show finally that $E^+[E^+(S) \cap S] = E^+(S)$, so that $E^+[E^+(S) \cap S]$ is compact. To that end, cover $S$ with convex normal neighbourhoods. As strong causality holds, they can be chosen such that every causal curve does not intersect them in a disconnected set. Further, as $S$ is spacelike, these neighbourhoods can be taken small enough so that the piece of $S$ contained in each of them is achronal. Now, as $S$ is compact, we can extract a finite sub-cover $\{\mathcal{U}_i\}_{i=1...n}$ of the previous cover. Take any point $q \in I^+(S)$, so that there exists $p_1 \in \mathcal{U}_1$ (say) with $q \in I^+(p_1)$. If some $p_1 \in E^+(S) \cap S$, then $q \in I^+[E^+(S) \cap S]$. If all such $p_1$ are not in $E^+(S) \cap S$, and as $p_1 \in S$, there must exist $p_2 \in S$ with $p_1 \in I^+(p_2)$. Also, as $\mathcal{U}_1 \cap S$ is achronal, $p_2 \notin \mathcal{U}_1$, so that $p_2 \in \mathcal{U}_2 - \mathcal{U}_1$ (say). If some such $p_2$ is in $E^+(S) \cap S$, then $q \in I^+[E^+(S) \cap S]$. If all such $p_2 \notin E^+(S) \cap S$, again there exists a $p_3 \in S$ with $p_2 \in I^+(p_3)$ and, given the way that the $\mathcal{U}_i$ were chosen, $p_3 \in \mathcal{U}_3 - (\mathcal{U}_2 \cap \mathcal{U}_1)$ (say). Proceeding in this manner, and as the number of the $\mathcal{U}_i$ is finite, the process must terminate, which means that there is a $p_j \in \mathcal{U}_j$ with $p_j \in E^+(S) \cap S$ and $q \in I^+[E^+(S) \cap S]$. In other words, for all $q \in I^+(S)$, it follows that $q \in I^+[E^+(S) \cap S]$. This obviously implies $I^+(S) = I^+[E^+(S) \cap S]$. Take now a point $r \in J^+(S)$. If $r \in I^+(S)$, then by the previous equality $r \in I^+[E^+(S) \cap S] \subseteq J^+[E^+(S) \cap S]$. If $r \notin I^+(S) = I^+[E^+(S) \cap S]$, then there is a $p \in S$ such that $r \in E^+(p)$. Obviously, $p \notin I^+(S)$ as otherwise $r$ would be in $I^+(S)$. Thus, $p \in S - I^+(S)$ so that $p \in E^+(S) \cap S$ and hence $r \in J^+[E^+(S) \cap S]$. This implies that $J^+(S) = J^+[E^+(S) \cap S]$. Hence, we have $E^+[E^+(S) \cap S] = J^+[E^+(S) \cap S] - I^+[E^+(S) \cap S] = J^+(S) - I^+(S) = E^+(S)$, as desired. ∎

In other words, if strong causality condition holds, the existence of a closed trapped surface implies the existence of a trapped set *unless* the spacetime is null geodesically incomplete. The following lemma shows a fundamental property of trapped sets [108].

**Lemma 4.1.** If the strong causality condition holds and there exists a future trapped closed set $\zeta$, then there is a future-endless timelike curve $\gamma$ contained in $D^+[E^+(\zeta)]$.

*Proof.* As strong causality holds, then $E^+(\zeta) \neq \varnothing$. Moreover, from the fourth point of Lemma 2.7 we know that $H^+[E^+(\zeta)]$ is non-compact or empty. If it is empty the result is trivial. If $H^+[E^+(\zeta)]$ is non-empty and non-compact, then choosing any smooth f-d timelike congruence in $V_4$ and given the achronality of $H^+[E^+(\zeta)]$, each curve of the congruence passing through $E^+(\zeta)$ can cross $H^+[E^+(\zeta)]$ at most once. If all of them crossed $H^+[E^+(\zeta)]$, there would be a one-to-one continuous map from $E^+(\zeta)$ to $H^+[E^+(\zeta)]$, which is impossible because $E^+(\zeta)$ is compact and $H^+[E^+(\zeta)]$ is not. Hence, there must be some curves of the congruence not intersecting $H^+[E^+(\zeta)]$ and thus remaining in $D^+[E^+(\zeta)]$. ∎

In fact, there are many curves with these properties, because given the $\gamma$ of the previous proposition, $I^-(\gamma) \cap D^+[E^+(\zeta)]$ has many future endless timelike curves contained in $D^+[E^+(\zeta)]$. An illustrative example of the property given by Lemma 4.1 can be easily seen in Fig. 12 of Example 4.3. There are closed future trapped surfaces $S$ as indicated, and then $E^+(S)$ is compact. However, $H^+[E^+(S)]$ is non-compact, as their generators are past-endless and approach the point at infinity $i^-$. Thus, there are curves contained in $D^+[E^+(S)]$ which are future-endless. A possibility is the curve $\gamma$ shown in Fig. 12. Another one would be the curve along $r = 0$ itself.

## 5. SINGULARITY THEOREMS

In this section a significant number of the singularity theorems is presented including, of course, those with a special relevance from the historical or the scientifical points of view. Apart from some brief comments and remarks, in this Section only the theorems and their proofs are presented. The reasonability of the assumptions, their physical significance, the possible exceptional cases and all illustrative examples are left for the next two Sections. A reader not interested in the proofs of the theorems but rather in their possible applications and the intuitive ideas behind them may go directly to the next Sections.

Perhaps the first singularity theorem was proven by Raychaudhuri [170] back in 1955, and independently by Komar [120,171]. Raychaudhuri's paper contained, of course, the fundamental equation (27), which lies behind *any* singularity theorem due to its important consequences concerning geodesic focusing (Propositions 2.3 and 2.4). Komar's paper contained basically the same ideas but the concept of strong energy condition[7] was introduced in a sense (Definition 2.11) as well as the use of Gaussian coordinates (28). In principle, the theorem is exactly the same as Proposition 2.3, so that it shows the focusing of a geodesic congruence [170] or the failure of the Gaussian coordinates [120]. However, *assuming* that the matter moves along the focused geodesic congruence a matter singularity can be obtained under some general and sometimes reasonable circumstances.

**Theorem 5.1 (Raychaudhuri and Komar).** Assume the matter content of the spacetime can be described by an energy-momentum tensor of the perfect fluid type and the velocity vector $\vec{u}$ of the fluid is geodesic and irrotational. If the expansion is positive at an instant of time and the energy condition $R_{\mu\nu}u^\mu u^\nu \geq 0$ holds, then there is a matter singularity in the finite past along every integral curve of $\vec{u}$.

Of course, the notion of 'an instant of time' is meaningful here and refers to the natural time defined in irrotational models by (24), or more precisely by (25) as we are also assuming that $\vec{a} = \vec{0}$. Thus, $\mathbf{u} = -dt$ and the assumption on the expansion is $\theta|_{t_0} > 0$. Notice that $\theta$ is a function of all four coordinates, as there is no restriction concerning the symmetry group of the spacetime — this happens for example in the Szekeres cosmologies [209,206,207,124]. Therefore, the condition $\theta|_{t_0} > 0$ means $\theta(t_0, x^i) > 0$ for all $x^i$ in a suitable coordinate system including $t$. When there is a three-dimensional symmetry group acting transitively on spacelike hypersurfaces, the theorem is also proved and was carefully analysed in [68,71]. In fact, the proof of the theorem implicitly assumes that the geodesic congruence can be extended up to where the focusing takes place, which may not happen sometimes [68,71].

*Proof.* The proof starts with a repetition of the proof of Proposition 2.3. Thus, defining $V$ as in (30), we get that $V$ vanishes and $\theta$ diverges in the past along every curve of the congruence before $t$ reaches the value $t_0 - 3/(\theta|_{t_0})$, provided that the congruence can be extended that far. Of course this focusing is not simultaneous in general (see Example 3.7 in Section 3). Now it is intuitively clear that there is going to be trouble with the assumption that matter moves along this congruence. To prove it rigourously, under the assumption of the perfect-fluid energy-momentum tensor

$$T_{\mu\upsilon} = \varrho u_\mu u_\upsilon + p h_{\mu\upsilon},$$

the conservation equations $\nabla_\mu T^{\mu\upsilon} = 0$ imply

$$u^\mu \nabla_\mu \varrho + (\varrho + p)\theta = 0, \qquad (\varrho + p)a^\mu + h^{\mu\upsilon}\nabla_\upsilon p = 0. \qquad (40)$$

---

[7] Raychaudhuri did not have to introduce SEC because he implicitly assumed a dust energy-momentum tensor, and took the condition $\varrho \geq 0$ for granted.

Setting $V^{1/3} \equiv a(t)$ along each curve, the first of these equations can be rewritten as

$$\frac{d}{da}(\varrho a^2) + a(\varrho + 3p) = 0 \implies \varrho = \frac{(\varrho a^2)|_{t_0}}{a^2} - \frac{1}{a^2}\int_{a_0}^{a} a(\varrho + 3p)da,$$

which under the conditions of the theorem implies $\varrho \to \infty$ as $a \to 0$. ∎

Of course, this theorem applies to the FLRW models, and also to most of the spatially homogeneous models [123,134,179]. The result holds for some spatially *inhomogeneous* models too, as for instance the general Szekeres family [209,206,207,124] which have no symmetry in general. The important assumption in Theorem 5.1 is the absence of acceleration and rotation so that, as mentioned before, acceleration (or rotation) of matter is somehow necessary to avoid singularities. This is a physically reasonable fact because, from the second equation in (40), acceleration is directly related to the existence of a gradient of pressure acting against gravitational attraction in perfect fluids. In fact, this holds for more general fluids [195]. The necessity of rotation for the avoidance of this type of singularities is usually accepted, and this was manifested by Gödel's revolutionary paper [95]. Also, this was supported by the Newtonian cosmologies, in which rotation prevents the appearance of the matter singularity (see, e.g., Ref. 197 and references therein). Nevertheless, *acceleration* with or without rotation can also do the job in General Relativity, and this has been forgotten sometimes. In Section 7, some interesting singularity-free perfect-fluid models without rotation will be presented.

Concerning the properties of the singularity, Theorem 5.1 has the virtue that predicts something more than just the existence of the singularity: it says where to locate it, that it is a matter singularity and that it is unavoidable for all the curves of the fluid congruence. This does not mean that the singularity is of big-bang type,[8] as was explicitly demonstrated in the various cases shown in Example 3.7, for which the conditions of the theorem are obviously satisfied. Moreover, the character of the singularity is not determined either, as was also seen in Examples 3.6 and 3.7. Of course, Theorem 5.1 can be equally proved under the condition of $\theta_{t_0} < 0$, and then the singularity is in the finite future. Some cases in Example 3.7 have the double future and past application of this theorem.

---

[8] There is an erroneous statement concerning this in [195].

A similar theorem is the following

**Theorem 5.2.** If $(V_4, g)$ contains a Cauchy hypersurface $\Sigma$ such that the timelike geodesic congruence emanating orthogonal to $\Sigma$ has an initial expansion $\theta|_\Sigma \geq c > 0$ and SEC holds along the congruence, then all the geodesics in the congruence are past incomplete.

*Proof.* The spacetime is globally hyperbolic (Proposition 2.31), so that by Proposition 2.33 there is a maximal curve from $\Sigma$ to any point $p \in V_4$. This maximal curve must be a timelike geodesic orthogonal to $\Sigma$ due to Proposition 2.12, and it cannot have any point focal to $\Sigma$ between $\Sigma$ and $p$. But Proposition 2.3 assures that these focal points should exist to the past at a proper time less than or equal $3/\theta|_\Sigma \leq 3/c$. Hence, no timelike geodesic orthogonal to $\Sigma$ has length greater than $3/c$ to the past.  ∎

Another similar theorem where the focusing of geodesics is proved under milder assumptions but strengthening the properties of $\Sigma$ was found in [215]. The proof is completely similar.

**Theorem 5.3.** If the spacetime contains a Cauchy hypersurface $\Sigma$ with vanishing second fundamental form (6) and there exist positive constants $b, c$ such that

$$\left| \int_{\tau_\Sigma}^{\tau_\Sigma + b} R_{\mu\upsilon} v^\mu v^\upsilon d\tau \right| \geq c > 0,$$

along every timelike geodesic orthogonal to $\Sigma$ with tangent vector $\vec{v}$ and proper time $\tau$, then all the geodesics in the congruence are both future and past incomplete.  ∎

The first 'modern' singularity theorem was published by Penrose in 1965 [162], and opened the door to the series of fundamental results obtained later by himself, Hawking, Geroch and others (see Ref. 220 for a review). In his important paper, the concept of closed trapped surface was introduced and the theorem was devised to be applicable to the general collapse of stars without spherical symmetry. The precise statement is

**Theorem 5.4 (Penrose).** If the null convergence condition holds and there are a non-compact Cauchy hypersurface $\Sigma$ and a closed trapped surface $S$, then the spacetime is null geodesically incomplete.

*Proof.* Suppose that $S$ is future trapped. If the spacetime were null geodesically complete, by Proposition 4.1 $E^+(S)$ would be compact. But the spacetime is globally hyperbolic (Proposition 2.31) and therefore causally simple (Proposition 2.26) so that $E^+(S) = \partial J^+(S)$, due to Proposition 2.24. Then, by Definition 2.16, $E^+(S)$ would be a *compact* proper achronal boundary. Choose any smooth timelike congruence in the spacetime. Every curve of the congruence intersects the Cauchy hypersurface $\Sigma$ exactly once, and meets $E^+(S)$ at most once (due to its achronality). Then, following the curves of the congruence a continuous map from $E^+(S)$ to $\Sigma$ is defined. If $E^+(S)$ were compact, and as $\Sigma$ is not, the image of $E^+(S)$ by this continuous map should have a boundary in $\Sigma$. But this is impossible due to Proposition 2.16. ∎

Thus, if the spacetime is to be globally hyperbolic and spatially *open*, the formation of closed trapped surface seems catastrophic. An improvement of this theorem can be found in [175], where the null convergence condition is assumed to hold only on average. Theorems 5.2, 5.3 and 5.4 assume the existence of a global Cauchy hypersurface so that the spacetime is globally hyperbolic. Nevertheless, it is obvious that all three theorems can be applied to the total Cauchy development int $D(\Sigma)$ of any *partial* Cauchy hypersurface. Of course, this would only prove the incompleteness of geodesics *within* int $D(\Sigma)$, and then possible regular extensions beyond the Cauchy horizon $H(\Sigma)$ might be performed to solve the incompleteness problem. As argued in Section 3, this extension is unsure, and there is no reason to assume, for instance, the same conditions as in the unextended region. So, the problem of singularities becomes a problem of extendibility again. Notice further that this can have physical relevance, as was claimed after Definition 3.3.

Penrose's result instigated a lot of work on the subject, and the main advances were found by Hawking in a series of papers with some initial inaccuracies, later corrected (see Ref. 220). Among the various results found there, let us present one which may apply to spatially closed universes [104]

**Theorem 5.5 (Hawking).** If SEC and the strong causality condition hold, and there exists an edgeless compact achronal spacelike hypersurface $\Sigma$ such that the timelike geodesic congruence emanating orthogonal to $\Sigma$ has a positive initial expansion, then the spacetime is past timelike geodesically incomplete.

In fact, this theorem can be proved without the assumption of strong causality [104]. Furthermore, there has appeared an alternative version in [28] where the assumption is the existence of a non-negative number $c$ such that $R_{\mu\nu}v^\mu v^\nu \geq -c^2/3$ for all timelike $\vec{v}$ and the initial expansion is greater than $c$. Finally, let us remark that the supposition of achronality for $\Sigma$ can be also dropped by working on the covering manifold of $V_4$ if necessary (see Refs. 88,104,107,165).

*Proof.* The idea is the same as in Theorems 5.2 and 5.3. As the initial expansion $\theta|_\Sigma$ is positive and $\Sigma$ is compact, there is a positive lower bound $K$ such that $\theta|_\Sigma \geq K$. If spacetime were timelike geodesically complete to the past, then there would be a point focal to $\Sigma$ along every past-directed timelike geodesic orthogonal to $\Sigma$ within a proper time $\tau \leq 3/K$ (Proposition 2.3). Following every such geodesic a proper time $\tau = 3/K$ and collecting the set of points so reached we would construct a compact set $\mathcal{K}$. This $\mathcal{K}$ would contain $D^-(\Sigma)$, due to Propositions 2.33 and 2.12, and thus it would also contain $H^-(\Sigma) \subset D^-(\Sigma) \subset \mathcal{K}$. It would follow that $H^-(\Sigma)$ is compact, which contradicts point (iv) in Lemma 2.7 because $H^-[I^-(\Sigma)] = H^-(\Sigma) \neq \emptyset$. ∎

The various developments were finally collected in a stronger theorem of much wider application by Hawking and Penrose [108]. In this fundamental paper most of the previous results were recovered under much weaker assumptions. The Hawking–Penrose theorem is the singularity theorem *par excellence*, even today.

**Lemma 5.1 (Hawking and Penrose).** The following three statements cannot hold simultaneously in any spacetime:
 (i) every endless causal geodesic has a pair of conjugate points,
 (ii) the chronology condition is satisfied,
 (iii) there is a trapped set $\zeta$.

*Proof.* First of all, Proposition 2.22 implies that the strong causality condition holds, so that Lemma 4.1 provides a future endless timelike curve $\gamma \subset \text{int } D^+[E^+(\zeta)]$. The set $J^-(\gamma) \cap E^+(\zeta)$ is achronal and closed, because $E^+(\zeta)$ is compact and achronal. As $\gamma \subset \text{int } D^+[E^+(\zeta)]$ every past-endless causal curve from $\gamma$ must cross $J^-(\gamma) \cap E^+(\zeta)$. Thus, $\partial J^-[J^-(\gamma) \cap E^+(\zeta)]$ consists of $J^-(\gamma) \cap E^+(\zeta)$ and the part of $\partial J^-(\gamma)$ to the past of $E^+(\zeta)$. In particular, consider the set $E^-[J^-(\gamma) \cap E^+(\zeta)] \subseteq \partial J^-[J^-(\gamma) \cap E^+(\zeta)]$. The null generators $\upsilon$ of $E^-[J^-(\gamma) \cap E^+(\zeta)]$ are thus portions of null generators $\mu \supset \upsilon$ of $\partial J^-(\gamma)$. But any $\mu$ is future endless, because $\gamma$ is future endless so that, extending $\mu$ to the past indefinitely, we obtain an endless null geodesic (still called $\mu$) which by condition (i) must have a pair of conjugate points. Proposition 2.14 implies that every $\mu$ enters into $I^-(\gamma)$ so that the null generator $\upsilon$ must have a past endpoint at or before the past endpoint of $\mu \cap \partial J^-(\gamma)$. Also, $\upsilon$ has a future endpoint at edge$[J^-(\gamma) \cap E^+(\zeta)]$, which is compact being a closed subset of the compact set $J^-(\gamma) \cap E^+(\zeta)$. Then, given that conjugate points vary continuously, the segment of $\mu$ from edge $\overline{[J^-(\gamma) \cap E^+(\zeta)]}$ to the past up to $\mu \cap \partial J^-(\gamma)$ is compact, hence the subsegments $v$ also generate a compact set $E^-[J^-(\gamma) \cap E^+(\zeta)]$. This means that $J^-(\gamma) \cap E^+(\zeta)$ is a past-trapped set, so that application of Lemma 4.1 provides another timelike curve $\lambda$ which is past endless and is contained in $D^-(E^-[J^-(\gamma) \cap E^+(\zeta)])$. In summary, there is a past endless timelike curve $\lambda$ and a future endless timelike curve $\gamma$, both contained in the set int $D(E^-[J^-(\gamma) \cap E^+(\zeta)])$ which is globally hyperbolic by Proposition 2.29, so that every causal curve from $\lambda$ to $\gamma$ intersects the compact set $E^-[J^-(\gamma) \cap E^+(\zeta)]$ (due to Lemma 2.7). The conditions of Corollary 2.8 are thus satisfied with $\mathcal{K} = E^-[J^-(\gamma) \cap E^+(\zeta)]$, and therefore there exists an endless maximal causal curve in int $D(E^-[J^-(\gamma) \cap E^+(\zeta)])$. But this is impossible, because by Proposition 2.10 this curve cannot have conjugate points, contradicting statement (i). ∎

**Theorem 5.6 (Hawking and Penrose).** If the chronology, generic and strong energy conditions hold, and there exists at least one of the following:
 (i) a compact achronal set $\Sigma$ without edge,
(ii) a closed trapped surface $S$,
(iii) a point $p$ such that the null geodesic families emanating from $p$ reconverge,
then the spacetime is causal geodesically incomplete.

*Proof.* The proof is very simple. By Corollary 2.5 either strong causality holds or $(V_4, g)$ is null geodesically incomplete. As remarked in the first part of Example 4.3, $E^+(\Sigma) = E^-(\Sigma) = \Sigma$, so that $\Sigma$ is both past and future trapped, in case $\Sigma$ exists. When $S$ exists, then Proposition 4.3 implies that $E^+(S) \cap S$ is a trapped set or the spacetime is null geodesically incomplete. Finally if $p$ exists Proposition 4.2 applies and either $\{p\}$ is a trapped set or $(V_4, g)$ is null geodesically incomplete. In summary, in all three possibilites there is a trapped set whenever the spacetime is null geodesically complete. But the generic condition and SEC imply, through Propositions 2.4 and 2.8, that every endless causal geodesic has a pair of conjugate points unless it is incomplete. The theorem then follows at once from the Lemma 5.1. ∎

A fourth alternative in Theorem 5.6 and some topological considerations were included in the case of non-simply connected spacetimes in [85,86]. Sometimes, the weakness of Theorem 5.6 is claimed to be the lack of information concerning the 'singularity', that is, whether the incompleteness is to the past or the future, for how many geodesics, the relation with the curvature, etcetera. Compare the situation with Theorems 5.1, 5.2 and 5.3, where the incompleteness is clearly placed either in the future or the past and on a significative number of geodesics. However, from the proof of Theorem 5.4 it is obvious that the incompleteness occurs in the future — if $S$ is future trapped — along null geodesics which are locally *maximal*. This follows because what really happens is that $E^+(S)$ is eventually non-compact due to the incompleteness of some of its null generating geodesic segments. This is in fact a general result and it can be adapted to all possibilities in Theorem 5.6 so that in all cases there is a *maximal* incomplete geodesic [205,118]. This is important because it allows one to show a limit on the severity of the singularity as measured by the curvature. The original limits on curvature growth for maximal geodesics are due to Tipler [214], later improved in [118,153,205]. The main result is that the first electric part (see subsection 2.1, just before Proposition 2.5) of the curvature with respect to the incomplete timelike geodesic, and similarly for null geodesics, cannot grow faster, in modulus, than $(\tau - \hat{\tau})^{-2}$ when approaching the singularity at $\hat{\tau}$.

It is certainly curious that one can put a limit on the curvature growth when approaching the end of an incomplete geodesic predicted by the theorems, but one does not know whether the curvature will diverge at all! Some results on this can be found in [43–46,48,212,220]. There are some expectations that the incomplete curves predicted by the theorems lead to singularities that, in general, if they are essential and not very specialized [43–45], will not be quasi-regular (see for instance Refs. 43–46,48,212,220).

Since the appearance of Lemma 5.1 and its corollary Theorem 5.6 a lot of work has been devoted to the weakening of its causality and energy assumptions [220]. Starting with the causality assumption (no closed timelike curves), there were soon indications that this condition can be relaxed substantially [213,216]. This work was later improved in [125], where a singularity theorem was proved if the null convergence and generic condition hold and the boundary of the set of points were casuality condition is violated is compact (remember that this is a proper achronal boundary; see subsection 2.3 before Definition 2.18). The latest on these series of results can be found in [137], which contains also all previous results. In this paper the final theorem is

**Theorem 5.7.** The following three statements cannot hold simultaneously in any spacetime:
 (i) every endless causal geodesic has a pair of conjugate points,
 (ii) the spacetime is not totally vicious and, if the chronology condition fails at the set $\mathcal{V} \subset V_4$, then either
   (a) there is a $p \in \partial \mathcal{V}$ such that for any arbitrarily small neighbourhood $\mathcal{U}_p$ of $p$ any closed timelike f-d curve passing through $\mathcal{V} \cap \mathcal{U}_p$ remains within a compact set $\mathcal{K}$, or

  (b) for any $p \in \mathcal{V}$ and all $q \notin I^+(p) \cap I^-(p)$, $[\partial J^+(q) \cup \partial J^-(q)] \cap [I^+(p) \cap I^-(p)] = \varnothing$;
(iii) there is a trapped set $\zeta$.

In other words, the causality assumption can be relaxed as far as there exists a region satisfying the causality condition and causally separated from the possible causality violating points. This, of course, would correspond to our own region in the Universe. The proof is standard [137]. ∎

Concerning the energy assumption, which is SEC plus generic condition in order to assure the existence of conjugate points, some results were found in [218], later corrected in [39], by using the averaged conditions referred to in subsection 2.1 just before Proposition 2.4. These results were later improved in [27,118]. Essentially, the result is that SEC in Theorem 5.6 can be replaced by WEC and/or integral conditions of type

$$\int R_{\mu\upsilon} v^\mu v^\upsilon d\tau \geq 0$$

(with equality holding only if $R_{\mu\upsilon} v^\mu v^\upsilon = 0$) along any geodesic with tangent vector $\vec{v}$ and affine parameter $\tau$. The relaxation of the strong energy condition in Theorem 5.6 is important from the physical point of view, as the typical inflationary models violate SEC (see Ref. 30 and references therein; see also Section 7). In this sense there have recently appeared new singularity theorems which are supposed to apply to the inflationary models [28,30,31]. In a first theorem SEC is dropped but some stronger causality assumption is assumed [28,30]:

**Theorem 5.8.** Suppose $(V_4, g)$ satisfies the null convergence condition, is (past) causally simple, and all edgeless achronal sets are non-compact. If there is a point $p \in V_4$ such that for every $q \in I^-(p)$ the volume of $J^-(p) - J^-(q)$ is finite, then the spacetime is null geodesically incomplete to the past.

*Proof.* The assumption on the edgeless achronal sets implies that the spacetime is spatially open (or infinite in space). The proof is simple considering the results of subsection 2.1. Specially, I refer to the comment after Definition 2.12, where the expansion $\vartheta$ of a null geodesic congruence with tangent vector $\vec{k}$ is equal to $k^\mu \partial_\mu(\log \mathcal{V})$, where $\mathcal{V}$ is the surface element in the surfaces orthogonal to $\vec{k}$. Thus, take any $q \in I^-(p)$, so that by point (i) in Proposition 2.15 there is a neighbourhood $\mathcal{U}_q \subset I^-(p)$ of $q$, and set up the null geodesic congruence orthogonal to a family of pieces of spacelike surfaces (including $q$ as a limit) contained in $\mathcal{U}_q \cap C_q^+$. The volume of the past of these surfaces is obviously proportional to the integral of $\mathcal{V}$ along the null geodesics, and as this must be finite by assumption because is contained in $J^-(p) - J^-(q)$, it follows that the expansion $\vartheta$ must be negative somewhere if the null geodesics are past complete. Then Propositions 2.7 and 2.14 imply that either the null generating segments of $E^-(q)$ are incomplete or they leave $E^-(q)$ entering into $I^-(q)$, so that $E^-(q)$ is compact. By causal simplicity $E^-(q) = \partial J^-(q)$, so that in the second case $E^-(q)$ would be a compact proper achronal boundary. This contradicts the other assumption of the theorem. Hence, spacetime must be null geodesically incomplete to the past. ∎

However, the above theorem applies only to open universes. A generalization was also presented in [31] with the added virtue of relaxing the causal simplicity assumption.

**Theorem 5.9.** Suppose the null convergence condition and strong causality hold. If there is a point $p \in V_4$ such that for every $q \in I^-(p)$ the volume of $J^-(p) - J^-(q)$ is finite, and there exists a past-endless timelike curve $\gamma$ with $\partial J^+(\gamma) \cap I^-(p) \neq \varnothing$, then the spacetime is null geodesically incomplete to the past.

*Proof.* The idea now is very simple. If the spacetime were null geodesically complete, by the proof of Theorem 5.8 the null geodesics generating $\partial J^-(q)$ must leave $E^-(q)$ and enter into $I^-(q)$. But as $\gamma$ is past-endless, the null generators of $\partial J^+(\gamma)$ are past endless. Take then $q \in \partial J^+(\gamma) \cap I^-(p)$, which leads immediately to a contradiction. ∎

Before ending this section, let us include a singularity theorem of limited application — it was devised for the problem of colliding plane waves — but which is important because it assures the existence of singularities in some rather simple and completely *vacuum* spacetimes. I will use this in the next Sections as an indication that the existence of singularities in General Relativity is really a *worrying problem*. The problem of singularities in colliding waves was first remarked with a simple example in [117] and shown to be somehow generic in [208]. There were some examples where the singularity might be substituted by a Cauchy horizon [77,100], but in the end the answer seems to be that singularities are generic in colliding plane wave spacetimes (see Ref. 100 for a complete review). The theorem was presented in [219] (see also Ref. 100).

**Theorem 5.10.** Suppose the spacetime contains two global spacelike commuting Killing vectors $\vec{\xi}_2$ and $\vec{\xi}_3$ acting on spacelike surfaces with $\mathbb{R}^2$ topology. Take a pseudo-orthonormal basis $\{\vec{k}, \vec{l}, \vec{\xi}_A\}$ as in subsection 2.1 and assume that at least one of the Newman–Penrose quantities $\sigma, \lambda, \Psi_0, \Psi_4, \Phi_{00}$ or $\Phi_{22}$ [100,123,150] is non-zero at some point $p \in V_4$. If $\vec{\xi}_A$ are tangent to a partial Cauchy hypersurface $\Sigma \ni p$ which is non-compact in the spacelike direction orthogonal to $\vec{\xi}_A$, and the null convergence condition holds, then $(V_4, \boldsymbol{g})$ is null geodesically incomplete.

The proof of this theorem is an adequate repetition of the proof of Theorem 5.4, because the conditions on the Newman–Penrose quantities assure the focal points along the null geodesics tangent to $\vec{k}$ or $\vec{l}$ (see Ref. 219 for the details). ∎

Finally, let us remark that the Einstein field equations have not been assumed in the above theorems, apart from Theorem 5.1. Thus, the theorems apply equally well to any theory where the spacetime is defined with a metric connection and satisfies the appropriate conditions (null convergence, generic, etcetera). However, I shall not enter into that here. Some possible references are [133,202]. Anyone interested in this may consult the excellent review [84].

# 6 ANALYSIS OF THE ASSUMPTIONS AND CONSEQUENCES OF THE SINGULARITY THEOREMS

In the previous section, various singularity theorems have been shown. As can be explicitly checked, all of them have the same structure, so that they can be summarized in the following *pattern* theorem:

**Theorem 6.1 (Pattern singularity "theorem").** If the spacetime satisfies
 (i) an energy condition,
 (ii) a causality condition, and
(iii) a boundary or initial condition
then it contains at least an incomplete causal geodesic.

Of course, the pattern theorem is meaningless by itself, because one has to specify the adequate energy, causality and boundary conditions for each case. Nevertheless, it is worth writing it down to keep the structure in mind and in order to look for new theorems or to improve the known ones. Furthermore, the pattern theorem allows one to perform a careful analysis of the true theorems as opposed to the widespread folklore concerning them.

All the assumptions and their reasonability will be analysed later on in this Section. Before that, let us make some intuitive comments on the proofs of the theorems. From a geometrical point of view, the proofs of the main theorems are based either on the construction of at least a *maximal* geodesic, therefore with no conjugate or focal points, so that this geodesic cannot be complete due to the energy condition (Propositions 2.3, 2.4, 2.7 and 2.8), or on the assumption of g-completeness leading to *compact* proper achronal boundaries which in turn is impossible if the spacetime is spatially open. From the physical point of view, the idea behind the proofs is different depending on the theorem, but in the main cases (Theorems 5.4 and 5.6) it goes like this: the basic assumption is the existence of a set *bound to be trapped*. This means that there is a certain region of the spacetime, be it a surface, or a point or a slice, which has its future or its past initially contained within a compact and decreasing spatial region. In other words, all possible matter contained within such a region cannot escape from a certain finite spatial decreasing zone at first. Of course, as long as gravity remains attractive and there is no way out back in time through violation of causality, the particles in such a region will collapse until one of two things happens: either all of them collapse to a region too small to contain such high quantity of particles, matter or radiation, whence the singularity, or eventually they reach untrapped regions and may try to escape their fatal fate. However, in this second possibility the whole future of the particles is contained within the future of the future light cone of the initial set, that is, within the future of a proper achronal boundary $B$, which is compact if no singularity is reached by itself. This can certainly happen in some situations, such as de Sitter or Reissner–Nordström spacetime [107] or the case in Fig. 12 of Example 4.3. The ideal case in order to avoid the singularity would be that all particles *necessarily* crossed the Cauchy horizon of the proper achronal boundary $B$, as then they will be freed from the catastrophic influence that gravity exerts on them, but this is impossible because this Cauchy horizon is non-compact or empty. Therefore, there are some particles that can travel indefinitely without ever leaving the Cauchy development of $B$, so that either they approach a singularity or go out to infinity. In this last possibility the argument becomes a little more complicated as one has to consider the event horizon of the curve which goes to infinity. Again this is a proper achronal boundary and its combination with the previous one produces, by taking an adequate subset, yet another proper achronal boundary which, by the energy conditions, is compact or reaches a singularity. The reasoning proceeds then as before but to the past, and in the case that there is another particle that could go out to infinity, its combination with the first one produces the possibility of travelling indefinitely from the past to the future, remaining within a finite spatial region and avoiding the focusing effect. But this is not possible if there is some matter or radiation around, the arrow of time is inevitable and gravity remains sufficiently attractive (SEC).

From the above reasoning or from the proof of the theorems themselves a drastic conclusion may be reached: if the adequate energy and causality conditions hold, then General Relativity favours the existence of incomplete curves rather than the realization of the trapping of a set bound to be trapped (unless in the trivial case of compact slices with $E^{\pm}(\Sigma) = \Sigma$). In other words, Lemma 5.1 assures that its point (i) and point (iii) cannot hold simultaneously and, loosely speaking, it seems as if the theory prefers to violate them both in a majority of physical situations containing a set bound to be trapped, because if point (i) held then point (iii) would also hold. Why this is so may seem logical in some intuitive situations depicted before, such as when a lot of matter reaches small enough regions of space, but in the general case it remains something of a mystery. Furthermore, for the sake of clarity I have repeatedly used, in the previous intuitive reasoning, the words particles or matter instead of the precise term 'causal curve', as there is no need for anything to move along the incomplete geodesics!. This poses a real problem, because the incomplete curves do appear in completely empty spacetimes. Thus, the intuitive idea that singularities may have something to do with the existence of matter, or due to a bad description of it, is false. The case of colliding plane waves and Theorem 5.10 allows no alternative. Spacetimes with simple pure gravitational waves and no matter whatsoever are destroyed by the existence of singularities. One may argue that there is some gravitational energy which can be localized in too small regions, and that the exact plane symmetry of the *infinite* waves is not realistic. It has been claimed, however, that the singularity is there even if one relaxes the assumption of plane symmetry and considers finite waves [246]. Hence, these examples show that the singularities are a really worrying problem in classical relativity [24].

The rest of this section is devoted to the analysis of the assumptions and conclusions of the theorems, their reasonability and the identification of each part of the above pattern "theorem" with the corresponding actual part of the theorems of Section 5.

### 6.1. Assumption of differentiability

As remarked in the paragraph after Definition 2.5, one of the usually unstated but fundamental assumptions of the singularity theorems is the $C^2$ differentiability of the metric, which may forbid some important cases of physical interest such as shock waves or the whole *matched* spacetime of a star. In these cases the metric will be $C^{2-}$ [132,143]. This problem was realized and briefly treated in a particular case in [107]. The claim is that the geodesic incompleteness predicted by the theorems cannot be avoided with a low differentiability. Nevertheless, trying to reproduce all the results required by the theorems leads usually to very big technical difficulties [45,46]. In order to realize the magnitude of the problem let us collect here a list of the *basic* places where the $C^2$ differentiability has been explicitly used:

**Normal neighbourhoods and coordinates.** The most basic place where the $C^2$ assumption is used is the definition of normal coordinates and normal neighbourhoods. These coordinates and neighbourhoods do exist under the $C^{2-}$ condition, but then the exponential map is a homeomorphism and the change to normal coordinates is only continuous. Of course, this has relevance later almost everywhere, in particular at places where the differentiability of quantities in normal coordinates or of the dependence of the geodesics on the initial positions is required.

**Gaussian coordinates.** Of course, the same happens with the existence of the coordinate systems (28) and (29). Thereby, this has influence also on the definition and interpretation of focal and conjugate points.

**Maximal curves.** The fundamental Corollary 2.2 and Propositions 2.10, 2.11 and 2.12 require a $C^2$ metric in order to perform the second derivative of the length. In principle, without this assumption there may be locally maximal curves *with* conjugate points.

**Existence of focal points.** Obviously, the Raychaudhuri equation (27) has a term proportional to the Ricci tensor, which may be discontinuous. In general this will not modify the results of Propositions 2.3, 2.4, 2.7 and 2.8, but it will change the improved results mentioned several times in which the geodesic focusing takes place assuming only an averaged SEC or similar, as they usually assume the continuity of the term involving the Ricci tensor [27,28,218].

**Curvature growth.** Similarly, the results on curvature growth and the limit of such along maximal geodesics require a $C^2$ metric [46,48,118, 153,205,212,214, 220].

**Trapped sets.** Propositions 4.1 and 4.2 rely upon the $C^2$ differentiability too, as otherwise it is not necessary that the corresponding $E^+$'s are compact because the null geodesic segments generating them may have focal points and still be maximal. This is used several times in the proof of the singularity theorems.

**Proofs of theorems.** Apart from all the above, there are some places where still the assumption is used in the proof of the theorems themselves, as for instance when using the continuous variation of conjugate points in the Hawking–Penrose Lemma 5.1.

The question arises of whether this is an important problem or not. It seems that there is a tacit agreement that most of the results will be recovered in the general case of $C^{2-}$ metrics; trying to prove them is quite another matter, though. Alternatively, one can try to set up a counter-example with a $C^{2-}$ metric. However, what is required is the analysis of the geodesics in this spacetime, so that it seems logical to assume some symmetry, or separability in certain coordinate systems, in order to integrate the geodesic equations. But this will lead to the absence of the problem, because if the symmetry is too high (say a 3-dimensional group), or the separability worthwhile that the geodesic equations can be partially integrated, then there will be constants — either because of the symmetries or of the separation constants — which will reduce the problem to an equivalent one with a system of ODE of type

$$\frac{dx^\mu}{d\tau} = F^\mu(p, g, A^i)$$

where the $A^i$ are the constants and the $F^\mu$ are well-behaved functions which do not depend on the derivatives of the metric, but only on the metric itself. Then the differentiability of the metric will be more than enough to assure the differentiability of the exponential map, for example. Thus, in general any try to construct a counter-example of the fundamental results listed before is highly non-trivial and requires a full analysis of complicated systems of second order differential equations.

Nevertheless, there are some simple cases in which the theorems can be proved allowing for $C^{2-}$ metrics. One explicit case will be presented in subsection 7.1 of the next section. This applies to spherically symmetric solutions which are static in some regions. There, allowing for an arbitrary finite number of matching hypersurfaces, so that the metric may fail to be $C^2$ at all these hyersurfaces, does not help in constructing a non-singular metric if SEC holds [141]. Whether or not this is an indication of the validity of the theorem in generic $C^{2-}$ spacetimes is difficult to decide, but it seems a very particular case to draw any definite conclusion from it.

### 6.2. Energy and generic conditions

The energy condition used in Theorems 5.1, 5.2, 5.5, 5.6 and 5.7 is SEC, in the last two cases together with the generic condition just to assure point (i) in Lemma 5.1 or Theorem 5.7, while it is the null convergence condition in Theorems 5.4, 5.8, 5.9 and 5.10 (in this last case together with the non-vanishing of some quantity, which is equivalent to a sort of 'generic' condition). The remaining Theorem 5.3 assumes a kind of SEC condition on average. Also the Penrose theorem can be proved with a null convergence on average [175]. As argued after the proof of Theorem 5.7, the focusing of geodesics will hold by assuming simply WEC, which implies the null convergence condition, and/or an averaged SEC (see Refs. 27,28 and references therein). Hence, in general it seems that either SEC on average or the null convergence condition or both are the necessary energy conditions. All in all, what is needed is the appearance of conjugate or focal points along the appropriate geodesic families. Any condition assuring this will do. The physical implication behind the energy conditions is simply this focusing, and can be stated loosely as 'gravity will remain attractive enough and thus it will favour geodesic focusing'. Obviously, the energy condition by itself does not guarantee any singularity.

A good review on energy conditions and where they are used can be found in [233] and references therein. The possible violation of the energy condition has been claimed several times on different grounds [14,27,32,84, 107,183,218,220,233]. As remarked above, usually the theorems will require SEC (on average) and/ or the null convergence condition. However, as argued after Proposition 2.6, SEC is the less physical energy condition of all: the energy density may be non-negative and causally propagated for every possible reference system and still SEC may be violated. Subsection 7.3 of the next section provides an interesting model where DEC holds, the spacetime satisfies all assumptions of Theorems 5.5 and 5.6, but still is non-singular because SEC is violated. In fact, important physical examples violate SEC: the most typical of all is the energy-momentum tensor of a massive scalar field [107,218,233]. Of course, the scalar field being a reasonable physical field satisfies *all* energy conditions but SEC. Pions or Higgs scalar fields have this type of energy-momentum tensor.

On the other hand, the violation of WEC (and hence of DEC) is impossible for physically realistic classical matter or radiation. However, there are some indications that these conditions may fail to hold in several semi-classical (or quantum) phenomena. A list of them is [233]: Casimir effect, which violates *all* energy conditions, Hawking radiation, wormholes, and cosmological inflation. This last case is of interest also from the classical point of view, and it has been recently claimed [32] that WEC may be violated in some inflationary scenarios, opening the door to non-singular models which are spatially homogeneous and isotropic at some scales.

A final few words concerning the generic condition (see also Ref. 13). Although this is not an energy condition strictly speaking, it is used for the same purposes than the energy condition: to assure the focusing effect on geodesics through Propositions 2.4 and 2.8. Furthermore, the generic condition is implied by the strict SEC, as was shown in the Propositions 2.5 and 2.6. The generic condition will be satisfied in every 'generic' spacetime, understanding by such any spacetime with not too much symmetry or other special properties such as a specialized Petrov type. Thus, the generic condition is used to assure the genericity of the spacetime, but it has the undesired minor side-effect that special cases of some relevance are left uncovered by the theorems. It may be thought that this is of no importance, as perhaps other simpler theorems will do. Unfortunately this is not the case. It is absolutely necessary to include the generic condition among the assumptions of the theorems, as is proved by many easily constructable counter-examples. One is the Reissner–Nordström spacetime, not covered by the Hawking–Penrose and other traditional theorems even though it has singularities [28], but much simpler spacetimes can be built. Some illustrative examples are presented in subsection 7.1.

### 6.3. Causality assumptions

The causality assumption assumed in Theorems 5.2, 5.3 and 5.4 is global hyperbolicity; strong causality in Theorems 5.5 (in this case this assumption is not necessary; Ref. 104), 5.6 and 5.9; causal simplicity in Theorem 5.8; a rather complicated assumption which can be replaced by the fulfilment of the strong causality condition in some open non-empty set, in Theorem 5.7; and finally, there is no explicit causality assumption in the Raychaudhuri–Komar Theorem 5.1 and in the last Theorem 5.10. However, in both cases as well as in the refined version of Theorem 5.5 proved in [104,107], the assumption is hidden behind other more physical conditions. In these cases there are spacelike achronal partial Cauchy hypersurfaces. Therefore, the theorems are proved in the Cauchy developments of these hypersurfaces within which global hyperbolicity holds. At first sight, it may seem that there is a wider variety of possibilities in the causality condition than in the energy assumption, but looking carefully through the proofs of the theorems, one always moves in (perhaps hypothetical) globally hyperbolic subsets of the spacetime. For instance, in the Hawking–Penrose Lemma 5.1 the whole reasoning is performed in int $D(E^-[J^-(\gamma) \cap E^+(\zeta)])$. It seems that the essential thing is the existence of non-empty $E^{\pm}$'s, which in turn needs the absence of closed timelike curves. The physical implication of the causality condition is of two types (not necessarily both of them have to be present in a particular theorem): the impossibility of particles travelling into the future and being able to influence their own past (roughly 'no-one can avoid the arrow of time'), and the existence of maximal geodesics within some subsets of the spacetime due to global hyperbolicity. Again, the causality condition, whether by itself or in combination with the energy condition, proves nothing. It is interesting to remark that, as seen in Proposition 2.22 and Corollary 2.5 of subsection 2.3 together with Propositions 2.5 and 2.6 of subsection 2.1, the energy and causality conditions are not completely independent in general.

The reasonability of the causality assumption is difficult to question in their milder versions (strong causality), but perhaps not so in their stronger ones (global hyperbolicity). Nevertheless, the main theorems, including Theorem 5.6, assume just the strong causality condition. Furthermore, the improved version of Theorem 5.7 assures that the result holds even when violation of causality is present as long as the spacetime is not totally vicious. The Gödel spacetime [95] shows that this condition is difficult to relax, for it has reconverging light cones and marginally trapped surfaces, satisfies the null convergence condition, and is geodesically complete [28]. Gödel's spacetime is totally vicious. In consequence, it seems that the causality violation cannot help in avoiding the singularity as long as it is not *universal*. The causality condition is therefore the least restrictive and most well-founded assumption of the singularity theorems.

**6.4. Boundary and initial assumptions**

The assumed boundary conditions are: the positivity of the expansion of the geodesics orthogonal to a hypersurface in Theorems 5.1, 5.2, and 5.5; the vanishing of the second fundamental form of the hypersurface in Theorem 5.3; the existence of a closed trapped surface and the non-compactness of the Cauchy hypersurface in Theorem 5.4; the existence of a trapped set in its various possible forms in Theorems 5.6, 5.7, 5.8 and 5.9; and the non-compactness of the partial Cauchy hypersurface together with the existence of the symmetry in Theorem 5.10. In summary, apart from the very special case of Theorem 5.1, the boundary condition assumes, or leads to, the existence of an appropriate compact set, usually through a trapped set, but not necessarily so, *provided* that the spacetime is g-complete. Of course, this is the *key* assumption in the theorems, and nothing could be proved without it.

It seems to me that one of the most surprising things concerning singularity theorems is the wide acceptance of the boundary condition. For example, in Ref. 220, p. 148 we can read: "The initial conditions — such as the existence of a trapped surface [...] — are rarely questioned.". In principle, however, it should have been the truly questioned assumption, for various reasons. Firstly, because it is *the essential* assumption; secondly, because it has observational consequences; thirdly, for sometimes one gets the impression that it is difficult to distinguish between the boundary *assumption* and the derived *conclusion*. And last but not least, because its reasonability and its realization in the actual spacetimes is far from assured. Thus, the boundary condition must be carefully analysed and tested.

Of course, the reasonability and realization of the boundary condition has been claimed several times in the literature [106–108]. For example, assuming the so-called Copernican principle (implying the spatial homogeneity of the Universe, or almost so) and using the existence and isotropy around us of the cosmic microwave background radiation (CMBR), a derivation of the FLRW line-element for the spacetime is achieved in [107] by using the results of [61]. Nevertheless, this derivation relies on two important things: the stated assumption of the spatial homogeneity of the Universe (is the Universe really spatially homogeneous?), and the assumption that matter must move along geodesics if solely under its own influence (Ref. 107, p. 350). This second point is not supported by any theoretical or physical observation whatsoever, and its modification leads to other possible models [80]. Concerning the first point, I leave it to the reader — but let me recommend Appendixes B and C in [124] and the contribution in [135]. Furthermore, once the FLRW spacetime is obtained, the existence of the trapped 2-spheres in these spacetimes is used (see Example 4.2). But then again, this assumes that the FLRW model holds for distances big enough so that the trapped 2-spheres do exist, and this is not obvious as the order of the radius of the trapped 2-sphere is the same as that of our horizon. This will be considered in some detail in subsections 7.2 and 7.7.

Similarly, it has been claimed that the spectrum and isotropy of the CMBR implies the existence of so much matter that our past-directed null geodesics must reconverge somewhere to the past [107,108]. Whether this reasoning is model-dependent is unclear to me, as it uses the Hubble 'constant', which in principle could be a function of spatial position, and also red-shifts to measure distances. Moreover, it is also implicitly assumed that *all* matter is, and has been for some large time, expanding around us, as no significant blue-shifts are allowed. Finally, it is also assumed that the isotropy of the radiation is due to the thermalization by repeated scattering. All this may be true, but there still is room for a reasonable doubt if for example the density of matter fell off rapidly enough when going away from us — it could then grow again, in some situations. This has observational implications and can be tested. Furthermore, the fact that the Hubble function can depend on the direction of observation has not been ruled out so far. As a final important point, it must be stressed that the existence of *big* inhomogeneities in the Universe does not necessarily lead to a *big* trace left over the spectrum of the CMBR, as has been demonstrated several times [6,157,169,181] (see the discussion in Ref. 124). Further analysis on these matters will be given in subsections 7.6 and 7.7. In any case, even if the whole reasoning were adequate, Theorem 5.6 would not say when and how the singularity is. For example, it could be timelike and to our future! Or it could simply lead to a removable singularity — there is no way to decide whether or not our past horismos $E^-$ is contained within an inextendible spacetime. These possibilities are forgotten sometimes.

With regard to the most specific Theorem 5.10, here the boundary conditions are appropriate for a very particular physical situation, and of course they are not reasonable if the waves are not exactly plane and of infinite extent. The same happens with Theorems 5.8 and 5.9, which were devised for particular inflationary situations, so that the assumptions are inapplicable to generic cases and there are many examples in which the difference of volumes of the past are not finite. Concerning Theorem 5.1 the main assumption is the absence of rotation and acceleration, which happens sometimes but fails in general. Of course, the conclusion here is simple: perfect fluids and similar matter needs either acceleration or rotation to avoid matter singularities. The possibility that acceleration (without rotation) may prevent the singularities has been dismissed too often, even though it works perfectly well; see subsections 7.6 and 7.7. With respect to Theorem 5.2, the assumption may seem reasonable for globally hyperbolic expanding universes, but there are two subtleties which invalidate this view, namely, that there is no way to know whether the *whole* Universe is expanding at a given instant and the assumption that the expansion is bigger than some *positive* constant. So, there certainly are globally hyperbolic spacetimes expanding *everywhere* at some instant of time and fulfilling the energy conditions which are singularity-free (subsection 7.6) or with non-big-bang singularities (subsection 7.5). The same criticism holds for Theorems 5.3 and 5.5, but in this last case the compact slice makes it more reasonable to assume the strictly positive expansion if the spacetime is spatially closed (the weakness of this theorem really lies on the assumption of SEC; see subsection 7.3). Finally, Penrose's Theorem 5.4 assumes a non-compact Cauchy hypersurface, which makes the theorem inapplicable to non-globally hyperbolic spacetimes, so that extensions of globally hyperbolic regions are needed again, and also to spacetimes with compact slices such as de Sitter, Reissner–Nordström, or the one shown in Fig. 12 of Example 4.3.

On the other hand, one can raise doubts as to whether the other assumptions in the theorems themselves allow for the actual appearance of the required boundary condition. For instance, in very simple cases the assumption of SEC (plus regular curvature) forbids the existence of closed trapped surfaces, as illustrated in subsection 7.1 of the next section. It is also unclear whether trapped sets are formed from generic initial data, specially because these data determine only the corresponding Cauchy development, and beyond this one has to resort to choosing extensions once again.

### 6.5. The conclusions of the singularity theorems

The conclusion of the theorems states the existence of *at least* one incomplete causal geodesic. In Theorem 5.1 the conclusion is much stronger, and states the existence of a matter singularity along every curve of a timelike congruence, if these curves do not find another singularity first. This is an exceptional case, but perhaps it has led to the wrong idea that all the theorems implied a singularity of this sort. The incompleteness in Theorems 5.2, 5.3 and 5.5 are also for a whole congruence, but in this case there is no indication of whether the singularity will be essential or not, and when essential if it may be quasi-regular, or what is the character of the singularity, so that in Cosmology it could well be timelike. Even worse, in the rest of the theorems the conclusion is milder and only the existence of incomplete causal geodesics can be stated for sure. Naturally, this still leaves two doors open: *extensions and singularities*.

In the case that the g-incompleteness signals a removable singularity and a regular extension is possible, the problem arises of which extension among the huge amount of possibilities. Furthermore, there is no reason to assume that the same conditions, or other equivalent ones, which held in the unextended spacetime must also hold throughout the extension. Thus, the energy, causality or boundary conditions which led to the existence of incomplete geodesics *do not have* to be satisfied necessarily in the extended spacetime. As discussed in Example 3.5 the extended region of an extendible spacetime may not be determined from the physics of the unextended region, so that the existence of a singularity cannot be asserted *unless* further hypotheses restricting the extended region are assumed. It is interesting in this respect to consider the case of Schwarzschild spacetime (Example 3.3). This spacetime has incomplete geodesics approaching $r = 2M$ and there are several alternatives: a singular extension, a regular extension with closed trapped surfaces and singularities (both shown in Example 3.3), and a regular extension with closed trapped surfaces but *no* singularities (Example 4.3). In this last possibility SEC is violated in the extended region, but this only happens at small regions and always within the zone $r < 2M$, from which we have no physical information whatsoever and no way to get it![9] This is surely the generic picture in gravitational collapse, and in this case the problem becomes a problem of choosing between singular, regular but singularly extendible or completely regular extensions beyond the horizon. Of course, the type of singularity (if any) is then completely unknown, unless a particular extension is chosen.

---

[9] It is fundamental to note here that there is no way to observe a black hole region, apart from entering into it if this is possible. If any star ever collapses to form a black hole, the only information or physical influence which an external observer will receive is from the star at instants previous to the crossing of the horizon.

In the second possibility, when g-incompleteness indicates an essential singularity, the problem of its character, severity and location remains. In Cosmology, for instance, one tends to think that the singularity will be in the past and of big-bang type (see Definition 3.6). Well, nothing of the sort. There are definite indications that this is not so in general spatially homogeneous models [50,67,69], but also the set of inhomogeneous spacetimes that have been obtained in the last decades show that there is no reason whatsoever to believe that the cosmological singularity has any definite property [53,54,56,79,124,139,178,194–6,198,204]. Further, there is the possibility of *singularity-free* spacetimes satisfying all possible energy and causality conditions in their stronger versions and being considerable as cosmological models because they have expanding slices and cannot be the interior of a star [40,53,54,56,101,136,138,158,178,193–6,247]. In the next Section, several explicit examples of well-behaved singularity-free models will be given. Thus, regarding cosmological questions the conclusion of the theorems is weak in two respects: it allows for localized singularities — just a timelike singularity has nothing to do with the idea of a beginning of the Universe — and there are good models not covered by the theorems and therefore non-singular and inextendible, so that there appear some doubts about the genericity of big-bang singularities in Cosmology. This last point is related to the old discussion [242] of whether or not the set of 'physical' spacetimes containing singularities is of zero measure or not. In my opinion, the set of explicitly known spacetimes is of zero measure so far, and there is no simple way to answer the previous question. I will come back to this in subsection 7.7.

## 6.6. Summary of main weaknesses of the singularity theorems

All in all, it seems to me that the main problems of the singularity theorems are:

- The question of the $C^{2-}$ differentiability of the metric to allow for shock waves and matching of spacetimes, such as the interior and exterior of collapsing stars.
- The assumption of the unphysical SEC instead of the physically well-motivated DEC (or WEC). In fact, SEC can certainly be violated by real physical systems.
- The very weak conclusion (at least an incomplete causal geodesic), which still permits the possibility of having regular, or singular but quasi-regular, extensions.
- The absence of physical motivation in order to favour some extension

in front of others.
- The lack of information about the properties of the singularities, concerning type, location, and character.
- The reasonableness and reality of the boundary or initial conditions.
- Last but perhaps greatest, the absolute necessity of the boundary or initial assumption in their stronger versions, which may seem to substitute effectively for the singularity, so that the theorems may look tautological. Of course, they are not so at all, but the effect of the boundary condition on the theorems resembles very much the action of a *time bomb*.

A few last words may be in order. In this section, the emphasis has been put on the criticism of the singularity theorems. The only reason for this is to instigate further developments which will improve or eliminate some of the above difficulties. Of course, the singularity theorems can also be considered from a positive viewpoint, remarking in particular their virtues, beauty, generality, applicability as well as the influence that the needed techniques have had in other fields. No doubt that would be a much easier task than the one presented herein.

## 7. EXAMPLES OF NON-SINGULAR SPACETIMES AND OTHER INTERESTING METRICS

In this section several interesting and illustrative examples of singularity-free spacetimes and related are presented. These examples show some of the doors left open by the theorems, and reinforce some of the claims of the previous section.

### 7.1. Regular static spherically symmetric spacetimes

The purpose of this example is three-fold: firstly, it serves to show that the generic condition is certainly needed in the singularity theorems. Secondly, it provides a simple example where the very same conditions of the singularity theorems (SEC), together with the absence of curvature singularities, forbid the existence of closed trapped surfaces. And thirdly, it also provides us with a theorem where the differentiability of the metric is only $C^{2-}$. The whole thing is based on [141].

The line-element is given by the most general *static* and spherically symmetric metric in coordinates $\{x^\mu\} = \{T, \mathcal{R}, \vartheta, \varphi\}$

$$ds^2 = -F^2(\mathcal{R})dT^2 + d\mathcal{R}^2 + r^2(\mathcal{R})(d\vartheta^2 + \sin^2\vartheta d\varphi^2), \quad (41)$$

where $r(\mathcal{R})$ and $F(\mathcal{R})$ are arbitrary functions. The main assumption is that the spacetime is regular at a centre of symmetry (without loss of generality this centre can be chosen at $\mathcal{R} = 0$ so that $r(0) = 0$). Of course, this does not avoid all possible singularities. The local chart is then well-defined in a neighbourhood of $\mathcal{R} = 0$. The assumption of regularity is equivalent to the following asymptotic behaviours for $F$ and $r$

$$r(\mathcal{R} \to 0) \approx \mathcal{R} - \frac{m_0}{3}\mathcal{R}^3, \qquad F(\mathcal{R} \to 0) \approx F_0 + F_1 \mathcal{R}^2, \quad F_0 > 0,$$

where $m_0$, $F_0$ and $F_1$ are constants and $F_0$ is strictly positive. The energy-momentum tensor of this spacetime can be computed without difficulty and takes the form $(T_{\mu\nu}) = \mathrm{diag}(\varrho, p_r, p_t, p_t)$ in the natural orthonormal cobasis $\theta^\mu \propto dx^\mu$, where

$$\varrho = \frac{1}{r^2}(1 - r'^2 - 2rr''),$$

$$p_r = \frac{1}{r^2}\left(-1 + r'^2 + 2rr'\frac{F'}{F}\right),$$

$$p_t = \frac{F''}{F} + \frac{F'}{F}\frac{r'}{r} + \frac{r''}{r},$$

and the prime denotes derivative with respect to $\mathcal{R}$.

The existence of closed trapped surfaces in this spacetime cannot happen in the local chart covered by the chosen coordinates, because those closed trapped surfaces require a timelike $dr = r'd\mathcal{R}$, which is manifestly impossible in these coordinates. The coordinates fail to describe the manifold if and only if the function $F(\mathcal{R})$ vanishes somewhere, which is also equivalent to the existence of incomplete geodesics approaching the zone where $F(\mathcal{R}) = 0$, because the vanishing of $r$ at values $\mathcal{R} > 0$ is either *another* regular centre of symmetry or a curvature singularity of the spacetime. Then, in [141] the following theorem is proved:

"Suppose the line-element (41) is $C^{2-}$, failing to be $C^2$ in a finite but arbitrary number of hypersurfaces. If there is a regular centre of symmetry, and the condition $\varrho + p_r + 2p_t \geq 0$ holds, then $F > 0$ everywhere, so that there are no closed trapped surfaces in this spacetime."

It must be stressed that SEC in our case is equivalent to

$$\varrho + p_r \geq 0, \qquad \varrho + p_t \geq 0, \qquad \varrho + p_r + 2p_t \geq 0,$$

so that the assumed condition $\varrho + p_r + 2p_t \geq 0$ is implied by, but is less restrictive than, SEC. Thus, the above theorem can be re-phrased by saying that the formation of closed trapped surfaces is forbidden by SEC for the case of regular, static and spherically symmetric spacetimes. This was one of the claimed purposes. However, there still is another way of looking at the theorem, namely, the violation of the $C^2$ condition for the metric does not help in constructing a singularity-free black hole (spacetimes with closed trapped surfaces and no singularity) if SEC holds. Consequently, in this very particular case the $C^{2-}$ differentiability seems unimportant. This was another goal.

It may be interesting to remark that the possible hypersurfaces with $r' = 0$ (or equivalently with a vanishing one-form $dr$) are completely ordinary *timelike* hypersurfaces as long as $F$ does not vanish there. Therefore, the possible hypersurfaces $r' = 0$ which appear in our treatment do not have anything to do with horizons or other null hypersurfaces, and the Killing vector $\partial/\partial t$ is timelike everywhere. Nevertheless, $r' < 0$ is the condition so that the null geodesic families emanating from any point at $\mathcal{R} = 0$ start to reconverge, and thus point (iii) in the Hawking–Penrose Theorem 5.6 holds when $r' < 0$ for some $\mathcal{R}$. These spacetimes are globally hyperbolic so that there is no problem with the causality conditions. If SEC is also assumed, so that $F > 0$ everywhere as just seen, then inevitably $r$ must vanish again at some other value $\bar{\mathcal{R}} > 0$ of $\mathcal{R}$. This will be a curvature singularity in most cases, except when $F = F_0 = $ const. In this case $\varrho + p_r + 2p_t = 0$. Thus, when SEC holds with constant $F$ and $r' < 0$ at some value of $\mathcal{R}$, then the spacetime is g-complete, globally hyperbolic and with reconverging horismos — in fact, it has compact achronal sets without edge too. The only possibility to avoid the Hawking–Penrose Theorem 5.6 is therefore the violation of the generic condition.[10] And this is, of course, exactly what happens because when $\varrho + p_r + 2p_t = 0$ the generic condition is violated for the timelike curves with constant $\mathcal{R}$, $\vartheta$ and $\varphi$, which are geodesics *now*. Therefore, the generic condition cannot be eliminated from the singularity theorems.

In order to illustrate these points, let us consider briefly the interior Schwarzschild solution, given by [123]

$$ds^2 = -[a - b\cos(\mathcal{R}/c)]^2 dT^2 + d\mathcal{R}^2 + c^2 \sin^2(\mathcal{R}/c)(d\vartheta^2 + \sin^2\vartheta d\varphi^2),$$

where $a, b$ and $c$ are constants. This is a solution of Einstein's equations for a perfect fluid ($p_r = p_t$) with density and pressure given by

$$\varrho = \frac{3}{c^2}, \qquad \varrho + 3p = \frac{6b\cos(\mathcal{R}/c)}{c^2[a - b\cos(\mathcal{R}/c)]}.$$

---

[10] This became apparent after writing [141] in a discussion with Marc Mars, to whom I am grateful.

As is well-known, the case with $b = 0$ is the Einstein static universe, while the case $a = 0$ gives the static part of de Sitter model. In all cases, we have that $r = c \sin(\mathcal{R}/c)$, and therefore $r' = \cos(\mathcal{R}/c)$. It follows that $r' = 0$ at $\mathcal{R} = c\pi/2$ and $r' < 0$ for $\mathcal{R} > c\pi/2$. The hypersurface $\mathcal{R} = c\pi/2$ is timelike and without any special relevance in general, apart from the fact of being the 'equator' from where null geodesics start reconverging and SEC may start to fail. The exceptions are the Einstein static universe $b = 0$, with $\varrho + 3p = 0$ everywhere, and the de Sitter model with $a = 0$, considered below. The Einstein static universe $b = 0$ is a simple example of the possibility shown above, as the spacetime is globally hyperbolic, satisfies SEC, it contains points with reconverging light cones and compact achronal edgeless hypersurfaces, and nevertheless it is g-complete. The generic condition is violated here for the timelike geodesics with constant values of $\mathcal{R}$, $\vartheta$ and $\varphi$.

The other special case $a = 0$ is the static part of de Sitter spacetime and has $\varrho + 3p < 0$, so that $F(\mathcal{R}) = |b| \cos(\mathcal{R}/c)$ is a decreasing function of $\mathcal{R}$ until it vanishes exactly at $\mathcal{R} = c\pi/2$. The coordinates are not valid there and the causal geodesics reach $\mathcal{R} = c\pi/2$ in finite affine parameter, so that the spacetime is g-incomplete. In this case there is a regular extension providing the full de Sitter metric in which $\mathcal{R} = c\pi/2$ is a true Killing horizon which separates the regions with and without trapped 2-spheres in the maximally extended de Sitter model. The extension is given by imbedding the above spacetime into $\mathbb{R} \times S^3$ with coordinates $\{t, \chi, \vartheta, \varphi\}$ by means of ($b = 1$)

$$T = c \log \left[ \frac{\sinh(t/c) + \cosh(t/c) \cos \chi}{\sqrt{1 - \cosh^2(t/c) \sin^2 \chi}} \right], \qquad \sin(\mathcal{R}/c) = \cosh(t/c) \sin \chi,$$

so that the metric takes the FLRW form (Example 3.1)

$$ds^2 = -dt^2 + c^2 \cosh^2(t/c) [d\chi^2 + \sin^2 \chi (d\vartheta^2 + \sin^2 \vartheta d\varphi^2)].$$

This spacetime is geodesically complete. It satisfies the null convergence condition (obviously, because $R_{\mu\nu} \propto g_{\mu\nu}$) and is globally hyperbolic, as the hypersurfaces $t = $ const. are Cauchy hypersurfaces. Further, there are closed trapped 2-spheres in the added regions $\cosh^2(t/c) \sin^2 \chi > 1$ which were not described by the original manifold. Thus, all the assumptions of Theorem 5.4 are fulfilled except for the non-compactness of the Cauchy hypersurface, which reveals itself as essential for the validity of Penrose's theorem.

### 7.2. FLRW plus Vaidya spacetimes

Here we present some simple spacetimes built by matching a Vaidya metric (Example 3.5) to the flat ($k = 0$) FLRW models (Example 3.1). These matched spacetimes will illustrate two important facts, namely, (i) it is possible to have FLRW expanding realistic universes in a big region but forming a subpart of a bigger non-FLRW spacetime such that the whole spacetime does not have closed trapped surfaces; and (ii) it is feasible to have spherically symmetric realistic stars which collapse and *evaporate* completely without ever forming a singularity or a black-hole region.

The matching of FLRW with Vaidya in the general case was succesfully achieved and interpreted in [73,74] (see also references therein). Some applications were later developed in [76], upon which this example is based. Here the flat FLRW is assumed with a barotropic equation of state $p = \gamma \varrho$, so that the scale factor takes the form shown in (37) of Example 3.6. We assume $\gamma \geq 0$, so that all energy conditions are satisfied. In this example, the coordinates used are those of Examples 3.1 and 3.5, but we have tilded those of Vaidya's when necessary to avoid confusion. Also, we have chosen $\varepsilon = 1$. The matching is performed across a timelike hypersurface $\sigma$ preserving the spherical symmetry and imbedded into each spacetime by means of [see (1)] $\vartheta = \Phi^2(u) = u^2$, $\varphi = \Phi^3(u) = u^3$, $\tilde{\vartheta} = \tilde{\Phi}^2(u) = u^2$, $\tilde{\varphi} = \tilde{\Phi}^3(u) = u^3$ together with

$$t = \Phi^0(u) = u^1, \qquad \chi = \Phi^1(u) = \frac{3\gamma}{C}\left(\frac{1+\gamma}{1+3\gamma}\right)(u^1)^{1+3\gamma/3(1+\gamma)} - \chi_0,$$

$$\tilde{t} = \tilde{\Phi}^0(u) = f(u^1), \qquad r = \tilde{\Phi}^1(u) = 3\gamma\left(\frac{1+\gamma}{1+3\gamma}\right)u^1 - \chi_0 C(u^1)^{2/3(1+\gamma)},$$

where $\{u^i\}$ are intrinsic coordinates of $\sigma$, $\chi_0$ is an arbitrary constant and $f$ is a function of the time coordinate $u^1$ defined by

$$\frac{df}{du^1} = \frac{3(\gamma+1)^3}{3(1+\gamma)^2 + 2C\chi_0(1+3\gamma)(u^1)^{-(1+3\gamma)/3(1+\gamma)}} > 0.$$

The mass function $M(\tilde{t})$ of the Vaidya spacetime is completely determined (in parametric form at least) from the previous formula and its expression on $\sigma$ given by

$$M[f(u^1)] = \frac{2}{9(1+\gamma)^2(u^1)^2}\left[3\gamma\left(\frac{1+\gamma}{1+3\gamma}\right)u^1 - \chi_0 C(u^1)^{2/3(1+\gamma)}\right]^3.$$

The case $\gamma = 0$ is the Einstein–Straus model [64] of a Schwarzschild cavity in a flat FLRW background universe, or alternatively a classical collapse similar to that of Oppenheimer–Snyder [155,241] with a Schwarzschild exterior ($M$ = const.) and a dust FLRW interior.[11] In these classical cases, the matching hypersurface has $\chi = -\chi_0$, so that $\chi_0$ must be negative and the mass is positive. Direct generalizations of the Einstein–Straus cavity (or its complementary matching describing the classical collapse) with a Vaidya interior (exterior) are thus given by the above formulae with $\chi_0 < 0$. The outcome in the collapse case is a black hole with its typical singularity [76].

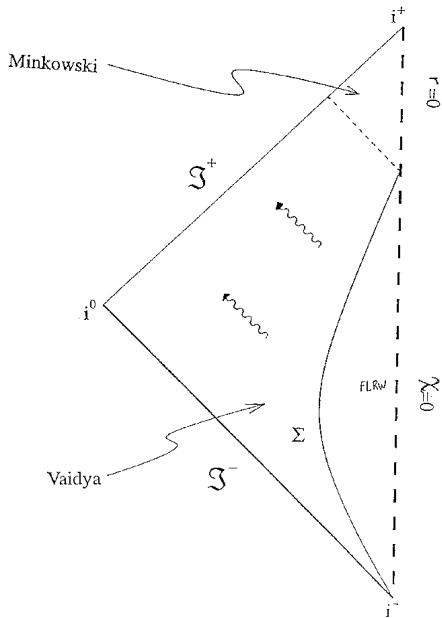

Figure 13. This is a Penrose diagram similar to that in Fig. 2(a), but now the exterior asymptotically flat region is a portion of Vaidya radiating spacetime. Thus, the interior body collapses indefinitely but at the same time radiates all of its mass away and disappears completely. The resulting final part of the spacetime is simply a portion of flat Minkowsi spacetime.

However, the possibility of $\chi_0 > 0$ is perfectly valid in the case $\gamma > 0$, which in particular forbids a vacuum exterior. In this case, it is easily checked that the mass function $M$ vanishes when $u^1$ reaches the value $(u^1)^{1+3\gamma/3(1+\gamma)} = C\chi_0(1 + 3\gamma)/3\gamma(1 + \gamma)$, where also $r$ and $\chi$ vanish.

---

[11] By the typical duality, we can interchange the 'interior' and 'exterior' metrics keeping the same matching conditions by simply joining the two complementary pieces left unused in one of the matchings. These two different junctions with the same matching equations are called complementary (see, e.g., Ref. 76).

Thus, the matching hypersurface starts at the origin of coordinates and with vanishing mass function in Vaidya, which implies the 'shrinking' of the FLRW model to a single point. So, the whole model is pure flat Minkowski spacetime before that value of $u^1$. The Penrose diagram of the spacetime is given by Figure 13 but 'upside down' [76], so that future and past are interchanged. The spacetime is initially flat, but some radiation comes from infinity and, by some unknown but perhaps possible processes, produces an expanding FLRW central region which gets bigger and bigger as the radiation keeps flowing in. The time-reversal of the model is presented in the actual Fig. 13, where it is shown that the spacetime represents a collapsing star with FLRW perfect-fluid interior, emitting radiation continuously until its mass is wholly radiated away and the spacetime becomes flat. Hence, the star 'evaporates' completely. In other words, there exists the possibility of a complete collapse to a single point without the appearance of a singularity as long as the star losses its mass at the appropriate rate. It is remarkable that this collapse preserves the spherical symmetry all along and is for a homogeneous and isotropic perfect fluid. Furthermore, all energy conditions are amply satisfied.

An important remark, however, is that before giving any credit to these models one should try to identify the physical fields involved in the interior and the exterior and to put the adequate junction conditions for them. Otherwise, the possibility of having transmutation of matter, ambiguous evolution and other similar absurdities (such as those criticized in Ref. 78) may appear. The pure matching of two spacetimes is too crude a model in general, and the specification of the proper boundary conditions for the physical fields involved is necessary and may forbid some situations. This is specially so when dealing with unspecified pure radiation, also called 'null dust', such as the case of the Vaidya metric (see Ref. 78). Of course, as argued in [78]: 'one may [...] disregard the matter field equations [and their boundary conditions] and consider only the phenomenological energy-momentum tensor. The results obtained by these means should not, in principle, be considered as physically tenable unless confirmed by the complete treatment of the problem'. Actually, all cases so far analysed have a good complete physical interpretation when carefully studied, and I am persuaded that the above FLRW+Vaidya model has an appropriate interpretation satisfying the corresponding field equations and boundary conditions (see also Refs. 73,74). Thus, the plausibility of the complete collapse of spherically symmetric realistic *radiating* stars without ever forming a singularity or a black-hole region seems feasible.

Coming back to the model with a FLRW expanding central region inside a Vaidya spacetime, it is evident that the universe will appear locally homogeneous and isotropic for any observer living inside the FLRW part, which might interpret the whole universe as such. Notice that the incoming radiation can be scattered when entering into the FLRW zone and thus an internal observer might be unaware of the exterior Vaidya part (see Ref. 73). Then, this is an expanding FLRW universe created out of some incoming radiation and with no big-bang in the past, and no singularity at all. Let us stress again that all energy and causality conditions are amply fulfilled. The avoidance of the singularity theorems is accomplished in this simple model because the boundary condition is not satisfied: there are no closed trapped surfaces, no points with reconverging horismos, no trapped sets, no closed edgeless slices, etcetera. This seems in contradiction with Example 4.2, where every FLRW model with $a_t \geq 0$ and $\varrho > 0$ was shown to contain closed trapped 2-spheres. However, as remarked in the cited example, this happens *provided that the* FLRW *model extends far enough*. In our case, the FLRW zone is sufficiently small that no closed trapped 2-sphere exists. Of course, this still allows for quite big FLRW zones, as explicitly shown in the next paragraph.

Even though this example may seem rather artificial, its importance lies in the fact that it proves the possibility of FLRW expanding realistic regions without having a big-bang in the past. In other words, we could very well live in a FLRW region inside a non-FLRW spacetime so that neither closed trapped surfaces nor other sets bound to be trapped are ever formed. Obviously, what is needed is the 'end' of the FLRW zone at values of $\chi$ *before* the apparent horizon is reached. Therefore, the question arises of whether these FLRW regions can be big enough to comply with our observations, and to accomodate the whole or most of the observed Universe. As proved in Example 4.2, the AH is a hypersurface located at distance $a(t_0)\Sigma(\chi_{AH}, k) = \sqrt{3/\varrho}$ at present time $t_0$. Putting units back and taking $\varrho_0 = 2.0 \times 10^{-31}$ g cm$^{-3}$ as the (luminous) mass density of the Universe we obtain a radius of about $29 \times 10^3$ Mpc. On the other hand, taking the critical density [241,159] $\varrho_c = 1.9 \times 10^{-29} h_0^2$ g cm$^{-3}$, the radius of the horizon now is of the order of $3.0/h_0 \times 10^3$ Mpc. In both cases, this is beyond the farest observed objects and there are no observations which may indicate that the Universe is FLRW that far. Hence, there is plenty of room to put our matching hypersurface without perturbing the observed cosmos. This means that, even in the case we could persuade ourselves that the Universe can be adequately described by a FLRW model, there is no definite evidence that a closed trapped 2-sphere exists around us. Of course, more realistic models are needed, but the door allowing for its construction is certainly open.

### 7.3. A spatially closed non-singular cosmology

Here we present a spatially closed singularity-free universe satisfying DEC and all other conditions everywhere. This shows that SEC may be violated in some regions while keeping the physically well-motivated DEC everywhere and producing singularity-free models. Of course, the de Sitter model seen in subsection 7.1 has also these properties. The purpose of this example is to prove that de Sitter's spacetime is not an exception, that this can happen in spatially inhomogeneous models, and that the expanding and contracting phases of the model can be infinite.

The spacetime belongs to the general class of spatially inhomogeneous non-rotating cosmologies [234], and more particularly to the special subclass admitting two commuting spacelike Killing vector fields — the so-called $G_2$ models. These cosmological models were classified in [235] and the particular one to be presented in this example belongs to the class B(i) in [235], that is to say, the $G_2$ group of motions acts orthogonally transitively but no Killing field is hypersurface-orthogonal. It was found with a systematic study of these metrics under an ansatz of separation of variables [145]. The manifold is $\mathbb{R} \times S^3$ with coordinates $\{T, \zeta, \vartheta, \phi\}$, and the line-element reads [145]

$$ds^2 = $$
$$e^{\beta \sin \vartheta + b^2 \sin^2(cT) - a^2} \left( -\frac{b^2 \sin^2(2cT)(1 - e^{-2b^2}) dT^2}{2 - 2e^{-2b^2 \sin^2(cT)} - 2(1 - e^{-2b^2})\sin^2(cT)} + \frac{d\vartheta^2}{c^2} \right)$$
$$+ \cos^2 \vartheta \, e^{b^2 \sin^2(cT) - a^2} d\phi^2 + e^{-b^2 \sin^2(cT) + a^2} (dz + \sqrt{2} a e^{-a^2} \sin \vartheta d\phi)^2,$$

where $\beta$, $b$ and $c$ are arbitrary constants and we use the abbreviation $a^2 \equiv b^2/(1 - e^{-2b^2})$. When $\beta = 0$, the spacetime possesses a four-dimensional group of isometries with a three-dimensional Bianchi type IX subgroup acting transitively on spacelike hypersurfaces. In the general case, the energy-momentum tensor is of perfect-fluid type and the coordinates are comoving. The energy density and pressure of the fluid are given by

$$\varrho = \frac{c^2(b^2 \sin^2(cT) - \beta \sin \vartheta + 2 - a^2)}{2 e^{b^2 \sin^2(cT) + \beta \sin \vartheta - a^2}},$$
$$p = \frac{c^2(b^2 \sin^2(cT) - \beta \sin \vartheta - a^2)}{2 e^{b^2 \sin^2(cT) + \beta \sin \vartheta - a^2}}.$$

As is apparent, these quantities are completely regular everywhere and thus there is no matter singularity. It can be also checked that there is no curvature singularity at all [145] and that the spacetime is g-complete. With the above formulas at hand, it is an easy matter to check that the DEC will be satisfied whenever

$$|\beta| \leq 1 - a^2 \quad \Longrightarrow \quad \frac{b^2}{1 - e^{-2b^2}} \leq 1$$

which permits a non-empty interval for $b$. Nevertheless, the SEC cannot be satisfied everywhere, because it requires

$$|\beta| \leq \frac{1}{2} - a^2 \quad \Longrightarrow \quad \frac{b^2}{1 - e^{-2b^2}} \leq \frac{1}{2},$$

which is fulfilled only in the very special case with $b = 0$ and $\beta = 0$. Actually, the limit $b \to 0$ with $\beta = 0$ leads to the Einstein static universe in the somewhat disguised form

$$ds^2 = -dT^2 + \frac{1}{c^2}(d\vartheta^2 + d\phi^2 + dz^2 + 2\cos\vartheta d\phi\, dz).$$

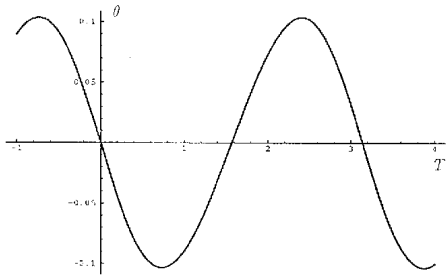

Figure 14. Plot for the expansion of the perfect-fluid congruence in the closed non-singular cosmological model of subsection 7.3. The expansion depends also on the coordinate $\vartheta$ through the overall factor $e^{-\beta \sin \vartheta/2}$, so that the essential dependence is on $T$. The sign of $\theta$ changes periodically and for ever. This is an alternatively expanding and contracting universe with no singularity. The particular values of the parameters in this plot are $\beta = 1/5$, $b = 1/2$ and $c = 1$.

Hence, this spacetime is a three-parametric singularity-free generalization of the Einstein static universe which satisfies the DEC everywhere. The expression of the expansion of the velocity vector of the perfect fluid is

$$\theta = \frac{cb}{\sqrt{2}} \sqrt{\frac{1 - e^{-2b^2 \sin^2(cT)} - (1 - e^{-2b^2})\sin^2(cT)}{(1 - e^{-2b^2})e^{b^2 \sin^2(cT) + \beta \sin\vartheta - a^2}}}.$$

In Figure 14, a typical plot of the expansion is shown for the parameters $\beta = \frac{1}{5}$, $b = \frac{1}{2}$ and $c = 1$ (so that DEC holds).

The nice thing about this spacetime is that it expands and contracts periodically *ad eternum* without ever collapsing catastrophically. This spacetime has compact hypersurfaces without edge: any slice $T$ = const. Some of these hypersurfaces have positive expansion of the timelike geodesic congruence emanating orthogonally from them as this expansion equals the value of $\theta$ displayed above at each hypersurface $T$ = const. Hence, this spacetime satisfies all the conditions of Theorem 5.5 except for the SEC, which reveals itself again as basic in the theorems. It has been already remarked in subsection 6.2 that the physically realistic energy condition is DEC, and that SEC can be violated by realistic physical fields. Therefore, this model provides another example of how to avoid the conclusions of the singularity theorems while keeping the physically compelling DEC.

### 7.4. Definition of classical cosmological model

The terms 'cosmology' and 'cosmological model' have already appeared in this paper. Is there, then, any accepted definition of cosmological model in General Relativity? Well, incredible as it may seem, the answer is no. Usually, the study of the 'Universe as a whole' starts with a discussion of the homogeneity and isotropy of the observations, or the Cosmological (Copernican) Principle (see e.g. Refs. 112,128,149,159,239,241). Of course, any good theory must admit at least a handful of cosmological models, and not *only* the unique one representing the actual Universe. The only definition I have found is that given in the Preface of [124], which defines a cosmological model as any spacetime which contains, as a particular case, a FLRW solution. Of course, this is a good criterion for avoiding many solutions which are rather awkward, but it has the side-effect of leaving aside many other reasonable solutions which should be termed as cosmological and cannot be classified in any of the following 'classes': stars, galaxies, clusters, vacuum solutions, combinations of the previous, and a very long etcetera.

In this sense, it seems to me that it is necessary to distinguish between *general* families of solutions, which should certainly contain FLRW metrics as particular cases, and *particular* spacetimes belonging to those general families but themselves not containing any FLRW model. As a trivial example, consider the Bianchi models [50,134,177,179], or the general spherically symmetric spacetime. Obviously, these general families include the FLRW models, and thus they are cosmological models by the previous definition. But take now a particular Bianchi model, or a definite spherically symmetric spacetime. It may well happen that this particular metric does not include any FLRW as a limit case. Does it have the right to be called cosmological? Well, it depends: sometimes it does, sometimes it does not. For example, if the spacetime has no matter one should hardly say that this is a cosmology, because we are pretty sure that the Universe contains matter (this rules out Schwarzschild spacetime as a cosmological model, even though it has spherical symmetry). But it may happen that a model has no obvious 'bad' property and nevertheless does not include the FLRW model. The question is whether these models have the right to be called cosmological *theoretically*. The problem of what is a *realistic* cosmological model is quite another different matter, and we should solve the former first.

The above discussion gives a pre-eminence to FLRW models which is, in my opinion, more than dubious. However, even accepting that FLRW spacetimes give a good first approximation to the description of the Universe, the question remains. Consider, for instance, the Lemaître–Tolman spacetime [20,130,223], which is the most general spherically symmetric solution for dust, given by [123,124]

$$ds^2 = -dt^2 + \frac{F_{,r}^2 dr^2}{1 + 2G(r)} + F^2(t,r)(d\vartheta^2 + \sin^2\vartheta d\varphi^2)$$

where $F(t,r)$ satisfies

$$F_{,t}^2 = 2G(r) + \frac{2\mu(r)}{F}$$

with arbitrary functions $\mu(r)$ and $G(r)$. The energy density reads

$$\varrho = \frac{2\mu_{,r}}{F^2 F_{,r}}.$$

Now, all dust FLRW models are included here for the particular choice $2G = -kr^2$, $\mu \propto r^3$, and $F = ra(t)$, where $k$ is the curvature index and $a(t)$ is the corresponding scale factor. However, the general solution includes many other metrics — for instance, Schwarzschild for $\mu =$ const. If we concentrate in the simplest case $G = 0$, the general solution is $2F^3 = 9\mu(r)[t - f(r)]^2$ with arbitrary $f(r)$. Now, which particular functions $f(r)$ and $\mu(r)$ will give a spacetime worth to be called cosmological?

Yet another example. Consider the following cylindrically symmetric model found in [79]

$$ds^2 = \sinh^4(at)\cosh^2(3a\rho)(-dt^2 + d\rho^2)$$
$$+ \sinh^4(at)\sinh^2(3a\rho)\frac{1}{9a^2}\cosh^{-2/3}(3a\rho)d\varphi^2$$
$$+ \sinh^{-2}(at)\cosh^{-2/3}(3a\rho)dz^2, \qquad (42)$$

which is a solution of Einstein's equations for a comoving perfect fluid with pressure and energy density given by

$$p = 5a^2 \sinh^{-4}(at) \cosh^{-4}(3a\rho), \qquad p = \varrho/3,$$

so that the barotropic equation of state is realistic for radiation-dominated matter. It is easy to see that there is a big-bang matter singularity in the finite past at $t = 0$. By the way, it can be easily seen that this solution violates the conjecture explained just before Definition 3.6, because

$$\lim_{t \to 0} \frac{C_{\mu\upsilon\rho\sigma} C^{\mu\upsilon\rho\sigma}}{R_{\mu\upsilon} R^{\mu\upsilon}} = \infty$$

so that the divergence of the matter quantities is less severe than that of the Weyl tensor near the big-bang. This model cannot be considered an exterior solution, as it has matter everywhere ($\varrho \neq 0$) and the pressure does not vanish anywhere, which is the condition for its possible matching with a vacuum exterior (see Ref. 132). It is globally hyperbolic, satisfies all possible energy conditions, and the fluid is expanding for ever and everywhere. Is this a cosmological solution? Well, it may be argued that the cylindrical symmetry will not give a good description of the *actual* Universe. This is controversial, because all FLRW models do have cylindrical symmetry (see e.g. Ref. 51,194)! But anyway, the problem is to decide what a *theoretical* cosmological model is, and *not* what *the* cosmological model is that adequately describes the Universe. As a matter of fact, the existence of *the* realistic cosmological model is not assured on theoretical grounds, because two very different models may give rise to the same observations. An explicit example can be found in [69]. Thus, I think the above model can be considered as a theoretical, and maybe unrealistic, cosmological model.

All in all, the definition of classical cosmological model to be used in this paper is [195]

**Definition 7.1.** A theoretical classical cosmological model is any space-time containing appropriate matter everywhere which is expanding in a region for a certain period of time.

This definition should be taken as tentative, and I hope my colleagues will help obtain a better one. The condition of having matter everywhere is invoked to avoid vacuum and similar models, or situations with *localized* objects, such as stars, galaxies, etcetera. For localized objects, there usually exists a limit hypersurface (usually timelike) up to which the matter extends and where the exterior solution (usually vacuum) starts. In these cases there is a precise component of the energy-momentum tensor (some normal pressure or tension) which has to vanish at this hypersurface [143]. Nevertheless, there could be situations where there is matter everywhere but describing a localized object if this mater falls off too rapidly away from the object. This is the main difficulty in defining and constructing cosmological models. One could impose the condition that the model is not asymptotically flat, but in fact there is no way to decide whether or not the actual Universe fulfils this condition, (an explicit example is provided by the spacetime in subsection 7.2). This is a crucial point: if there is a way to impose severe restrictions on the decaying of the density of matter away from us, the possibility of non-singular cosmological models would be much restricted. It is interesting to keep this in mind when studying the non-singular models to be presented in the following subsections. As a matter a fact, an interesting result very recently found by Raychaudhuri [172] shows that irrotational non-singular models (satisfying SEC and other reasonable conditions) must have a vanishing *spacetime* average of the energy density and of other scalar quantities. This new theorem applies to most non-singular models known hitherto, in particular to those to be presented in subsections 7.6 and 7.7. However, the relevance of this result is dubious, mainly because the averages are taken over the whole spacetime, as opposed to pure *spatial* averages which would certainly be much more physical. Actually, the analogous quantity (a spacetime averaged density) computed in open FLRW models is also vanishing.

Going back to the definition, by 'appropriate' matter is meant general fluids, or a gas of galaxies, or similar things, but not, for instance, a pure combination of electromagnetic plane waves. The requirement of expansion of matter is a little bit more complex. The idea is to eliminate static models, which might have been good cosmological models in principle. However, it seems that there is a basic consensus[12] that the Universe is expanding due to the observation of red-shifts and the simplest predictions of the theory. Nevertheless, the step from the measured red-shifts to the expansion of the Universe is not immediate and depends on the model and on the observer. To begin with, in a given spacetime some observer may see no red-shifts while some other observers do see them. Somehow, it is tacitly accepted that one always refers to the observer defined by the cosmological matter itself. Whether this is reasonable or not will not be analysed here. Furthermore, let us remember that the red-shift $z$ measured by an observer described by a timelike congruence with unit tangent vector

---

[12] If this is not so, this condition might be dropped from the definition.

$\vec{u}$ is defined by

$$1 + z \equiv \frac{(u_\mu k^\mu)_{\text{emitter}}}{(u_\mu k^\mu)_{\text{observer}}},$$

where $\vec{k}$ is the tangent vector to the null geodesic $x^\mu = \gamma^\mu(\tau)$ going from the emitter to the observer and $\tau$ its affine parameter. The red-shift $z$ is then given by an integral along the null geodesic through the formula [66]

$$\frac{d(u_\mu k^\mu)}{d\tau} = \frac{\theta}{3}(u_\mu k^\mu)^2 + \sigma_{\mu\upsilon}k^\mu k^\upsilon - (u_\upsilon k^\upsilon)a_\mu k^\mu. \qquad (43)$$

Of course, the expansion $\theta$ of the congruence has an important effect, but $z$ depends also on the shear and the acceleration of the observer. Thus, very complicated models with acceleration and shear and perhaps contracting regions may give rise to a particular place in which the measurements produce the observed red-shifts. In other words, there is no easy way to translate the red-shifts to the kinematic quantities of the matter described here by $\vec{u}$. In fact, all this should be refined with an adequate treatment of the data on the past light rays of a particle of the cosmological matter (see Refs. 57,58,70). In general, for irrotational models [234], there is a well-defined time function (24), and thus one can simplify matters by requiring that $\theta$ be positive for some spatial region at some instant of this preferred time. This is the condition to be used in what follows, even though it may not quite collect all that is behind Definition 7.1.

No energy or causality conditions have been assumed in the definition. Of course, SEC could be added but then de Sitter spacetime and the general inflationary models would be ruled out as classical cosmological models. One can also assume DEC but not SEC, which would be physically acceptable, but it seems that the fashionable modern inflationary models may violate even WEC [32]. Of course, completely unrealistic cosmological models may fit with Definition 7.1, but the question of what is a *realistic* cosmological model is even more tricky and delicate. It seems that, apart from the red-shifts, the main observational evidences concerning the real Universe are (see also Refs. 135,159): the distribution and abundance of chemical elements such as Helium; the existence, spectrum and degree of isotropy of the CMBR; and the clumping of matter. This last point is one of the greatest difficulties Cosmology has to face nowadays and is obviously not explained by the pure FLRW models — there is some hope that FLRW perturbations may give an answer, though. Concerning the existence of the CMBR and the primordial character of chemical elements all that is needed is a very hot and dense epoch in our causal past. Of course, many models will satisfy this. Perhaps the most demanding point is the small anisotropy showed by the CMBR around us. Requiring the existence of a *geodesic* congruence observing the isotropic distribution of a collisionless gas of photons, Ehlers, Geren and Sachs proved their classical theorem that the model must be a FLRW spacetime [61]. However, if the geodesic assumption is dropped, many other possible models survive [80]. Moreover, the assumption of exact isotropy is too strong and there are definite indications that the observations on the CMBR can be almost isotropic with very big inhomogeneities in the Universe, as already remarked [6,124,157,169,181]. Therefore, just to be on the safe side, let us simply define [195]

**Definition 7.2.** A realistic cosmological model is any classical cosmological model which explains as much observational evidence as possible.

One could refine the above by saying that the model must explain *all* observational evidences at hand. Unfortunately, in that case FLRW models would not be realistic because they do not explain the lumpiness or clumping of matter. Thus, in order to keep FLRW models as realistic, one must admit the softer definition above.

There are plenty of papers supporting the above definitions (see for a selection Refs. 4,35,36,50,59,66,67,69,79,95,97,98,112,123,124,128,134,135, 147–149,177,179,199,206,207,209,211,221,222,227,234–238,240,241    and many references therein), where they were used either explicitly or tacitly without any problem. However, since the appearance of the singularity-free models and its repercussion [136], there has been some controversy about whether or not they should be considered as cosmological models. The claim that the singularity-free models known hitherto may be unrealistic [40,178,194,195] has not helped and sometimes they are disregarded even as *unrealistic* cosmologies. Unfortunately, this view may lead to the tautological existence of the singularity in cosmological models. In order to avoid this problem and to provide more relevant examples concerning the singularity theorems, some cosmological models are presented in the next subsections 7.5, 7.6, 7.7 and 7.8 where, in particular, most reasonable criticism on the non-singular models will be addressed.

### 7.5. Big-bang-free cosmological models

Here we present some cosmological models without big-bang singularities and satisfying all energy, causality and simplicity conditions. The traditional way to achieve these properties was the addition of electric charge to the matter describing spherically symmetric spacetimes. This usually produces a repulsive force preventing collapses (see Refs. 29,124,200,229 and references therein). There is no need for that, however, and there are some explicitly known perfect-fluid models with these properties. Some examples have been already presented here, such as for instance the case $a < 0$ of Fig. 6 and the case of Fig. 9 in Example 3.7 of Section 3. Another explicit very simple example is given by the spherically symmetric solution presented in the first row ($k = 1$) of Table 5 in [36], with line-element

$$ds^2 = e^{2ax}(-dt^2 + dx^2) + e^{2ax}(e^{2at} + \tfrac{1}{2})(d\vartheta^2 + \sin^2\vartheta d\varphi^2),$$

where $a > 0$ is a constant. This is a solution for a stiff fluid (perfect fluid with $p = \varrho$), and the energy density and pressure read

$$p = \varrho = a^2 e^{-2ax}(e^{2at} + \tfrac{1}{2})^{-4}.$$

The coordinate system is comoving, but it is not well-adapted to the global structure of the spacetime. A better coordinate system is defined by

$$T = \frac{e^{ax}}{a}\sinh(at), \qquad R = \frac{e^{ax}}{a}\cosh(at)$$

so that the metric takes the form

$$ds^2 = -dT^2 + dR^2 + \tfrac{1}{2}(T+R)(T+3R)(d\vartheta^2 + \sin^2\vartheta d\varphi^2)$$

and the density and pressure become

$$p = \varrho = 8(R-T)^3(T+R)^{-1}(T+3R)^{-4}.$$

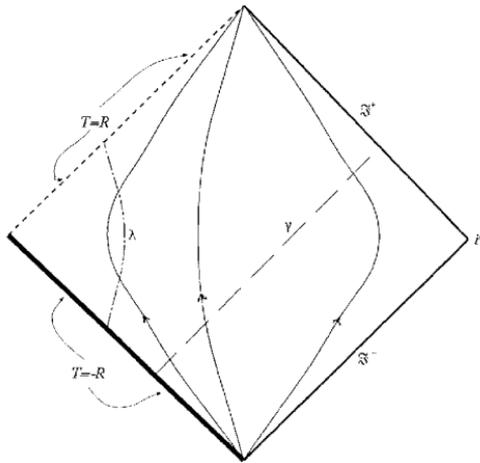

Figure 15. This is the conformal diagram of the first spacetime presented in subsection 7.5. There is a null singularity at $R = -T$ which is not a big bang because many causal curves start at $\mathscr{I}^-$. In particular, the whole perfect fluid congruence, which is represented by arrowed lines, does not originate at the singularity. In this sense, the cosmological matter avoids the past null singularity. Null geodesics and timelike geodesics, such as $\gamma$ and $\lambda$ shown in the figure, start at the null singularity. The null geodesic $\gamma$ ends at $\mathscr{I}^+$, while the timelike geodesic $\lambda$ reaches $T = R$ with finite proper time. Thus, this spacetime is extendible across $T = R$.

The manifold is defined by the portion of $\mathbb{R}^4$ defined by $R > |T|$ in spherical coordinates $\{T, R, \vartheta, \varphi\}$. Then, the energy density and pressure are positive everywhere so that the energy conditions are fulfilled. The spacetime is globally hyperbolic too and there are no closed trapped 2-spheres. Now it is easily seen that $T = -R$ is a null singularity corresponding to $x \to -\infty$ and $t \to -\infty$. No curve of the fluid comes from this singularity as they have $x =$ const., so that this singularity is null and *not* of big-bang type. However, null geodesics, timelike geodesics and other causal curves start at this singularity (see Figure 15). Similarly, some null and timelike geodesics as well as other causal curves reach $R = T$ at finite proper time. This is not a curvature singularity of the spacetime and in fact spacetime is regularly extendible through $R = T$, which corresponds to $x \to -\infty$ and $t \to \infty$.

In principle, this spacetime does not describe a localized object as the pressure is non-zero all over the spacetime and the fluid congruence is expanding everywhere and for ever, the expansion being

$$\theta = 2e^{2at}(e^{2at} + \tfrac{1}{2})^{-1} = 4(T + R)(T + 3R)^{-1}.$$

However, the energy-momentum tensor tends to zero when approaching $T = R$, through where the spacetime is extendible. Thus, whether or not this is a cosmological model depends on the choice of the matter content in the extended region.

Other perhaps more interesting explicit perfect-fluid models were found in [178], such as the following line-element given in cylindrical coordinates $\{t, \rho, \varphi, z\}$ ($a \neq 0$ is a constant)

$$\begin{aligned}ds^2 = {}& \cosh^{2(5+2\sqrt{5})}(at)\, e^{2(3+\sqrt{5})a\rho} \sinh^{2(2+\sqrt{5})}[(\sqrt{5}-1)a\rho](-dt^2 + d\rho^2) \\ & + \cosh(at)\, e^{(2-\sqrt{5})a\rho} \sinh[(\sqrt{5}-1)a\rho] \times \\ & \times [\cosh^{4+\sqrt{5}}(at)\, e^{a\rho} \sinh^{2+\sqrt{5}}[(\sqrt{5}-1)a\rho]d\varphi^2 \\ & + \cosh^{-(4+\sqrt{5})}(at)\, e^{-a\rho} \sinh^{-(2+\sqrt{5})}[(\sqrt{5}-1)a\rho]dz^2].\end{aligned}$$

The manifold is $\mathbb{R}^4$ without the $\rho = 0$–axis and the velocity vector of the perfect fluid is proportional to $\partial_t$ (comoving coordinates). The pressure of the perfect fluid is

$$p = c \cosh^{-2(5+2\sqrt{5})}(at)\, e^{-(5+3\sqrt{5})a\rho} \sinh^{-(5+2\sqrt{5})}[(\sqrt{5}-1)a\rho],$$

where $c$ is a positive constant and the barotropic equation of state reads

$$p = \frac{1 + \sqrt{5}}{4} \varrho.$$

Both the pressure and the energy density are positive everywhere, so that all energy conditions hold. The line-element is causally stable and there is no big-bang singularity. Nonetheless, there is a *matter* timelike singularity at $\rho = 0$ as the density and pressure diverge there. This is a big-bang-free classical cosmological model according to Definition 7.1, because the expansion of the fluid congruence is positive for all $t > 0$ [178] and the pressure does not vanish anywhere. Therefore, the classical cosmological models can have a timelike singularity and no big-bang. Notice that this singularity is avoided by most particles as it resembles a singular axis, but most particles can travel into the past and future indefinitely.

Another even more curious example is given by the same manifold as before but with the line-element [178]

$$ds^2 = \cosh^{1+\sqrt{2}}(at)\tanh^{1-\sqrt{2}}(\sqrt{4+3\sqrt{2}}\,a\rho)(-dt^2 + d\rho^2)$$
$$+ \cosh(at)\sinh(\sqrt{4+3\sqrt{2}}\,a\rho)\cosh^{(\sqrt{2}-2)/2}(\sqrt{4+3\sqrt{2}}\,a\rho) \times$$
$$\times [\cosh^{1+\sqrt{2}}(at)\sinh^{\sqrt{2}-1}(\sqrt{4+3\sqrt{2}}\,a\rho)d\varphi^2$$
$$+ \cosh^{-(1+\sqrt{2})}(at)\sinh^{1-\sqrt{2}}(\sqrt{4+3\sqrt{2}}\,a\rho)dz^2].$$

This is a perfect-fluid solution in comoving coordinates with a pressure and an energy density given by

$$p = \text{const}\, \frac{\tanh^{1+\sqrt{2}}(\sqrt{4+3\sqrt{2}}\,a\rho)}{\cosh^{1+\sqrt{2}}(at)}, \qquad p = \frac{4\sqrt{2}-5}{7}\varrho,$$

so that the equation of state takes again the simple linear form $p = \gamma \varrho$, now with $\gamma < \frac{1}{3}$. As easily checked, pressure and energy density are both regular everywhere, so that there is no matter singularity in this case. The expansion is positive for all $t > 0$ and the pressure is non-zero everywhere, so that the spacetime is a classical cosmological model according to Definition 7.1 and satisfies all energy conditions. On the other hand, it can be shown that $\rho = 0$ is a true singularity for the Weyl tensor [178]. Thus, there are well-behaved classical cosmological models with *pure Weyl timelike* singularities and no big-bang. This singularity resembles again a singular axis but now with matter quantities regular there.[13]

---

[13] These spacetimes are not cylindrically symmetric because the candidate to axis is a singularity (see Ref. 144).

All spacetimes shown in this subsection prove that reasonable cosmological models can have timelike (or null) singularities and that there is no 'singular beginning' for them. In fact, in the next subsections we are going to see that there are completely non-singular cosmological models too.

### 7.6. A simple singularity-free spacetime

Here we present the first singularity-free models to be found satisfying all energy, causality and simplicity conditions. The manifold is $\mathbb{R}^4$ with cylindrical coordinates and the line-element can be written as [178,193,227]

$$ds^2 =$$
$$\cosh^4(at)\cosh^2(3a\rho)\left(-dt^2 + \frac{\sinh^2(3a\rho)}{\cosh^2(3a\rho) + (K-1)\cosh^{5/3}(3a\rho) - K} d\rho^2\right)$$
$$+ 4\cosh^4(at)\frac{\cosh^2(3a\rho) + (K-1)\cosh^{5/3}(3a\rho) - K}{(5K+1)^2 a^2 \cosh^{2/3}(3a\rho)} d\varphi^2$$
$$+ \cosh^{-2}(at)\cosh^{-2/3}(3a\rho)dz^2, \tag{44}$$

where $a > 0$ and $K > 0$ are arbitrary constants. This metric has well-defined cylindrical symmetry and the elementary flatness condition [123, 142] is satisfied in the vicinity of the axis $\rho = 0$. The energy-momentum tensor is of perfect-fluid type with comoving velocity vector and the energy density and pressure take the form

$$\varrho = 15a^2 K \cosh^{-4}(at)\cosh^{-4}(3a\rho), \qquad p = \varrho/3,$$

so that again the equation of state is realistic for radiation-dominated matter. All energy conditions are amply fulfilled everywhere. Furthermore, it is easy to check that $R_{\mu\nu}k^\mu k^\nu > 0$ for all causal vectors $\vec{k}$ so that the generic condition holds too due to Propositions 2.5 and 2.6. The metric is obviously causally stable ($t$ is a time function), and in fact it is globally hyperbolic, the hypersurfaces $t = $ const. being Cauchy hypersurfaces [40]. The particular case $K = 1$ was found in [193] and fully analysed in [40]. The conclusions hold for the general case with arbitrary $K$ too. In particular, there is no matter singularity, as is apparent from the above. It can be also checked that there is no curvature singularity at all [178,193]. Actually, this spacetime is g-complete (whence inextendible, Proposition 3.1) and singularity-free. The question arises: how can this simple and well-behaved metric avoid the singularity theorems? It is very interesting to look into this spacetime closer to answer this question.

To begin with, the expansion of the perfect fluid is

$$\theta = 3a\cosh^{-3}(at)\cosh^{-1}(3a\rho)\sinh(at), \tag{45}$$

so that the fluid undergoes a contracting phase for $t < 0$, a rebound at $t = 0$ and an expanding phase for all $t > 0$. This is a classical cosmological model according to Definition 7.1 because it cannot be interpreted as a finite body and is expanding half of its history. Let us start with Theorem 5.1. Evidently, all conditions of this theorem are satisfied *except* for the geodesic motion of matter. The acceleration (20) of the perfect fluid congruence is

$$\mathbf{a} = -3a\tanh(3a\rho)d\rho$$

so that there appears a gradient of pressure [see (40)] which acts opposing gravitational attraction, and thus the model can contract for $t < 0$ and nevertheless re-bound at $t = 0$ without reaching any singularity. Consequently, acceleration is enough to avoid the singularities predicted by the Raychaudhuri–Komar theorem.

Consider then Theorems 5.2 and 5.3. The former requires SEC (fulfilled), a Cauchy hypersurface (fulfilled) with expansion bounded away from zero (almost fulfilled, but not quite!). As we can see, the expansion of any $t = $ const. Cauchy hypersurface is *positive everywhere*, but it is *not bounded below by a positive number*, as $\theta \to 0$ for $\rho \to \infty$ at every $t$. This subtle difference makes Theorem 5.2 inapplicable to this spacetime. Let us remark, therefore, that non-singular cosmological models can have matter expanding everywhere at some instant of time. For the singularity theorems to apply it is not enough with an expansion which is strictly positive everywhere, it must be bigger than a fixed positive constant. Similarly, with regard to Theorem 5.3, the condition of a Cauchy hypersurface with vanishing second fundamental form is fulfilled by the hypersurface $\Sigma : t = 0$, as can be checked from the expression for the shear [178,193] and the expansion. Further, it is easily checked that

$$\left| \int_{\tau_\Sigma}^{\tau_\Sigma + b} R_{\mu\nu} v^\mu v^\nu d\tau \right| > 0$$

for every timelike $\vec{v}$ and every $b > 0$. But then again, the subtlety appears: this is strictly positive everywhere, but not greater than a given positive constant because it goes to zero for $\rho \to \infty$. Thus, this theorem is inapplicable to the spacetime in consideration. In both Theorems 5.2 and 5.3 the escape is the same. Any Cauchy hypersurface $t = t_0 > 0$ has a positive expansion $\theta_0 > 0$. Hence, every single geodesic orthogonal to the hypersurface must have a past-conjugate point within a finite proper time $3/\theta_0$. However, $\theta_0$ is positive but not bounded from below by a positive number, so that these proper times do not have a finite upper bound. This allows for the existence of a maximal geodesic from $t = t_0$ up to any $p \in J^-(t = t_0)$, as is compulsory for any Cauchy hypersurface.

Theorems 5.4, 5.5 and 5.7 can be included in Theorem 5.6 for our purposes now. All conditions in the Hawking–Penrose theorem are fulfilled except for the boundary condition. The spacetime has no compact achronal edgeless set (for a formal proof, see Ref. 40), no closed trapped surfaces,[14] and no point with reconverging horismos. To see this last point, the easiest thing is to consider the f-d radial null geodesics $\gamma$, given by constant $\varphi$ and $z$ together with [40] ($\tau$ is the affine parameter)

$$\frac{d\gamma^0}{d\tau} = \left|\frac{d\gamma^1}{d\tau}\right|, \qquad \frac{d\gamma^1}{d\tau} = c\cosh^{-4}(a\gamma^0)\cosh^{-2}(3a\gamma^1),$$

where $c$ is a constant. Therefore, given any point $p \in V_4$, there are future- and past-directed radial null geodesics passing through $p$ and going to infinity ($\rho \to \infty$), whence there cannot be reconvergence of the light cones. Finally, the boundary conditions of the other Theorems of Section 5 are obviously not fulfilled either.

To gain further insight into the properties of this model, let us consider briefly what is the behaviour of causal curves. The spacetime is globally hyperbolic, and therefore there is a maximal causal curve between any pair of causally related points by Proposition 2.32. However, the spacetime satifies DEC and the strict SEC, whence the generic condition too, and thus every endless causal curve has a pair of conjugate points due to Proposition 2.4 so that it cannot be maximal for points 'beyond' these conjugate points by Corollary 2.2 and Proposition 2.10. Therefore, given any compact set $\mathcal{K}$, there must be points to the future and past of $\mathcal{K}$ that can be joined by a maximal causal curve not passing through $\mathcal{K}$. Otherwise, the result of Proposition 2.34 would hold and, *a posteriori*, some singularity theorem would apply. Let us then see this property explicitly. Take the family of 'circular' null curves defined by

$$z = z_0 = \text{const.}, \qquad \rho = \rho_0 = \text{const.}, \qquad t = \varphi \frac{\sinh(3a\rho_0)}{3a\cosh^{4/3}(3a\rho_0)} + b,$$

where $b$ is a constant. These are not geodesics except for the particular case with $\cosh(3a\rho_0) = 2$. As is evident, the coordinate time $t$ elapsed between *any* two values of $\varphi$ can be chosen as small as desired by taking a big enough $\rho_0$. This property may seem surprising, but it is a consequence of the fact that distances along the $\varphi$-circles become extremely small in comparison with the radial distances for big $\rho$. Consider also the family of null curves parallel to the axis, that is

$$\varphi = \varphi_0 = \text{const.}, \quad \rho = \rho_0 = \text{const.}, \quad \frac{dt}{dz} = \cosh^{-3}(at)\cosh^{-4/3}(3a\rho_0),$$

so that along these curves $t \leq z\cosh^{-4/3}(3a\rho_0) + h$, for some constant $h$. All these curves are not geodesics except for the axis $\rho_0 = 0$ itself. Again, the time $t$ between *any* two values of $z$ can be made as small as required by choosing $\rho_0$ big enough. This is a consequence of the fact that spatial distances along $z$ decrease as $\rho$ increases.

---

[14] The proof of this in [40] is not completely correct, but the result holds.

Now, take *any* compact set $\mathcal{K}$. Then, the coordinates $t, \rho$ and $z$ will attain their maximum and minimum values on $\mathcal{K}$. Now, going sufficiently back in the past it is always possible to choose a point $p \in J^-(\mathcal{K})$ and an f-d radial null geodesic from $p$ such that, for some $t = t_1$, it will reach values of $\rho$ bigger than the maximum of $\rho$ at $\mathcal{K}$; continue then with an appropriate combination of circular and parallel to the axis null curves so that $z$ and $\rho$ reach the desired values without ever entering into $\mathcal{K}$ and, what is more important, without increasing too much the value of $t$. Finally, go from this point to a $q \in J^+(\mathcal{K})$ along the radial null geodesic or the adequate curve. This always gives a causal curve from the past of $\mathcal{K}$ to the future of $\mathcal{K}$ without crossing $\mathcal{K}$. As a final interesting remark, let us notice that the constructed curve is a combination of null curves, so that there must exist another timelike curve from $p$ to $q$ which is maximal. Therefore, the conclusion of Proposition 2.34 can be avoided because for any pair of sequences $\{p_n\}$ and $\{q_n\}$ as in the proposition, there always is an $n$ such that a maximal curve from $p_n$ to $q_n$ does not meet $\mathcal{K}$. In summary, the focusing effect on geodesics takes place fully in this solution, but there is no 'time-bomb' (boundary condition) leading to the catastrophe.

All in all, this simple example shows that there are well-founded, well-behaved classical cosmological models expanding everywhere for an infinite period of time (half of their history), with reasonable and realistic matter content, satisfying all possible energy, generic and causality conditions and, nevertheless, singularity-free. The boundary condition is therefore the essential assumption in the singularity theorems and its realization in some physical models can be certainly eluded. This model shows *explicitly* that the door to construct inhomogeneous non-singular cosmological models is wide open despite the singularity theorems. Unfortunately, the above simple model seems not to be realistic (see Definition 7.2). Hence, the question of whether the singularity theorems still allow for *realistic* singularity-free models is quite another problem. Elsewhere, by using very simple arguments I have argued that this door is also open [194]. The main ideas are collected in the next subsection.

### 7.7. FLRW plus non-singular models. Inflation and little bang

The singularity-free model of the previous subsection is not isolated, and in fact it belongs to a more general family of perfect-fluid non-singular cylindrically symmetric spacetimes [178]. Since their discovery in [193], other non-singular models have been found, and the list is increasing slowly but firmly [4,53–56,101,138,139,145,158,178,227,247]. A single family combining the FLRW models with some of these non-singular spacetimes can be built so that the parameters select from the complete spatial homogeneity and isotropy to the $G_2$-inhomogeneity.

The explicit family is given in cylindrical coordinates $\{t, \rho, \varphi, z\}$ by the line-element [194]

$$ds^2 = T^{2(1+n)} \Xi^{2n(n-1)} (-dt^2 + d\rho^2) \\ + T^{2(1+n)} \Xi^{2n} \Xi_{,\rho}^2 d\varphi^2 + T^{2(1-n)} \Xi^{2(1-n)} dz^2,$$

where $T(t)$ is an arbitrary function, $n \geq 0$ is a constant and $\Xi(\rho)$ is the solution of

$$\Xi_{,\rho}^2 = M\Xi^2 + N - nK\Xi^{2(1-2n)},$$

where $M$, $N$ and $K$ are arbitrary constants. Notice that this is not a remaining equation to be solved (a simple change of coordinate $\rho$ would eliminate this equation); it is preferable to maintain the equation because the behaviour of the solution depends on the sign of $M$ and the values of the other constants (see Ref. 194). Essentially, $\Xi$ behaves like a trigonometric function if $M < 0$, and like a hyperbolic function if $M > 0$. The case $M = 0$ is more involved. The general spacetime has well-defined cylindrical symmetry with a regular axis.

In the natural orthonormal cobasis $\theta^\mu \propto dx^\mu$, the energy-momentum tensor takes the diagonal form $T_{\mu\nu} = \text{diag}(\varrho, p_r, p_r, p_z)$, where the energy density $\varrho$ and pressures $p_r$, $p_z$ read explicitly

$$\varrho =$$
$$\frac{(2n-1)(n-1)(n+3)nK + \Xi^{4n}(n+1)(n-3)(M-(T_{,t}^2/T^2))}{T^{2(1+n)}\Xi^{2n(n+1)}}$$

$$p_r =$$
$$\frac{(2n-1)(n-1)^2 nK + \Xi^{4n}((n-1)^2 M - [(n+1)(n-3)+2](T_{,t}^2/T^2) - 2(T_{,tt}/T))}{T^{2(1+n)}\Xi^{2n(n+1)}}$$

$$p_z =$$
$$\frac{(2n-1)(n-1)^2 nK + \Xi^{4n}((n+1)^2 M - (n+1)(n-1)(T_{,t}^2/T^2) - 2(n+1)(T_{,tt}/T))}{T^{2(1+n)}\Xi^{2n(n+1)}}$$

The condition for a perfect fluid is

$$p_r = p_z \equiv p \quad \Longleftrightarrow \quad n\left(\frac{T_{,tt}}{T} + \frac{T_{,t}^2}{T^2} - 2M\right) = 0$$

so that, for perfect fluids, either $n = 0$ with arbitrary $T(t)$ or $T(t)$ is a known simple function [194] with arbitrary $n$. However, this condition is *not* assumed in what follows.

The non-zero components of the kinematic quantities (20) and (22) of the fluid congruence in the chosen cobasis are

$$\theta = (n+3)\frac{\Xi^{n(1-n)}T_{,t}}{T^{n+2}}, \qquad a_1 = -n(n-1)\frac{\Xi_{,\rho}\Xi^{n(1-n)}}{T^{1+n}\Xi},$$
$$\sigma_{11} = \sigma_{22} = -\frac{\sigma_{33}}{2} = \frac{2n\theta}{3(n+3)}.$$

These general spacetimes are classical cosmological models in the sense of Definition 7.1. The acceleration and shear vanish when $n = 0$, in which case the fluid is perfect, from where it follows (see e.g. Ref. 124) that the metric for $n = 0$ is a FLRW model. Actually, all FLRW models are included here, and they are invariantly characterized within the general class by the simple condition $n = 0$.[15] In this FLRW case, $T$ coincides with the FLRW scale factor $a(t)$ and $-\mathrm{sign}(M)$ is the usual curvature index $k$. This interpretation of $M$ and $T$ is valid for the whole general class. Moreover, the singularity-free solution of the previous subsection is included in this general family as the particular case with $M = a^2/4 > 0$, $K > 0$, $n = 3$, $T^2 = \cosh(at)$, $N = 3K - M$.

---

[15] The line-element obtained by setting $n = 0$ is not that given in Example 3.1, rather it is the cylindrically symmetric expression of the FLRW models [51,124,194].

As a matter of fact, there are also many singularity-free metrics included in this family. Restricting our attention to those models satisfying both SEC and DEC, the singularity-free subclass is uniquely characterized by the following properties:

$$M > 0, \qquad K \geq 0, \qquad n \geq 3, \qquad T \neq 0,$$
$$0 \leq \frac{T_{,tt}}{T} + \frac{T_{,t}^2}{T^2} - 2M \leq (n-3)\left(M - \frac{T_{,t}^2}{T^2}\right).$$

The first of these conditions says that these non-singular models are 'spatially open'. This was to be expected due to Theorems 5.5 and 5.1 as SEC has been assumed. If SEC were not taken into account other more general non-singular models could appear, as already shown in subsection 7.3. The discussion of why these spacetimes are g-complete and singularity-free and how they avoid the general conclusions of the singularity theorems is essentially the same as that given in the previous subsection and therefore will be omitted here. More interesting is that, from the last condition displayed above it follows the remarkable fact that the general non-singular subclass contained here and satisfying both SEC and DEC *must have one and only one single rebound time*, which is defined by $T_{,t} = 0$, see the expression for the expansion. Whether this is a generic property of singularity-free well-behaved models is not sure, but seems certainly plausible. For the case under consideration this provides an invariant characterization of the 'bang' happening at the rebound, where expansion changes from negative to positive values. Therefore, the expanding era *starts* at the rebound and will continue for ever. This rebound may be called the *little bang* because it is completely regular.

From the above discussion is known that the metrics with $n = 0$ are the FLRW spacetimes, while the non-singular models satisfying SEC and DEC must have $n \geq 3$. Hence, an interpretation of the fundamental parameter $n$ is needed. A straightforward calculation using the above formulas gives the relative shear

$$\frac{\sigma}{\theta} = \frac{2n}{\sqrt{3}(n+3)},$$

from which it follows that $n$ measures the anisotropy and inhomogeneity of the model. Therefore, for this particular class of models, the regularity of the spacetime requires a minimum amount of inhomogeneity in the matter content. This does not necessarily lead to contradiction with observations (see Refs. 124,135). A more detailed analysis shows that the relative distance $D$ between neighbouring world-lines of the fluid congruence leads to the following generalized Hubble law for these models [194] ($\vec{u}$ is the unit velocity vector of the fluid)

$$u^\mu \partial_\mu D = D[1 - n \cos(2\Theta)] \frac{3}{n+3} \mathcal{H}, \qquad (46)$$

where $\mathcal{H}$ stands for the *Hubble function* $\mathcal{H} \equiv \theta/3$, as usual. Here $\Theta$ is a polar angle selecting the particular direction of observation for each particle as follows: take the orthonormal spacelike triad $\{\vec{e}_1, \vec{e}_2, \vec{e}_3\}$ with $\vec{e}_1 \propto \partial_\rho$, $\vec{e}_2 \propto \partial_\varphi$ and $\vec{e}_3 \propto \partial_z$ and define the spatial direction of observation by means of the unit spatial vector

$$\sin \Theta \cos \Phi \vec{e}_1 + \sin \Theta \sin \Phi \vec{e}_2 + \cos \Theta \vec{e}_3,$$

where $\Phi$ is the other angle needed to define uniquely the direction. From formula (46) the receding velocity of typical particles in the fluid congruence is independent of the direction iff $n = 0$, that is to say, iff the model is FLRW. This is a logical result. For $n \neq 0$ the receding velocity does not depend on $\Phi$. This is a very interesting property which is *not* obvious in principle, because particles outside the axis of symmetry could see different receding rates depending on whether they look towards the axis or away from it. However, this does not happen, which means that, at any possible event, all directions with the same $\Theta$ are equivalent from this particular *local* point of view. On the other hand, neighbouring particles recede in expanding epochs for a given direction iff $1 - n \cos(2\Theta) > 0$, so that for $0 \leq n \leq 1$ all fluid particles recede independently of the direction. But, when $n > 1$, there always exist a set of directions in which the particles do not recede but rather come closer to each other. This set of directions constitute the solid interior of a double cone with vertex at each point, $\vec{e}_3$ as axis, and an opening angle of $\frac{1}{2} \arccos(1/n)$ [194]. The relation of this with the observations of red- or blue-shifts is not immediate, because the shift parameter should be computed using the null geodesics of the metric and formula (43). In any case, the value of $n$ has definite and precise implications concerning observation, so that these models are *testable* in order to see if they are realistic or not.

Another remark is important concerning this family and regarding inflation. A simple calculation for the deceleration parameter $q$ defined by

$$u^\mu \partial_\mu \left(\frac{1}{\theta}\right) \equiv \frac{1}{3}(1+q),$$

gives

$$q(t) = \frac{1}{n+3}\left(2n + 3 - 3\frac{T\,T_{,tt}}{T_{,t}^2}\right).$$

Then, it is readily proved that $q$ is always negative for a period after the little bang of the non-singular subclass. The duration of inflation depends on the anisotropy of the model through $n$. For FLRW models ($n = 0$), inflation typically lasts forever and SEC is violated. The non-FLRW non-singular models have a *finite* duration for inflation, which is shorter for bigger $n$, and they can also satisfy all energy conditions. For instance, the singularity-free model of subsection 7.6 is inflationary after the little bang and for a finite period given by $0 < \sinh(at) < 1$. This seems to be a generic property of singularity-free models [4,53,54,56,101,138,139,145,158,178,247] which perhaps has not received the attention it deserves, as there is no need to violate SEC: resort to the spatially inhomogeneous models so that a finite duration of inflation with smooth exit can be achieved without violating SEC nor DEC and, for the same token, one can also have no singularity at all.

The general family of this subsection supports the view that the singularity-free spacetimes deserve the name of classical cosmological models. The *only* basic concepts of standard cosmology, that is to say, the scale factor and the curvature index, are here fully represented by $M$ and the arbitrary function $T$. Other parameters appear, specially $n$ measuring the 'separation' of any particular model in the family with respect to its FLRW analogue, which may be defined by the same $T$ and $M$ with $n = 0$. Questions such as the instability of the singularity-free models or their hypothetical zero-measure in the space of metrics loss their meaning here *unless* they are raised against FLRW models too.

Can these classical non-singular cosmological models be realistic? This is the fundamental question, of course, and a possible answer has been schematically put forward in [194]. The idea is that the little bang (that is, the *regular* rebound) may be used as a substitute for the traditional big bang. Thus, the history of the Universe can be explained exactly in the same way as usual, because all that is needed is a very hot and highly dense expanding era. The parameters of the models can be so chosen that all matter decomposes into its elementary constituents in the collapsing epoch previous to the little bang, and then the formation of light nuclei as well as the existence of the CMBR can be explained without any difficulty. It remains, of course, the small anisotropy in the spectrum of the CMBR and the formation of structure. Concerning the former, it seems that the models are not realistic, but this depends on the integral (43) and is not clear in principle. Concerning the latter, these simple models can only help partially, as matter at least concentrates near the axis of symmetry. In any case, one can always speculate as follows: generalize the family presented in this section by letting $n(t, \rho)$. Then, the only new component of $T_{\mu\nu}$ is $T_{01}$, which can be interpreted as an energy flux in the radial direction. These processes of energy transport may lead to the homogeneization of some regions ($n = 0$ there) and to the bigger inhomogeneity of some other zones ($n > 3$) at present time. In this case, one can achieve a FLRW region which is big enough to accomodate present observations, and at the same time sufficiently small that no closed trapped surface appears. This possibility was shown to be feasible explicitly in subsection 7.2. The profile for $n$ near present time does not have to be maintained indefinitely into the past because $n$ depends also on $t$, and thus one can change $n$ smoothly such that $n \geq 3$ at the regular little bang. In this way, one can also have the realistic equation of state $p = \varrho/3$ near the little bang by choosing $n = 3$ there. Furthermore, the model will be automatically inflationary for a finite period after the little bang. And, of course, it will be singularity-free.

Therefore, in my opinion these models prove that the possibility of constructing singularity-free *realistic* cosmological models has not been ruled out at all by the singularity theorems nor by the combination of observations with them. Of course, other more general and sophisticated models should do the job, but this simple family shows the way one must follow in searching for those realistic non-singular models. A final objection can be and has been raised concerning these explicit models. Hitherto, all of them have cylindrical symmetry so that it might seem that the non-singular character of the spacetimes is a consequence of this symmetry. To answer this difficulty, a simple cosmological model with spherical symmetry is presented in the next subsection. In my view, this may help to settle this final problem definitely.

### 7.8. A spherically symmetric singularity-free model

Some spherically symmetric perfect-fluid models with no big-bang singularity have already been presented in subsection 7.5. In this subsection, a spherically symmetric cosmological singularity-free model recently found in [53] will be shown. Of course, there many known regular spherically symmetric spacetimes [24–26,123], but some of them are static and the rest may be interpreted as the interior of finite bodies for they can be smoothly matched to the Schwarzschild vacuum exterior. However, there was no reason for the non-existence of classical cosmological spherically symmetric (non-FLRW) models without singularity, because the Raychadhuri–Komar Theorem 5.1 cannot be applied to them as those models have non-zero acceleration in general. The possibility of existence of such models was emphasized several times (see e.g. Ref. 195), and Dadhich has been recently successful in constructing a very simple one [53]. It seems that this work is — at the time of writing this report — being generalized to a general family of solutions with better properties [55].

The manifold is $\mathbb{R}^4$ in spherical coordinates $\{t, r, \vartheta, \varphi\}$ and the line-element takes the remarkable simple form

$$ds^2 = -[1 + a(t)r^2]dt^2 + \frac{1 + 2a(t)r^2}{1 + a(t)r^2}dr^2 + r^2(d\vartheta^2 + \sin^2\vartheta d\varphi^2),$$

where $a(t)$ is an arbitrary function of $t$ which is assumed to be positive everywhere. The computation of the energy-momentum tensor in the orthonormal cobasis $\theta^\mu \propto dx^\mu$ gives the following non-zero components

$$T_{00} = a\frac{3 + 2ar^2}{(1 + 2ar^2)^2}, \quad T_{01} = \frac{ra_{,t}}{(1 + ar^2)(1 + 2ar^2)^{3/2}}, \quad T_{11} = \frac{a}{1 + 2ar^2},$$

$$T_{22} = T_{33} = T_{11} + r^2\frac{(3 + 5ar^2)r^2a_{,t}^2 - a_{,tt}(1 + ar^2)(1 + 2ar^2)}{2(1 + ar^2)^3(1 + 2ar^2)^2}.$$

This matter content can be interpreted in several different ways, such as a fluid with anisotropic pressures and no energy flux, or a perfect fluid plus heat flux, or a combination of perfect fluid plus null radiation (see e.g. Ref. 74). The particular interpretation is of no importance for our purposes as long as the energy-momentum tensor satisfies the energy conditions. The spacetime is static for $a$ = const., and in this case it is a perfect-fluid solution found by Tolman [224]. It can be seen that SEC will hold whenever the following inequalities are fulfilled:

$$\frac{a_{,t}^2}{a^2} \leq \frac{4(1 + ar^2)^4}{r^2(1 + 2ar^2)},$$

$$\frac{2}{(1 + 2ar^2)^2(1 + ar^2)}\sqrt{4a^2(1 + ar^2)^4 - a_{,t}^2r^2(1 + 2ar^2)} + 2T_{22} \geq 0.$$

There are many possible choices for $a$ such that these conditions are satisfied. In fact, $a$ can be so chosen that these inequalities hold and at the same time the metric has no matter singularity. Inspection of the above displayed formulas show that this is feasible for a wide variety of possibilities as long as $a$ remains positive, bounded and well-behaved. Some particular choices can be found in [53,55]. Still, there could appear some curvature singularity in the Weyl tensor. However, this is not so because the only surviving independent component of the Weyl tensor ($\Psi_2 = C_{0202}$, Refs. 100,123,150) has the form

$$\Psi_2 = -\frac{2a^2 r^2}{3(1+2ar^2)^2} + r^2 \frac{(3+5ar^2)r^2 a_{,t}^2 - a_{,tt}(1+ar^2)(1+2ar^2)}{2(1+ar^2)^3(1+2ar^2)^2},$$

which obviously remains also finite if the energy-momentum is finite. The curvature components reach their maximum values at $r = 0$ and fall off like at least $r^{-2}$ for big values of $r$.

Therefore, the spacetime satisfies the strict SEC (whence the generic condition by Propostions 2.5 and 2.6), is globally hyperbolic and, as can be checked, is g-complete. This spacetime has no closed trapped surfaces, no point with reconverging horismos and no edgeless compact slices. This last part is obvious because $r$ reaches arbitrarily large values. The absence of reconverging horismos follows because the family of radial null geodesics has an expansion proportional to $1/r$, which cannot change sign. Finally, the easiest way to see the non-existence of closed trapped 2-spheres is by computing the mass function $M(t,r)$, which should be less than $r/2$. The explicit computation gives

$$2M(t,r) = \frac{ar^3}{1+2ar^2} \implies 1 - \frac{2M}{r} = \frac{1+ar^2}{1+2ar^2} > 0.$$

Consequently, the singularity theorems are not applicable again because of the failure of the boundary condition.

A final question remains. It could happen that this model were not cosmological in the sense that it could describe a finite compact object. For this to happen, it would be necessary the existence of a timelike hypersurface $\sigma$ across which the spacetime could be matched to the Schwarzschild exterior solution. However, the Israel conditions [113,143] would demand then that the energy-momentum tensor on $\sigma$ were such that $T_{\mu\nu}n^\mu|_\sigma = 0$, where **n** is the spacelike normal one-form to $\sigma$. Taking into account WEC and the above properties of the spacetime, this would require inevitably the following condition on $\sigma$:

$$T_{00}T_{11} - T_{01}^2|_\sigma = 0 \implies \left(\frac{a_{,t}^2}{a^2} - \frac{(3+2ar^2)(1+ar^2)^2}{r^2}\right)\bigg|_\sigma = 0.$$

Fortunately, this still leaves some room to choose $a(t)$ such that this condition is impossible and all the above conditions are maintained.

Therefore, this is a family of classical cosmological models with spherical symmetry and no singularity. It has some nice features, as for instance its perfect fluid character *at the centre* $r = 0$ with a realistic equation of state $p = \varrho/3$ there. Other improved versions of this type of spacetimes are currently under investigation [55]. It would perhaps be interesting to study their properties, because they can certainly be also realistic, the spacetime being isotropic around $r = 0$. Furthermore, it would also be nice to have such a model with a perfect fluid content all over. In principle, they are allowed, but there seems to be no example available so far in the literature.

### 7.9. A final curious example

As a final curiosity, let us give *yet another* possible extension of Schwarzschild spacetime (Example 3.3). This extension is completely regular and has closed trapped 2-spheres (it violates SEC, as must happen due to the result in subsection 7.1), but nonetheless illustrates the fact that the structure of the extensions may be quite complicated even in simple cases. The model was found in [140] and its global maximal extension and Penrose conformal diagram cannot be shown in a single sheet of paper.[16]

In Eddington–Finkelstein-like coordinates, and using the same line element of the non-singular black hole of Example 4.3, one can choose the functions $m(r)$ and $\beta(r)$ such that $\beta = 0$ and $m = M$ for all $r \geq 2M$, so that the spacetime coincides with Schwarzschild vacuum solution for all $r \geq 2M$. This was already done in the explicit Example 4.3. However, the continuation of $m(r)$ and $\beta(r)$ can be chosen such that now there appear two (or more!) disconnected regions containing trapped 2-spheres. A particular completely explicit case satisfying WEC and with two such regions can be found in [140]. In this case, there exist four values $r_3 < r_2 < r_1 < 2M$ where $2m(r) = r$ and the spacetime is regular everywhere, in particular at $r = 0$, so that the 2-spheres with constant $t$ and $r$ are trapped if either $r_1 < r < 2M$ or $r_3 < r < r_2$. The corresponding Penrose diagram is drawn in Figure 16, where there still appear incomplete null geodesics reaching $t \to \pm\infty$ with finite proper time and $r$ tending to one of the values $2M$, $r_1$, $r_2$ or $r_3$. One of its possible maximal extensions is presented in Figure 17, so that the resulting spacetime is b-complete. Again, there is topology change but the extension is not so simple as before, given the more complicated structure of horizons of the interior black hole region and, in fact, the complete Penrose diagram cannot be drawn in a single sheet of paper.

In summary, in this paper there are at least *eleven* completely inequivalent extensions of the original Schwarzschild vacuum spacetime of

---

[16] This global regular extension was constructed together with my former student M. Mercè Martín-Prats.

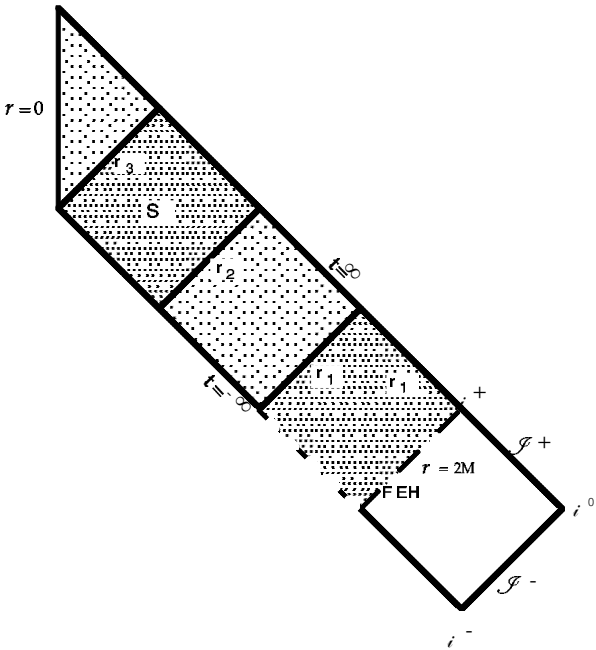

Figure 16. Penrose's conformal diagram of the spherically symmetric spacetime of subsection 7.9. There is an asymptotically flat Schwarzschild region with event horizon at $r = 2M$, and a matter-filled interior (shadowed) zone with three other null hypersurfaces $r = r_1, r_2, r_3$ such that $2m(r_i) = r_i$. Also, $r = 0$ is the origin of coordinates and there is no singularity there. Every point of the doubly-shadowed 'squares' represents a closed trapped 2-sphere, such as the $S$ shown. The null geodesics reach the $r_i$ with $t \to \pm\infty$ but finite affine parameter, and thus they are incomplete. This spacetime is extendible to the future and past across $t \to \pm\infty$, and a possible maximal extension is given in the next figure.

Example 3.3, namely, the Eddington–Finkelstein extensions (advanced and retarded), which are themselves extendible (Example 3.3); the Kruskal extension combining the two previous ones (Example 3.3, Fig. 1); two extensions by matching with an interior star, one with singularities and the other regular (Example 3.3, Fig. 2); the unorthodox singular extension of Example 3.3; another one by joining Eddington–Finkelstein diagram with Vaidya spacetime (Example 3.5, Fig. 4); the extendible extension satisfying WEC (Example 4.3, Fig. 11) and its own maximal and singularity-free extension (Fig. 12); and finally, the two new extensions shown in this subsection (Figs. 16 and 17). This should make it completely clear that the problem of singularities is unavoidably mixed with the question of how, why, when and to where a given incomplete spacetime can be extended.

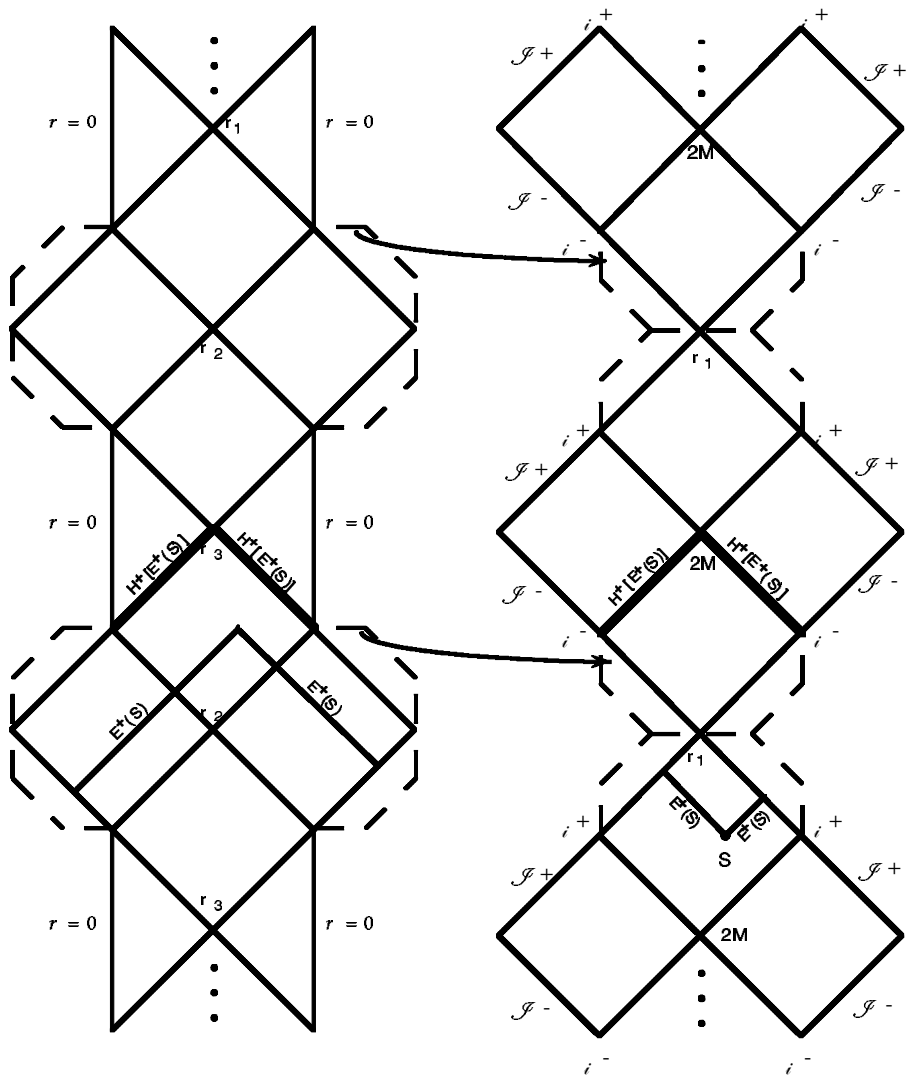

Figure 17. This is the Penrose conformal diagram of a maximal extension of the spacetime in Fig. 16. This conformal diagram cannot be shown in a single sheet of paper, but a 3-dimensional model with the identifications as indicated is given. The whole spacetime is b-complete and SEC is violated. The 2-sphere $S$ is a closed trapped surface and its $E^+(S)$ is shown. This $E^+(S)$ is compact. Its Cauchy horizon, which is not connected in this case, is also explicitly shown. This is yet another inequivalent extension of the asymptotically flat Schwarzschild spacetime.

## ACKNOWLEDGEMENTS

First of all, I would like to thank Marc Mars for many interesting suggestions and for several conversations which have been very useful for this work. I am also grateful to him for a careful and detailed reading of the manuscript and for correcting many errors. He has also supplied some important references. Comments from Lluís Bel and Jesús Martín have also been helpful. Other valuable suggestions came from Naresh Dadhich, Paco Fayos and Malcolm MacCallum. I am indebted to Leonardo Fernández-Jambrina for providing a copy of [12]. Some help from Raül Vera, Ruth Lazkoz and Carlos F. Sopuerta is acknowledged. I thank the referees for many comments and remarks.